# Lecture notes in fluid mechanics

Laurent Schoeffel, CEA Saclay

These lecture notes have been prepared as a first course in fluid mechanics up to the presentation of the millennium problem listed by the Clay Mathematical Institute. Only a good knowledge of classical Newtonian mechanics is assumed. We start the course with an elementary derivation of the equations of ideal fluid mechanics and end up with a discussion of the millennium problem related to real fluids. With this document, our primary goal is to debunk this beautiful problem as much as possible, without assuming any previous knowledge neither in fluid mechanics of real fluids nor in the mathematical formalism of Sobolev's inequalities. All these items are introduced progressively through the document with a linear increase in the difficulty. Some rigorous proofs of important partial results concerning the millennium problem are presented. At the end, a very modern aspect of fluid mechanics is covered concerning the subtle issue of its application to high energetic hadronic collisions.

## §1. Introduction

Fluid mechanics concerns the study of the motion of fluids (in general liquids and gases) and the forces acting on them. Like any mathematical model of the real world, fluid mechanics makes some basic assumptions about the materials being studied. These assumptions are turned into equations that must be satisfied if the assumptions are to be held true. Modern fluid mechanics, in a well-posed mathematical form, was first formulated in 1755 by Euler for ideal fluids.

Interestingly, it can be shown that the laws of fluid mechanics cover more materials than standard liquid and gases. Indeed, the idea of exploiting the laws of ideal fluid mechanics to describe the expansion of the strongly interacting nuclear matter that is formed in high energetic hadronic collisions was proposed in 1953 by Landau. This theory has been developed extensively in the last 60 years and is still an active field of research. This gives a very simple 3-steps picture of a non-trivial phenomenon observed in hot nuclear matter after the collision of high energetic heavy ions, composed of a large collection of charged particles.

(i)

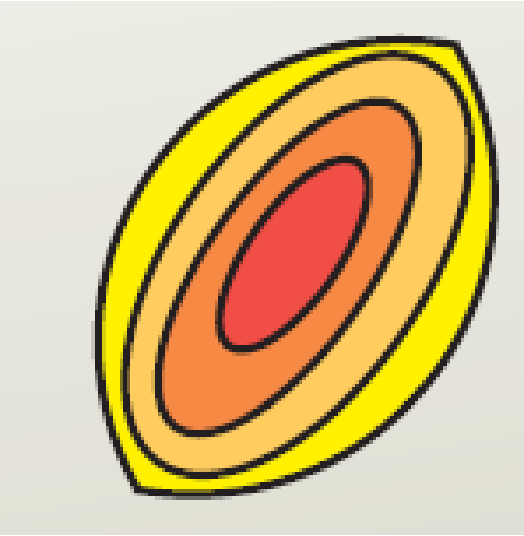

After the collision a nuclear medium, a zone of high density of charges, is formed with high pressure in the middle (center of the collision).

(ii)

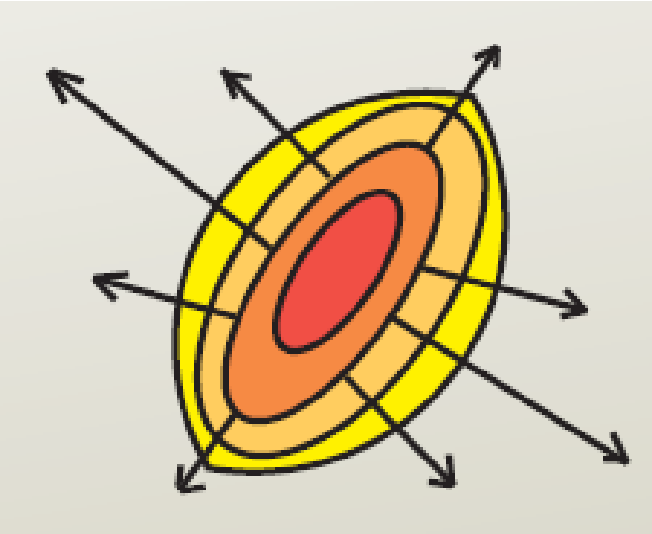

According to the laws of fluid mechanics, as we shall prove them, this implies that an acceleration field is generated from high pressures to low pressures.

(iii) 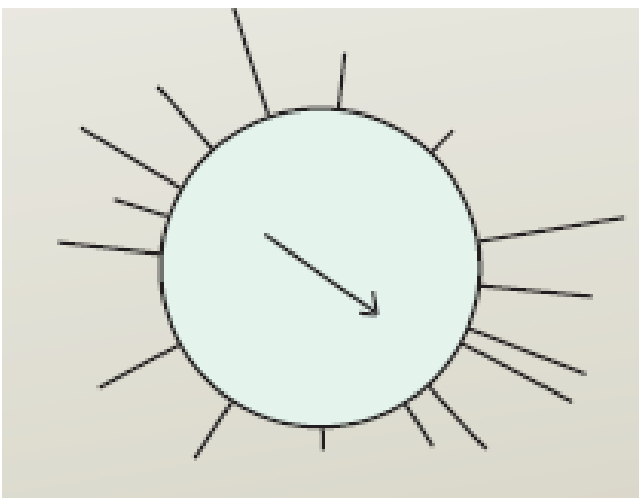 This implies that particles will flow in a certain transverse direction, as indicated on the figure. This is known as the transverse flow property, well established experimentally.

We come back on these ideas and their developments in the last section of this document. It requires a relativistic formulation of fluid mechanics. Up to this section, we always assume that the dynamics is non-relativistic.

**§2. Continuum hypothesis**

Fluid mechanics is supposed to describe motion of fluids and related phenomena at macroscopic scales, which assumes that a fluid can be regarded as a continuous medium. This means that any small volume element in the fluid is always supposed so large that it still contains a very great number of molecules. Accordingly, when we consider infinitely small elements of volume, we mean very small compared with the volume of the body under consideration, but large compared with the distances between the molecules. The expressions fluid particle and point in a fluid are to be understood in this sense. That is, properties such as density, pressure, temperature, and velocity are taken to be well-defined at infinitely small points.

These properties are then assumed to vary continuously and smoothly from one point to another. Consequently, the fact that the fluid is made up of discrete molecules is ignored. If, for example, we deal with the displacement of some fluid particle, we do mean not the displacement of an individual molecule, but that of a volume element containing many molecules, though still regarded as a point in space. That's why fluid mechanics is a branch of continuum mechanics, which models matter from a macroscopic viewpoint without using the information that it is made out of molecules (microscopic viewpoint).

## §3. Mathematical functions that define the fluid state

Following the continuous assumption, the mathematical description of the state of a moving fluid can be characterized by functions of the coordinates x, y, z and of the time t. These functions of (x, y, z, t) are related to the quantities defined for the fluid at a given point (x, y, z) in space and at a given time t, which refers to fixed points in space and not to fixed particles of the fluid. For example, we can consider the mean local velocity **v**(x, y, z, t) of fluid particles or fluid points, called the fluid velocity, and two thermodynamic quantities that characterize the fluid state, the pressure p(x, y, z, t) and the mass density ρ(x, y, z, t), the mass per unit volume of fluid. Following the discussion of §2, two remarks can be done at this stage:

i. The fluid is assumed to be a continuum. This implies that all space-time derivatives of all dependent variables exist in some reasonable sense. In other words, local properties such as density pressure and velocity are defined as averages over elements large compared with the microscopic structure of the fluid but small enough in comparison with the scale of the macroscopic. This allows the use of differential calculus to describe such a system.

ii. All the thermodynamic quantities are determined by the values of any two of them, together with the equation of state. Therefore, if we are given five quantities, namely the three components of the velocity **v**, the pressure p and the mass density ρ, the state of the moving fluid is completely determined. We recall that only if the fluid is close to thermodynamic equilibrium, its thermodynamic properties, such as pressure, density, temperature are well-defined. This requires (as a very former hypothesis) that local relaxation times towards thermal equilibrium are much shorter than any macroscopic dynamical time scale. In particular, microscopic collision time scale (between elementary constituents of the fluid) needs to be much shorter than any macroscopic evolution time scales. This hypothesis is almost a tautology for standard fluids build up by molecules at reasonable density, but becomes not trivial in the case of some hot dense matter state created in high energetic hadronic collisions.

In the following, we prove that these five unknown quantities describe completely the case of what we define as **ideal fluids**, in which we take no account of processes of energy dissipation. Energy dissipation may occur in a moving fluid as a consequence of internal friction (or viscosity) within the fluid and heat exchange between different parts of it. Neglecting this phenomenon, we can find a set of five equations that are sufficient obtain a closed system: 5 equations for 5 unknown quantities. Interestingly, we can gain some intuition about the behavior of the ideal flow by expressing in more details its pressure field. An ideal fluid, in particular, is characterized by the assumption that each particle pushes its neighbors equally in every direction. This is why a single scalar quantity, the pressure, is sufficient to describe the force per unit area that a particle exerts on all its neighbors at a given time. Also, we know that a fluid particle is not accelerated if its neighbors push back with equal force, which means that the acceleration of the fluid particle results from the pressure differences. In short, the pressure force can be seen as a global interaction of all fluid particles.

When the energy dissipation inside the fluid is not neglected, we need to consider also the internal energy density $e(x, y, z, t)$ and heat flux density $q(x, y, z, t)$ as four additional unknown functions to be determined by a proper set of closed equations: nine equations are needed in such cases. This concerns what we define as **real fluids**. We discuss the case of real fluids in more details later in the document. However, a few intuitive arguments can be made with no mathematical formalism. When the energy dissipation is not neglected, this means that we take into account frictional forces inside the fluid. Their main effect is that they enhance the local coherence of the flow. They counteract at each point the deviation of the velocity field from its local average. This means that if a fluid particle moves faster than the average of its neighbors, then friction slows it down.

When fluid quantities are defined at given fixed points $(x, y, z)$ in space and at a given time $t$, we speak of the *Eulerian* description of the fluid. When fluid quantities are defined as associated to a (moving) particle of fluid, followed along its trajectory, we speak of the *Lagrangian* point of view. An important notion can be derived from this last view point: the flow map.

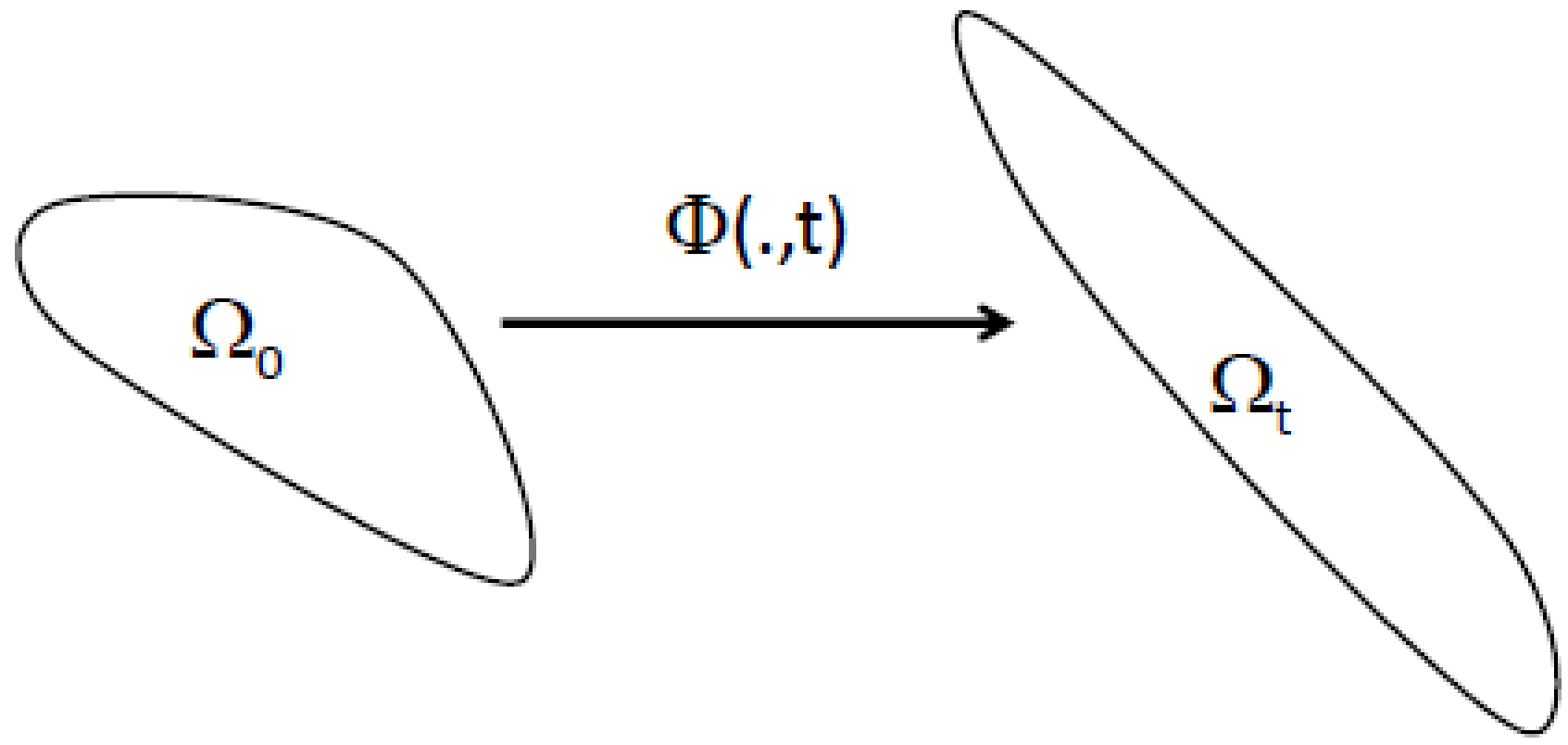

A fluid particle moving through the fluid volume is labeled by the (vector) variable **X** (defined at t=0). At some initial time, we define a subset of fluid particles (of the entire fluid volume) $\Omega_0$. The fluid particles of this subset will move through the fluid within time. We introduce a function $\boldsymbol{\Phi}(\mathbf{X}, t)$ that describes the change of the particle position from the initial time up to $t > 0$. The function $\boldsymbol{\Phi}(.,t)$ is itself a vector, with 3 coordinates corresponding to the 3 coordinates of space $(\Phi_x, \Phi_y, \Phi_z)$. This means that we can denote the position of any particle of fluid at time t by $\boldsymbol{\Phi}(\mathbf{X}, t)$ which starts at the position **X** at t=0. Then, we can label the new subset of fluid particles at time t, originally localized in $\Omega_0$, as:

$$\Omega_t = \{\boldsymbol{\Phi}(\mathbf{X}, t) : \mathbf{X} \text{ belongs to } \Omega_0\}.$$

Stated otherwise, the function $t \rightarrow \boldsymbol{\Phi}(\mathbf{X}, t)$, represents the trajectory of particles: this is what we also call the flow map. In particular, the particle velocity is given by:

$$\mathbf{v}(x, y, z, t) = \frac{\partial \boldsymbol{\Phi}}{\partial t}(\mathbf{X}, t) \text{ with } \boldsymbol{\Phi}(\mathbf{X}, t = 0) = \mathbf{X}.$$

We can think of $\Omega_t$ as the volume moving with the fluid.

## §4. Limits of the continuum hypothesis

According to the continuous assumption (§2), the physical quantities (like velocity, pressure and density) are supposed to vary smoothly on macroscopic scales. However, this may not be the case everywhere in the flow. For example, if a shock front of the density appears at some values of the coordinates at a given time, the flow would vary very rapidly at that point, over a length of order the collision mean free path of the molecules. In this case, the continuum approximation would be only piecewise valid and we would need to perform a matching at the shock front. Also, if we are interesting by scale invariant properties of fluid in some particular cases, we need to keep in mind that there is a scale at which the equations of fluid mechanics break up, which is the molecular scale characterized by the mean free path of molecules between collisions. For example, for flows where spatial scales are not larger than the mean distance between the fluid molecules, as for example the case of highly rarefied gazes, the continuum assumption does not apply.

## §5. Closed set of equations for ideal fluids

The derivation of equations underlying the dynamics of ideal fluids is based on three conservation principles:

i. Conservation of matter. Matter is neither created nor destroyed provided there is no source or sink of matter;
ii. Newton's second law or balance of momentum. For a fluid particle, the rate of change of momentum equals the force applied to it;
iii. Conservation of energy.

In turn, these principles generate the five equations we need to describe the motion of an ideal fluid: (i) Continuity equation, which governs how the density of the fluid evolves locally and thus indicates compressibility properties of the fluid; (ii) Euler's equations of motion for a fluid element which indicates how this fluid element moves from regions of high pressure to those of low pressure; (iii) Equation of state which indicates the mechanism of energy exchange inside the fluid.

We derive first the expression for the conservation of mass (i). Consider a fluid of mass density ρ, fluid particle velocity $\mathbf{v}$ and some volume Ω, fixed in space (i.e. fixed in some Newtonian reference frame). The mass of the fluid in this volume is $\int \rho dV$, where the integration is taken on the volume Ω. If the fluid moves, then there is a flow of mass across each element of surface $d\mathbf{\Sigma}$ on the boundary of the volume Ω, where the magnitude of the vector $d\mathbf{\Sigma}$ is equal to the area of the surface element and its direction is along the normal to the surface. Provided that there are no sources or sinks of fluid, the elementary mass of fluid flowing in unit time through an element $d\mathbf{\Sigma}$ of the surface bounding Ω is $\rho\mathbf{v}.d\mathbf{\Sigma}$. By convention $d\mathbf{\Sigma}$ is taken along the outward normal, which means that $\rho\mathbf{v}.d\mathbf{\Sigma}$ is positive when the fluid is flowing out of the volume and negative for a flow into the volume Ω.

The total mass of fluid flowing out of the volume in unit time is thus: $\oint \rho\mathbf{v}.d\mathbf{\Sigma}$, where the integration is taken over the whole of the closed surface bounding Ω. Therefore, we can write:

$$\frac{\partial}{\partial t}\int \rho dV = -\oint \rho\mathbf{v}.d\mathbf{\Sigma},$$

where $-\frac{\partial}{\partial t}\int \rho dV$ is the net decrease of the mass of fluid in Ω per unit time. Using the Green's formula to express $\oint \rho\mathbf{v}.d\mathbf{\Sigma}$ as a volume integral over Ω: $\int \mathrm{div}(\rho\mathbf{v})dV$. We obtain:

$$\int [\frac{\partial \rho}{\partial t} + \mathrm{div}(\rho\mathbf{v})]dV = 0.$$

Since this equation must hold for any volume, the integrand must vanish. This gives:

$$\frac{\partial \rho}{\partial t} + \mathrm{div}(\rho\mathbf{v}) = 0.$$

(1)

This is the continuity equation, the first fundamental equation of fluid mechanics. The vector $\mathbf{j} = \rho\mathbf{v}$ is called the mass flux density. Its direction is along the motion of the fluid and its magnitude equals the mass of the fluid flowing per unit time though a unit area perpendicular to the velocity of the fluid.

By expanding $\mathrm{div}(\rho\mathbf{v})$ as $(\mathbf{v}.\mathbf{grad})(\rho) + \rho\mathrm{div}(\mathbf{v})$, we can also write this equation as:

$$\frac{\partial \rho}{\partial t} + (\mathbf{v}.\mathbf{grad})(\rho) = \left[\frac{\partial}{\partial t} + (\mathbf{v}.\mathbf{grad})\right](\rho) = -\rho\mathrm{div}(\mathbf{v}).$$

(1′)

We identify the operator $[\frac{\partial}{\partial t} + (\mathbf{v}.\mathbf{grad})]$ that we define below as the material derivative, or the derivative following the flow.

Before developing some consequences of the continuity equation, we establish the Newton's second law (ii) for some volume $\Omega$ of the fluid, assumed to be ideal, characterized by its mass density, pressure and fluid particle velocity. In the absence of external force, by definition of the pressure, the total force acting on the ideal fluid in volume $\Omega$ is equal to: $-\oint \mathrm{p}d\mathbf{\Sigma}$.

This last formula represents the integral of the pressure taken over the surface bounding the volume, with similar conventions as previously defined for the surface element $d\mathbf{\Sigma}$. This surface integral can be transformed to a volume integral over $\Omega$: $-\oint \mathrm{p}d\mathbf{\Sigma} = -\int \mathbf{grad}(\mathrm{p})dV$.

Thus, the fluid surrounding any elementary volume dV exerts a force $-dV\,\mathbf{grad}(\mathrm{p})$ on that element. Moreover, from Newton's second law applied to this elementary volume dV, the mass times the acceleration equals: $\rho dV \frac{d\mathbf{v}}{dt}$.

Then, the Newton's second law of motion for the fluid per unit volume reads:

$$\rho \frac{d\mathbf{v}}{dt} = -\mathbf{grad}(\mathrm{p}).$$

(2)

Here, we need to be careful with the mathematical expression $d\mathbf{v}/dt$. It does not represent (only) the rate of change of the fluid velocity at a fixed point in space, which would be mathematically written as $\partial\mathbf{v}/\partial t$.

Rather, d$\mathbf{v}$/dt is the rate of change of the velocity of a fluid particle as it moves in space (see §2), called the material derivative, namely:

$$\frac{d\mathbf{v}}{dt} = \frac{[\mathbf{v}(x + dx, y + dy, z + dz, t + dt) - \mathbf{v}(x, y, z, t)]}{dt}.$$

This can be developed as:

$$\frac{d\mathbf{v}}{dt} = \frac{dx}{dt}\frac{\partial \mathbf{v}}{\partial x} + \frac{dy}{dt}\frac{\partial \mathbf{v}}{\partial y} + \frac{dz}{dt}\frac{\partial \mathbf{v}}{\partial z} + \frac{\partial \mathbf{v}}{\partial t} = (\mathbf{v}.\mathbf{grad})(\mathbf{v}) + \frac{\partial \mathbf{v}}{\partial t}.$$

There are two terms in the expression of $d\mathbf{v} = \frac{d\mathbf{v}}{dt}.dt$: the difference between the velocities of the fluid particle at the same instant in time at two points distant of (dx, dy, dz), which is the distance moved by the fluid particle during dt, and the change during dt in the velocity at a fixed point in space (x, y, z). Combining the vector equation (2) and the expression of the material derivative, we get:

$$\frac{\partial \mathbf{v}}{\partial t} + (\mathbf{v}.\mathbf{grad})(\mathbf{v}) = -\frac{1}{\rho}\mathbf{grad}(p).$$

(3)

This vector equation (3) represents a set of three equations (in three dimensions of space) that describe the motion of an ideal fluid, first obtained by Euler in 1755. That's why it is called the Euler's equations, the second fundamental set of equations of fluid mechanics (ii). It is trivial to expand the vector equation (3) on the 3 Cartesian coordinates of space (x, y, z) as a set of 3 equations (in a compact form):

$$\left(\frac{\partial}{\partial t} + v_x\frac{\partial}{\partial x} + v_y\frac{\partial}{\partial y} + v_z\frac{\partial}{\partial z}\right)\begin{bmatrix} v_x \\ v_y \\ v_z \end{bmatrix} = -\frac{1}{\rho}\begin{bmatrix} \partial p/\partial x \\ \partial p/\partial y \\ \partial p/\partial z \end{bmatrix}.$$

If external forces have to be considered these equations become:

$$\frac{\partial \mathbf{v}}{\partial t} + (\mathbf{v}.\mathbf{grad})(\mathbf{v}) = -\frac{1}{\rho}\mathbf{grad}(p) + \frac{1}{\rho}\mathbf{F}_{ext}.$$

Here, $\mathbf{F_{ext}}$ can be for example the gravity force $\rho\mathbf{g}$.

Before continuing the derivation of fundamental equations of fluid mechanics, we give some hints on how Euler's and continuity equations can be derived using the notion of flow map (§4). As we have seen, the flow map is a function $\mathbf{\Phi}(\mathbf{X}, t)$ that describes the change of the fluid particle position $\mathbf{X}$ at initial time (t=0) up to $t > 0$. Then, we have established the following relations (§4):

$$\mathbf{v}(x, y, z, t) = \frac{\partial \mathbf{\Phi}}{\partial t}(\mathbf{X}, t) \text{ with } \mathbf{\Phi}(\mathbf{X}, 0) = \mathbf{X}.$$

This means that the acceleration of a fluid particle can be written as:

$$\frac{d}{dt}\mathbf{v}(\mathbf{x}, t) = \frac{d}{dt}\mathbf{v}(\mathbf{\Phi}(\mathbf{X}, t), t) = \frac{\partial \mathbf{v}}{\partial t} + \left\{\frac{\partial v_x}{\partial x}\frac{\partial \phi_x}{\partial t} + \cdots\right\}.$$

We rewrite the last term as:

$$\left\{\frac{\partial v_x}{\partial x}\frac{\partial \phi_x}{\partial t} + \cdots\right\} = \left\{\frac{\partial \phi_x}{\partial t}\frac{\partial}{\partial x}v_x + \frac{\partial \phi_y}{\partial t}\frac{\partial}{\partial y}v_y + \frac{\partial \phi_z}{\partial t}\frac{\partial}{\partial z}v_z\right\} = (\mathbf{v}.\mathbf{grad})(\mathbf{v}).$$

We obtain:

$$\frac{d}{dt}\mathbf{v}(\mathbf{x}, t) = \frac{d}{dt}\mathbf{v}(\mathbf{\Phi}(\mathbf{X}, t), t) = \frac{\partial \mathbf{v}}{\partial t} + (\mathbf{v}.\mathbf{grad})(\mathbf{v}).$$

We find again the derivative of the velocity following the flow that leads to Euler's vector equations. Similarly, we can use the notion of flow map to write the conservation of mass and then the continuity equation. We consider fluid particles $\mathbf{X}$ initially localized in a subset $\Omega_0$ of the entire fluid volume. At time $t > 0$, they are contained in the subset (volume moving with the fluid):

$$\Omega_t = \{\mathbf{\Phi}(\mathbf{X}, t) : \mathbf{X} \text{ belongs to } \Omega_0\}.$$

Note that $\Omega_t$ is not fixed in space (i.e. not fixed in some Newtonian reference frame), but moving with the flow according to the flow map function.

The conservation of mass can then be written as:

$$\int_{\Omega_t} \rho(\mathbf{x}, t)dV = \int_{\Omega_0} \rho(\mathbf{X}, 0)dV.$$

Since the right-hand side is independent of the time, we can write:

$$\frac{d}{dt}\int_{\Omega_t} \rho(\mathbf{x}, t)dV = \frac{d}{dt}\int_{\Omega_t} \rho(\boldsymbol{\Phi}(\mathbf{X}, t), t)dV = 0.$$

In this expression, the time derivative represents the material derivative, following the movement of fluid particles. The situation is not that easy *a priori* as the domain of integration depends on the time in the above formula. It can be shown that the following relation holds:

$$\frac{d}{dt}\int_{\Omega_t} \rho(\boldsymbol{\Phi}(\mathbf{X}, t), t)dV = \int_{\Omega_t} [\frac{\partial \rho}{\partial t} + \mathrm{div}(\rho \mathbf{v})]dV.$$

Then, this leads to the continuity equation.

We come back to the Euler's equations (3). An important vector identity is the following:

$$\frac{1}{2}\mathbf{grad}(\mathbf{v}^2) = \mathbf{v} \times \mathbf{curl}(\mathbf{v}) + (\mathbf{v}.\mathbf{grad})(\mathbf{v}).$$

Then, equations (3) can be rewritten as:

$$\frac{\partial \mathbf{v}}{\partial t} + \frac{1}{2}\mathbf{grad}(\mathbf{v}^2) - \mathbf{v} \times \mathbf{curl}(\mathbf{v}) = -\frac{1}{\rho}\mathbf{grad}(p)$$

This expression of equations (3) has many interests that we shall see later. For example, in the case of constant mass density, taking the **curl** of this equation makes the gradients vanishing and we obtain a differential equation involving only the velocity field. Another interesting physical case appears when $\mathbf{curl}(\mathbf{v}) = 0$.

Then the velocity field can be written as the gradient of a scalar function and the above expression leads to an interesting simple equation.

In this section we have ignored all processes related to energy dissipation, which may occur in a moving fluid as a consequence of internal friction (or viscosity) in the fluid as well as heat exchange between different parts of the fluid. Thus we have treated only the case of **ideal fluids**, for which thermal conductivity and viscosity can be neglected. With the continuity equation, the Euler's equations make a set of 4 equations, for five quantities that characterize the ideal fluid (§2).

This means that we are missing one equation, which is coming with the last conservation principle, namely the conservation of energy (iii). The absence of heat exchange between the different parts of the fluid implies that the motion is adiabatic: thus the motion of an ideal fluid is by definition considered as adiabatic. In other words, the entropy of any fluid particle remains constant as that particle moves in space inside the fluid. We label the entropy per unit mass as s. We can easily write the condition for an adiabatic motion as:

$$\mathrm{ds}/\mathrm{dt} = 0.$$

This represents the rate of change of the entropy (per unit mass) of a fluid particle as it moves in the fluid. This can be reformulated as:

$$\frac{\partial \mathrm{s}}{\partial \mathrm{t}} + (\mathbf{v}.\mathbf{grad})(\mathrm{s}) = 0. \tag{4}$$

This expression is the general condition for adiabatic motion of an ideal fluid. This condition usually takes a much simpler form. Indeed, as it usually happens, the entropy is constant throughout some volume element of the fluid at some initial time, then it retains the same constant value everywhere in the fluid volume, at all times for any subsequent motion of the fluid. In this case, equation (4) becomes simply: s=constant.

Such a motion is said to be isentropic, which is what we assume in general for an ideal fluid, unless stated otherwise. This condition, together with an equation of state for the fluid provides a relation between the pressure and the mass density, and then the fifth equation: $p=p(\rho,s)$.

This allows us to know what happens to the density when pressure changes and consequently to close the system of equations describing the mechanics of ideal fluids: 5 equations for 5 variables (2 thermodynamic variables and the 3 coordinates of the velocity).

We rewrite the Euler's equations (3) in case of steady flow, for which the velocity is constant in time at any point occupied by the fluid:

$$\frac{\partial \mathbf{v}}{\partial t} = 0.$$

This means that the velocity field is a function only of the coordinates. Taking $\rho$ = constant for simplicity, we obtain:

$$\frac{1}{2}\mathbf{grad}(\mathbf{v^2}) - \mathbf{v} \times \mathbf{curl}(\mathbf{v}) = -\mathbf{grad}(p/\rho)$$

(3′)

Then, we define streamlines: the tangent to a streamline at any point gives the direction of the velocity at that point. Streamlines are thus defined by the set of equations:

$$\frac{dx}{v_x} = \frac{dy}{v_y} = \frac{dz}{v_z}.$$

One interest of steady flow is that the streamlines do not vary with time and thus coincide with the paths of the fluid particles. Obviously, this coincidence between streamlines and trajectory of fluid particles does not hold in non-steady flow: indeed, the tangents to the streamlines give the directions of the velocities of fluid particles as a function of the coordinates in space at a given instant, whereas the tangents to the path (trajectory) of a given fluid particle provide the direction of the velocities as a function of time.

It follows that the quantity $\mathbf{v^2}/2 + P/\rho$ is constant along a streamline, which coincides with the particle trajectory for steady flow. In general this constant is different with different streamlines: this is what is called the Bernoulli's equation.

Ideal fluids present an interesting property: mass and momentum conservation principles are uncoupled from energy conservation. Indeed, if we consider the entropy to be constant throughout the fluid, it is not required to consider explicitly the energy conservation to describe the motion of the fluid. We can show that the relation corresponding to energy conservation is a consequence of the continuity and Euler's equations under the condition of isentropic flow. We consider some volume element $\Omega$ of the fluid, fixed in space, and we find how the energy of the fluid contained in this element varies with time. In the absence of external force, the energy density, per unit volume of the fluid, can be written as:

$$U = \frac{1}{2}\rho\mathbf{v}^2 + \rho\epsilon.$$

The first term is the kinetic energy density and the second the internal energy density, noting $\epsilon$ the internal energy per unit mass.

The change in time of the energy contained in the volume element $\Omega$ is then given by the partial derivative with respect to time $\frac{\partial \int \mathrm{U}\mathrm{d}V}{\partial t}$, where the integration is taken over $\Omega$. Then, following a similar reasoning as for the continuity equation, we can write a general expression of energy conservation for the fluid in that volume element, namely:

$$\frac{\partial \int \mathrm{U}\mathrm{d}V}{\partial t} = -\oint \mathbf{F}.\mathrm{d}\mathbf{\Sigma} = -\int \mathrm{div}(\mathbf{F})\mathrm{d}V,$$

where **F** represents the energy flux density. We let as an exercise (below) to show that **F** is not equal to U**v**, but:

$$\mathbf{F} = \rho\mathbf{v}\left(\frac{1}{2}\mathbf{v}^2 + \epsilon + \frac{P}{\rho}\right) = \rho\mathbf{v}\left(\frac{1}{2}\mathbf{v}^2 + h\right).$$

With the notation $h = \epsilon + P/\rho$, which corresponds to the enthalpy of the fluid per unit mass. The proof uses only the continuity and Euler's equations. Since the equation of energy conservation must hold for any volume element, the integrand must vanish in:

$$\int [\frac{\partial U}{\partial t} + \mathrm{div}(\mathbf{F})] dV = 0.$$

We end up with the local expression of energy conservation:

$$\frac{\partial \rho(\frac{1}{2}\mathbf{v}^2 + \epsilon)}{\partial t} + \mathrm{div}\left(\rho \mathbf{v}(\frac{1}{2}\mathbf{v}^2 + h)\right) = 0.$$

The zero on the right hand side of this relation comes from the condition for adiabatic motion: $ds/dt = 0$, which is a necessary condition for an ideal fluid (see above).

Stated differently, if $ds/dt$ would be non-vanishing, this right hand side term would be necessarily proportional to: $ds/dt$. In fact, in the presence of heat flow within the fluid, which means that the fluid is not supposed to be ideal, the rate of heat density change reads: $\rho T \frac{ds}{dt}$, which leads to the general equation for non-vanishing $ds/dt$:

$$\frac{\partial \rho(\frac{1}{2}\mathbf{v}^2 + \epsilon)}{\partial t} + \mathrm{div}\left(\rho \mathbf{v}(\frac{1}{2}\mathbf{v}^2 + h)\right) = \rho T \frac{ds}{dt}.$$

<u>Exercise:</u> For an ideal fluid. Prove that the energy flux density can be written as: $\rho \mathbf{v}(\frac{1}{2}\mathbf{v}^2 + h)$ with $h = \epsilon + P/\rho$. Then, derive the local expression of the conservation of energy (per unit mass):

$$\frac{\partial \rho(\frac{1}{2}\mathbf{v}^2 + \epsilon)}{\partial t} + \mathrm{div}\left(\rho \mathbf{v}(\frac{1}{2}\mathbf{v}^2 + h)\right) = 0.$$

<u>Solution:</u> The idea is to compute the partial derivative $\frac{\partial \rho(\frac{1}{2}\mathbf{v}^2 + \epsilon)}{\partial t}$ using equations of fluid mechanics that we have established together with a thermodynamic relation involving the internal energy (per unit mass).

We write:

$$\frac{\partial \rho(\frac{1}{2}\mathbf{v}^2)}{\partial t} = \rho\mathbf{v}\frac{\partial \mathbf{v}}{\partial t} + \frac{1}{2}\mathbf{v}^2\frac{\partial \rho}{\partial t}.$$

In this identity, $\frac{\partial \rho}{\partial t}$ can be replaced by $-\mathrm{div}(\rho\mathbf{v})$ using the continuity equation and $\frac{\partial \mathbf{v}}{\partial t}$ is given by the Euler's equations. We obtain:

$$\frac{\partial \rho(\frac{1}{2}\mathbf{v}^2)}{\partial t} = \rho\mathbf{v}.[(\mathbf{v}.\mathbf{grad})\mathbf{v}] - \mathbf{v}.\mathbf{grad}(\mathrm{p}) - \frac{1}{2}\mathbf{v}^2\mathrm{div}(\rho\mathbf{v}).$$

With the vector identity: $\mathbf{v}.[(\mathbf{v}.\mathbf{grad})\mathbf{v}] = \frac{1}{2}\mathbf{v}.\mathbf{grad}(\mathbf{v^2})$ and

$\mathbf{grad}(\mathrm{p}) = \rho\mathbf{grad}(\mathrm{h}) - \rho\mathrm{T}\mathbf{grad}(\mathrm{s})$ (since dh=Tds+1/ρ dp), we obtain:

$$\frac{\partial (\frac{1}{2}\rho\mathbf{v}^2)}{\partial t} = -\rho\mathbf{v}.\left[\mathbf{grad}(\frac{1}{2}\mathbf{v^2} + \mathrm{h})\right] + \rho\mathrm{T}\mathbf{v}.\mathbf{grad}(\mathrm{s}) - \frac{1}{2}\mathbf{v}^2\mathrm{div}(\rho\mathbf{v}).$$

Also, using the thermodynamic relation: $d\epsilon = Tds + (p/\rho^2)d\rho$, we can write: $d(\rho\epsilon) = hd\rho + \rho Tds$. Using the adiabatic condition of motion, this leads to:

$$\frac{\partial (\rho\epsilon)}{\partial t} = -\mathrm{h}.\mathrm{div}(\rho\mathbf{v}) - \rho\mathrm{T}\mathbf{v}.\mathbf{grad}(\mathrm{s}).$$

Combining the results, we find the expected result:

$$\frac{\partial \rho(\frac{1}{2}\mathbf{v}^2 + \epsilon)}{\partial t} = -\mathrm{div}\left(\rho\mathbf{v}\left(\frac{1}{2}\mathbf{v}^2 + \mathrm{h}\right)\right).$$

In integral form, it reads:

$$\frac{\partial}{\partial t}\int \rho(\frac{1}{2}\mathbf{v}^2 + \epsilon)\mathrm{dV} = -\oint \rho\mathbf{v}\left(\frac{1}{2}\mathbf{v}^2 + \mathrm{h}\right)\mathrm{d}\mathbf{\Sigma}.$$

The left hand side is the rate of change of energy of the fluid in some given volume. The right hand side is therefore the amount of energy flowing out of this volume per unit time. Hence, $\rho\mathbf{v}\left(\frac{1}{2}\mathbf{v}^2 + \mathrm{h}\right)$ is the energy flux density vector. Its magnitude is the amount of energy passing per unit time through a unit area perpendicular to the direction of velocity. This means that any unit mass of fluid carries with it during its motion the amount of energy $\frac{1}{2}\mathbf{v}^2 + \mathrm{h}$ (and not $\frac{1}{2}\mathbf{v}^2 + \epsilon$).

The fact that enthalpy appears and not internal energy simply comes from the relation:

$$-\oint \rho\mathbf{v}\left(\frac{1}{2}\mathbf{v}^2 + \mathrm{h}\right) d\mathbf{\Sigma} = -\oint \rho\mathbf{v}\left(\frac{1}{2}\mathbf{v}^2 + \epsilon\right) d\mathbf{\Sigma} - \oint \mathrm{p}\mathbf{v} d\mathbf{\Sigma}.$$

The first term is the total energy transported through the surface in unit time by the fluid and the second term is the work done by pressure forces on the fluid within the surface.

Summary of the three conservation principles for ideal fluids is provided in the table below:

| | |
|---|---|
| (i) Conservation of matter | Continuity equation: $$\frac{\partial \rho}{\partial t} + (\mathbf{v}.\mathbf{grad})(\rho) = \left[\frac{\partial}{\partial t} + (\mathbf{v}.\mathbf{grad})\right](\rho) = -\rho \mathrm{div}(\mathbf{v}).$$ |
| (ii) Balance of Momentum (Newton's second law) | Euler's (vector) equation: $$\frac{\partial \mathbf{v}}{\partial t} + (\mathbf{v}.\mathbf{grad})(\mathbf{v}) = -\frac{1}{\rho}\mathbf{grad}(p) + \frac{1}{\rho}\mathbf{F}_{\mathrm{ext}}.$$ Useful identity: $$\frac{1}{2}\mathbf{grad}(\mathbf{v}^2) = \mathbf{v} \times \mathbf{curl}(\mathbf{v}) + (\mathbf{v}.\mathbf{grad})(\mathbf{v}).$$ |
| (iii) Conservation of energy or absence of heat exchange between the different parts of the fluid which implies that the motion is adiabatic | Local form of the conservation of energy reads: $$\frac{\partial \rho(\frac{1}{2}\mathbf{v}^2 + \epsilon)}{\partial t} + \mathrm{div}\left(\rho \mathbf{v}(\frac{1}{2}\mathbf{v}^2 + h)\right) = 0.$$ For ideal fluids, this is equivalent to the condition: $$\frac{ds}{dt} = \frac{\partial s}{\partial t} + (\mathbf{v}.\mathbf{grad})(s) = 0 \text{ or } s = \text{constant}.$$ Together with an equation of state this provides a relation of the form: $p=p(\rho,s)$. |

Exercise: Momentum flux. We label the spatial component x, y, z of vectors by one index i. Prove that there exists a quantity $\Pi_{ik}$ that depends on two indices that verifies the relation:

$$\frac{\partial}{\partial t}\int \rho v_i dV = -\sum_{k=x,y,z}\oint \Pi_{ik} d\Sigma_k \text{ for each i=x, y, z.}$$

Solution: It can easily be shown that: $\Pi_{ik} = p\delta_{ik} + \rho v_i v_k$, where $\delta_{ik} = 1$ if $i = k$ and 0 otherwise. The integral relation above can be interpreted as usual. The left hand side is the rate of change of the component i of the momentum contained in the volume considered. The right hand side is therefore the amount of momentum flowing out through the bonding surface per unit time.

Thus, $\Pi_{ik}$ corresponds to the component i of the momentum flowing in unit time through unit area perpendicular to the axis labeled by k. The energy flux is given by a vector (depending on one index), the energy itself being a scalar. Here the momentum flux is given by a quantity depending on two indices, a tensor of rank 2, the momentum itself being a vector.

We consider now the kinetic energy contained in a volume $\Omega_t$ defined by the flow map ($\Omega_t$ is moving with the fluid), namely $E_K = \int_{\Omega_t}\frac{1}{2}\rho\mathbf{v}^2 dV$. Note that $\Omega_t$ is not fixed in space (i.e. not fixed in some Newtonian reference frame), but moving with the flow according to the flow map function. We are interested by the variation along the flow of the kinetic energy contained in this volume $E_K$. To get this information, we need to compute the (material) derivative $dE_K/dt$, knowing that $E_K$ is defined as an integral over the moving volume $\Omega_t$. As discussed previously, this is not obvious to commute the derivative and the integral as the integral domain depends on time. In fact, it can be shown that:

$$\frac{dE_K}{dt} = \frac{d}{dt}\left[\int_{\Omega_t}\frac{1}{2}\rho\mathbf{v}^2 dV\right] = \int_{\Omega_t}\mathbf{v}\rho.\frac{d\mathbf{v}}{dt}dV\,.$$

i. As a first case, we assume that all the energy is kinetic. This means that we consider the internal energy as a constant which does not matter in the expression related to energy balance. The principle of conservation of energy states that the rate of change in time of the kinetic energy in a portion of fluid (following the flow) equals the rate at which the pressure forces work. For simplicity, we neglect other external forces that may apply. Mathematically, this gives:

$$\frac{dE_K}{dt} = \frac{d}{dt}[\int_{\Omega_t} \frac{1}{2}\rho \mathbf{v}^2 dV] = \int_{\Omega_t} \mathbf{v}\rho.\frac{d\mathbf{v}}{dt}dV = -\oint_{S_t} p\mathbf{v}.d\mathbf{\Sigma}.$$

The last integral is taken on a closed surface bounding the volume $\Omega_t$. Also, the quantity $-\oint_{S_t} p\mathbf{v}.d\mathbf{\Sigma}$ equals $-\int_{\Omega_t} \mathrm{div}(p\mathbf{v})dV = -\int_{\Omega_t} \mathbf{v}.\mathbf{grad}(p)dV - \int_{\Omega_t} p\,\mathrm{div}(\mathbf{v})dV$. We can replace $-\mathbf{grad}(p)$ by $\rho.\frac{d\mathbf{v}}{\mathbf{dt}}$ in this identity using the Euler's equations, which leads to:

$$\frac{dE_K}{dt} = \int_{\Omega_t} \mathbf{v}\rho.\frac{d\mathbf{v}}{dt}dV = \int_{\Omega_t} \mathbf{v}\rho.\frac{d\mathbf{v}}{dt}dV - \int_{\Omega_t} p\,\mathrm{div}(\mathbf{v})dV.$$

This equality can only be realized if $\mathrm{div}(\mathbf{v}) = 0$, or (using the continuity equation) $d\rho/dt=0$. This corresponds to the condition of incompressible fluid (see §8).

ii. As a second case, we consider that the internal energy is not constant. Following the previous discussion, we can write in this more general case:

$$\frac{d(E_K + \rho\epsilon)}{dt} = -\oint_{S_t} p\mathbf{v}.d\mathbf{\Sigma}.$$

After developing the above expression as in the first case, we can easily show that this leads to the relation $\rho\frac{d\epsilon}{dt} = \frac{p}{\rho}\frac{d\rho}{dt}$ or equivalently $\rho^2\frac{d\epsilon}{d\rho} = p$. This corresponds to a re-writing of the variation of enthalpy as: $dh = dp/\rho$, which is also the condition for an isentropic motion.

## §6. Boundary conditions for ideal fluids

The equations of motion have to be supplemented by the boundary conditions that must be satisfied at the surfaces bounding the fluid. For an ideal fluid, the boundary condition is simply that the fluid cannot penetrate a solid surface. This means that the component of the fluid velocity normal to the bounding surface must vanish if that surface is at rest: $\mathbf{v}.\mathbf{n} = 0$.

In the general case of a moving surface, $\mathbf{v}.\mathbf{n}$ must be equal to the corresponding component of the velocity of the surface. At a boundary between two immiscible fluids, the condition is that the pressure and the velocity component normal to the surface of separation must be the same for the two fluids, and each of these velocity components must be equal to the corresponding component of the velocity of the surface.

Given boundary conditions, the continuity and Euler's equations, together with the relation for adiabatic motion, established in §5 form a closed set of equations necessary to determine the 5 unknown quantities, once initial conditions are assumed. Solving this problem means that, if we consider a moving fluid contained in a volume $\Omega$ at any instant t ($\Omega_t$ the corresponding moving volume), at each point $\mathbf{x}$ of $\Omega_t$, we can find a well-defined solution for the 5 quantities to be determined. Such that also the equations of fluid mechanics must contain non-diverging terms at all points of $\Omega_t$. In particular, the total kinetic energy density integrated over the moving volume $\Omega_t$, $\int \frac{1}{2}\rho\mathbf{v}^2 dV$, must remain finite, not diverging to infinity at any time. Stated differently, once we define initial conditions, a fluid mechanics problem is solved (globally) if we can show the existence and unicity of smooth solutions of equations (§5) for the velocity, pressure and mass density, for all points in the fluid volume and at all times, with dedicated boundary conditions. This is highly non trivial as we shall see next.

## §7. Introduction to nonlinear differential equations

A central issue in the study of nonlinear differential equations, as the Euler's equations, is that solutions may exist locally in time (that is, for short periods of time) but not globally in time. This is caused by a phenomenon called blow-up, illustrated in this section.

We first discuss this phenomenon with three simple ordinary differential equations for a real function $u(t)$:

$$(a)\ \frac{du}{dt} = u,\ (b)\ \frac{du}{dt} = u^2,\ (c)\ \frac{du}{dt} = -u + u^2.$$

Equation (a) is linear and its solution is:

$$u(t) = Ae^t.$$

This solution is defined globally in time and grows exponentially as time becomes infinite. Generally, global existence and exponential growth are typical features of linear differential equations. Equation (b) is nonlinear. Its solution is:

$$u(t) = 1/(\tau - t),$$

where $\tau$ is a parameter. The solution of (b) is then diverging when $t$ is approaching $\tau$. This example (b) shows that nonlinearities which grow super-linearly in $u(t)$ can lead to blow-up and a loss of global existence. Equation (c) is also nonlinear and its solution is:

$$u(t) = 1/(1 - Ae^t).$$

If $A \leq 0$, which corresponds to $0 < u(0) < 1$, then the solution exists globally in time.
If $0 < A < 1$, which corresponds $u(0) > 1$, then the solution blows up at $t = \log(\frac{1}{A})$.

Thus, there is a global existence of solutions with small initial data and local existence (in time) of solutions with large initial data. This type of behavior also occurs in many partial differential equations of more general functions of space and time: for small initial data, linear damping terms can dominate the nonlinear terms, and one obtains global solutions whereas for large initial data, the nonlinear blow-up dominates and only local solutions may exist.

Consider now a less simple example of nonlinear partial differential equation. A real function of $x$ and $t$, $u(x,t)$, is a solution of the equation: $\frac{\partial u}{\partial t} + \frac{\partial(\frac{u^2}{2})}{\partial x} = 0$, with the initial condition: $u(x,0) = u_0(x)$.

We can show that this equation cannot have a global smooth solution if $\frac{\partial u_0(x)}{\partial x} < 0$ at any point. The proof is simple, based on the previous discussion. Suppose that $u(x, t)$ is a smooth solution. We take the x derivative of the partial differential equation: $\frac{\partial u}{\partial t} + \frac{\partial (\frac{u^2}{2})}{\partial x} = 0$.

We obtain:

$$\frac{\partial u_x}{\partial t} + uu_{xx} + u_x^2 = 0.$$

The subscript x represents a derivative with respect to x: $u_x = \frac{\partial u}{\partial x}$. We can rewrite this expression as:

$$[\frac{\partial}{\partial t} + u\frac{\partial}{\partial x}](u_x) + u_x^2 = 0.$$

It is interesting to note as: $\frac{d}{dt} = [\frac{\partial}{\partial t} + u\frac{\partial}{\partial x}]$ the derivative operator appearing in the last formula (already seen in §5), which is the derivative along the characteristic curves associated with the function u. Then, we obtain an equation close to equation (b) above:

$$\frac{du_x}{dt} = -u_x^2.$$

Therefore, if $u_x < 0$ at initial time, the solution of this equation follows exactly what we have computed for equation (b) and it blows up at some positive time. A global smooth solution cannot exist.

The interest of this (partial) differential equation: $\frac{\partial u}{\partial t} + \frac{\partial (\frac{u^2}{2})}{\partial x} = \frac{\partial u}{\partial t} + u\frac{\partial (u)}{\partial x} = 0$ lies in the fact that it is equivalent to a one dimensional Euler's equation, u being the velocity field, with p=0 and in the absence of external forces. Already with this simple form, we remark that a (unique) global smooth solution may not exist in general. We discuss further these issues next in the particular (but so important) case of incompressible fluids.

## §8. Euler's equations for incompressible ideal fluids

For many (ideal) flows of liquids (and even gases), mass density (ρ) can be supposed constant throughout the volume of the fluid and along its motion. This is equivalent to neglecting compression and expansion of these fluids. We speak of incompressible fluids:

$$\rho = \text{constant}.$$

Equations of fluid mechanics are much simplified for an incompressible fluid. The continuity equation becomes:

$$\mathrm{div}(\mathbf{v}) = 0.$$

Euler's equations in the presence of a gravitational field become:

$$\frac{\partial \mathbf{v}}{\partial \mathrm{t}} + (\mathbf{v}.\mathbf{grad})(\mathbf{v}) = \frac{\partial \mathbf{v}}{\partial \mathrm{t}} + \frac{1}{2}\mathbf{grad}(\mathbf{v}^2) - \mathbf{v} \times \mathbf{curl}(\mathbf{v}) = -\mathbf{grad}\left(\frac{\mathrm{p}}{\rho}\right) + \mathbf{g}.$$

Obviously, we can take the **curl** of the above formula, which leads to an expression involving only the velocity field:

$$\frac{\partial \mathbf{curl}(\mathbf{v})}{\partial \mathrm{t}} = \mathbf{curl}[\mathbf{v} \times \mathbf{curl}(\mathbf{v})].$$

Interestingly, as the mass density is not an unknown function any longer for incompressible fluids, the closed set of equations for (ideal) fluids can be reduced to equations related to velocity only.

The vector $\boldsymbol{\omega} = \mathbf{curl}(\mathbf{v})$ is called the vorticity. The equation for the vorticity can be re-written after a proper development of:

$$\mathbf{curl}[\mathbf{v} \times \boldsymbol{\omega}] = \mathbf{v}.\big(\mathrm{div}(\boldsymbol{\omega})\big) - \boldsymbol{\omega}\big(\mathrm{div}(\mathbf{v})\big) + (\boldsymbol{\omega}.\mathbf{grad})(\mathbf{v}) - (\mathbf{v}.\mathbf{grad})(\boldsymbol{\omega}).$$

Using the condition $\mathrm{div}(\mathbf{v}) = 0$ and the identity: $\mathrm{div}(\boldsymbol{\omega}) = \mathbf{0}$, we obtain:

$$\partial \boldsymbol{\omega}/\, \partial \mathrm{t} + (\mathbf{v}.\mathbf{grad})(\boldsymbol{\omega}) = (\boldsymbol{\omega}.\mathbf{grad})(\mathbf{v}).$$

(5)

Together with the definition: $\boldsymbol{\omega} = \mathbf{curl}(\mathbf{v})$, these equations completely determine the velocity in terms of the vorticity. The vector equation (5) is of fundamental importance. To understand it, we need first to give a hint of what vorticity is physically.

We intend to show that vorticity encodes the magnitude and direction of the axis about which a fluid parcel rotates locally. For simplicity, we consider a 2-dimensional case in the (xOy) plane. We observe the deformation along the flow of a rectangular fluid parcel ABCD parameterized at time t by A(x,y), B(x+dx,y), C(x+dx,y+dy) and D(x,y+dy). Its surface is $\Sigma$=dx.dy and after the time interval dt, the points ABCD at time t have evolved to A'B'C'D' at time t+dt. It can easily be shown that: $d\Sigma/dt=\Sigma(\partial v_x/\partial x+\partial v_y/\partial y)$. Hence, the relative variation of the surface of the fluid parcel is given by the divergence of the velocity.

We label the angles generated by the flow as: $d\theta$=(**AB**,**A'B'**) and $d\varphi$=(**AD**,**A'D'**). The global rotation of the fluid parcel is given by the rotation of the diagonal of the rectangle, which we define as $\Gamma$dt. By construction, it is equal to: ½( $d\theta$+ $d\varphi$) (see figure).

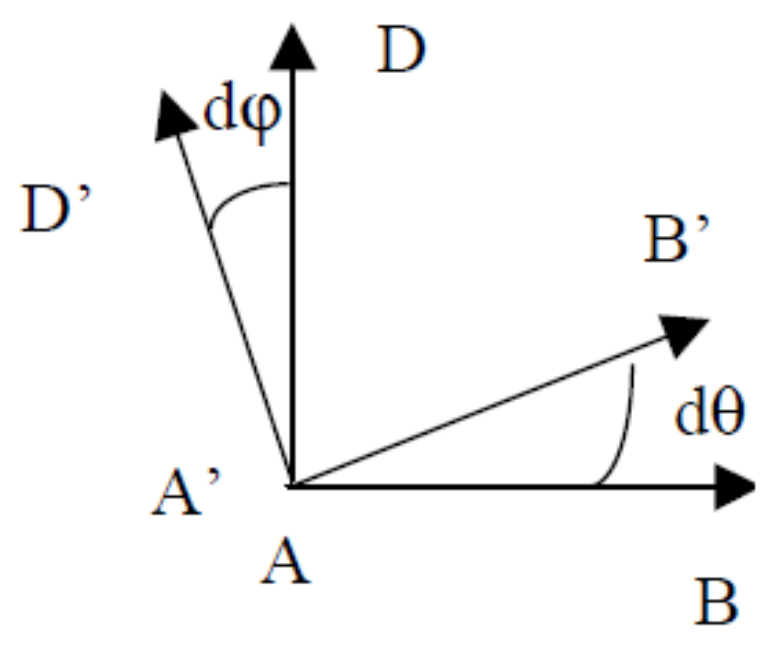

Again, this is easy to show that $d\theta+d\varphi=(\partial v_y/\partial x-\partial v_x/\partial y)dt$, or equivalently:

$$\Gamma=\tfrac{1}{2}(\partial v_y/\partial x-\partial v_x/\partial y).$$

Then the quantity $\Gamma$, which is characteristic of the rotation, is half the vorticity (in 2-dimensional space). This result can be generalized to 3-dimensional space with the vector result: $\boldsymbol{\Gamma}=\tfrac{1}{2}\boldsymbol{\omega}$. Vorticity is thus directly related to the magnitude and direction of the axis about which a fluid parcel rotates.

We now come back on the vector equation (5) characterizing the evolution in space and time of the vorticity. It can be written in another very important form using the flow map function. We can show that it is equivalent to:

$$\boldsymbol{\omega}(\boldsymbol{\Phi}(\mathbf{X},t),t) = \mathbf{grad}_{\mathbf{X}}\big(\boldsymbol{\Phi}(\mathbf{X},t)\big).\boldsymbol{\omega}_{\mathbf{0}}(\mathbf{X}) \text{ with } \boldsymbol{\omega}_{\mathbf{0}}(\mathbf{X}) = \boldsymbol{\omega}(\boldsymbol{\Phi}(\mathbf{X},0),0) = \boldsymbol{\omega}(\mathbf{X},0). \quad (5')$$

This expression is a bit unusual. Indeed, $\mathbf{grad}_{\mathbf{X}}\big(\boldsymbol{\Phi}(\mathbf{X},t)\big)$ can be expended under the 3 components of the gradient with respect to **X** (in 3-dimensional space), but each component is itself a vector due to the presence of **Φ**. That's why the quantity defined by $\boldsymbol{\omega}(\boldsymbol{\Phi}(\mathbf{X},t),t) = \mathbf{grad}_{\mathbf{X}}\big(\boldsymbol{\Phi}(\mathbf{X},t)\big).\boldsymbol{\omega}_{\mathbf{0}}(\mathbf{X})$ is a vector, where the scalar product is taken between the 3 components of the gradient and the 3 components of the vector $\boldsymbol{\omega}_{\mathbf{0}}(\mathbf{X})$.

In order to prove the equivalence between expressions (5) and (5'), we differentiate the relation (5') with respect to time (material derivative). We obtain:

$$\frac{d\boldsymbol{\omega}}{dt} = \frac{\partial\boldsymbol{\omega}}{\partial t} + (\mathbf{v}.\mathbf{grad})(\boldsymbol{\omega}) = \mathbf{grad}_{\mathbf{X}}(\mathbf{v}[\boldsymbol{\Phi}(\mathbf{X},t),t])\boldsymbol{\omega}_{\mathbf{0}}(\mathbf{X}).$$

On the right hand side, we have the derivative of a composition of functions which can be easily computed (exercise below). Knowing that $\mathbf{grad}_{\mathbf{X}}\big(\boldsymbol{\Phi}(\mathbf{X},t)\big)\boldsymbol{\omega}_{\mathbf{0}}(\mathbf{X})$, is equal to $\boldsymbol{\omega}(\boldsymbol{\Phi}(\mathbf{X},t),t)$, we end up with:

$$\frac{\partial\boldsymbol{\omega}}{\partial t} + (\mathbf{v}.\mathbf{grad})(\boldsymbol{\omega}) = (\boldsymbol{\omega}.\mathbf{grad})(\mathbf{v}).$$

This is the vector equation (5), which completes the proof.

Exercise: Prove that: $\mathbf{grad}_{\mathbf{X}}(\mathbf{v}[\boldsymbol{\Phi}(\mathbf{X},t),t])\boldsymbol{\omega}_{\mathbf{0}}(\mathbf{X}) = (\boldsymbol{\omega}.\mathbf{grad})(\mathbf{v})$.

Solution: we first expand explicitly the first term, using the notation $X=(X_1, X_2, X_3)$ and $\boldsymbol{\omega}_{\mathbf{0}}(\mathbf{X}) = (\omega_{0,1}, \omega_{0,2}, \omega_{0,3})$. We obtain:

$$\mathbf{grad}_{\mathbf{X}}(\mathbf{v}[\boldsymbol{\Phi}(\mathbf{X},t),t]).\boldsymbol{\omega}_{\mathbf{0}}(\mathbf{X}) = \frac{\partial\mathbf{v}[\boldsymbol{\Phi}(\mathbf{X},t),t]}{\partial X_1}\omega_{0,1} + \frac{\partial\mathbf{v}[\boldsymbol{\Phi}(\mathbf{X},t),t]}{\partial X_2}\omega_{0,2} + \frac{\partial\mathbf{v}[\boldsymbol{\Phi}(\mathbf{X},t),t]}{\partial X_3}\omega_{0,3}.$$

Also, in compact notations: $\frac{\partial\mathbf{v}[\boldsymbol{\Phi}(\mathbf{X},t),t]}{\partial X_1} = \frac{\partial\boldsymbol{\Phi}(.)}{\partial X_1}\frac{\partial\mathbf{v}}{\partial x}$.

The sum above can thus be rearranged as:

$$\left[\frac{\partial \boldsymbol{\Phi}(.)}{\partial X_1}\omega_{0,1}\frac{\partial}{\partial x}+\frac{\partial \boldsymbol{\Phi}(.)}{\partial X_2}\omega_{0,2}\frac{\partial}{\partial y}+\frac{\partial \boldsymbol{\Phi}(.)}{\partial X_3}\omega_{0,3}\frac{\partial}{\partial z}\right]\mathbf{v}.$$

This gives the result.

<u>Exercise</u>: Prove that if we are considering a space with only 2 dimensions, the vector equation (5) reads:

$$\frac{\partial \boldsymbol{\omega}}{\partial t}+(\mathbf{v}.\mathbf{grad})(\boldsymbol{\omega})=0.$$

Similarly, prove that in 2-dimensional space, the equivalent formula:

$$\boldsymbol{\omega}(\boldsymbol{\Phi}(\mathbf{X},t),t)=\mathbf{grad}_{\mathbf{X}}\big(\boldsymbol{\Phi}(\mathbf{X},t)\big).\boldsymbol{\omega}_{\mathbf{0}}(\mathbf{X}).$$

Which can be simplified into:

$$\boldsymbol{\omega}(\boldsymbol{\Phi}(\mathbf{X},t),t)=\boldsymbol{\omega}_{\mathbf{0}}(\mathbf{X}).$$

Comment these last two expressions.

<u>Solution</u>: When we consider a flow in a plane, which means in 2-dimensional space (2D) labeled in Cartesian coordinates (x,y), the velocity field can be represented as a 2-dimensional vector ($v_x$, $v_y$). Then, only the z-component of $\boldsymbol{\omega}=\mathbf{curl}(\mathbf{v})$ is non-zero by definition. This implies that the scalar product between $\boldsymbol{\omega}=(0,0,\omega_z)$ and the gradient in 2D is zero. Therefore, the vector equation (5) reduces to the relation:

$\frac{\partial \boldsymbol{\omega}}{\partial t}+(\mathbf{v}.\mathbf{grad})(\boldsymbol{\omega})=0$. The vorticity is thus a scalar, we write: $\omega_z=\omega$. The last expression becomes a one dimensional partial differential equation: $\frac{\partial}{\partial t}\omega+\left(v_x\frac{\partial}{\partial x}+v_y\frac{\partial}{\partial y}\right)(\omega)=0$.

Applying a similar method as in the text above, we can show that:

$$\omega\big(\Phi_x(X_1,X_2,t),\Phi_y(X_1,X_2,t),t\big)=\omega_0(X_1,X_2).$$

This proves the second relations. These two equivalent formulae mean that vorticity is conserved along paths of fluid particles in 2-dimensional flows.

Exercise: Conservation of circulation. We consider a closed curve build up by fluid particles at initial time $C_0$. This closed curve will move with the fluid, and at time t, we can represent this curve using the flow map notation as: $C_t = \{\boldsymbol{\Phi}(\mathbf{X}, t) : \mathbf{X} \text{ belongs to } \Omega_0\}$.

For an ideal incompressible fluid prove that the circulation of the velocity along the closed curve $C_t$ is conserved along the flow:

$$\oint_{C_t} \mathbf{v}.\mathrm{d}\mathbf{l} = \oint_{C_0} \mathbf{v}.\mathrm{d}\mathbf{l}$$

What can you conclude concerning the vorticity?

Solution: We compute the material derivative (following the flow of particles): $\frac{\mathrm{d}}{\mathrm{dt}}\left[\oint_{C_t} \mathbf{v}.\mathrm{d}\mathbf{l}\right]$.

We have already discussed this kind of calculus in §5. The difficulty is that the boundary of the integration depends on time. However, as explained in §5, we can write:

$$\frac{\mathrm{d}}{\mathrm{dt}}\left[\oint_{C_t} \mathbf{v}.\mathrm{d}\mathbf{l}\right] = \oint_{C_t} \frac{\mathrm{d}\mathbf{v}}{\mathrm{dt}}.\mathrm{d}\mathbf{l}.$$

Then, using the equations of motion: $\frac{\mathrm{d}\mathbf{v}}{\mathrm{dt}} = -\mathbf{grad}(\frac{p}{\varrho})$ with the fact the contour of integration is closed, we end up with the desired relation:

$$\frac{\mathrm{d}}{\mathrm{dt}}\left[\oint_{C_t} \mathbf{v}.\mathrm{d}\mathbf{l}\right] = 0.$$

This completes the proof. The circulation of the velocity along the closed curve $C_t$ is conserved along the flow.

Consequences for the vorticity: Using the circulation (Stokes) theorem, we can write: $\oint_{C_t} \mathbf{v}.\mathrm{d}\mathbf{l} = \int_{S_t} \mathbf{curl}(\mathbf{v}).\mathrm{d}\boldsymbol{\Sigma} = \int_{S_t} \boldsymbol{\omega}.\mathrm{d}\boldsymbol{\Sigma}$, where $S_t$ represents any surface moving with the flow bounded by the closed contour $C_t$. This means that the $i^{th}$ component of the vorticity vector can be seen as the limit circulation per unit area in the plane perpendicular to the ($x_i$)-direction. Intuitively, it measures how much a little leaf carried by the flow would spin about the ($x_i$)-direction.

Also, we can understand physically what happens if we are restricted to 2-dimensional space (2D). In 2D, incompressibility implies that $S_t$ is a constant of motion: it derives from the continuity equation which is equivalent to a volume preserving condition (surface preserving in 2D). Then, the condition that $\int_{S_t} \boldsymbol{\omega}.\,\mathrm{d}\,\boldsymbol{\Sigma}$ is conserved along the flow, in the limit of very small area implies that the vorticity is conserved along the flow lines.

This can be formulated as: $\frac{\mathrm{d}\boldsymbol{\omega}}{\mathrm{dt}} = 0$. This corresponds to what we have shown mathematically in the previous exercise.

Differently, for 3-dimensional space (3D), there is no constraint on $S_t$ following the continuity equation (the volume preserving condition for incompressible fluids). Thus, conservation of the flux of vorticity cannot control the magnitude of the vorticity vector.

To conclude briefly this discussion, we have understood intuitively some differences between the 2D and 3D physics cases. As we shall see later, these differences can explain why 2D equations of fluid mechanics cannot have singularities why 3D equations might.

We consider the right hand side term $((\boldsymbol{\omega}.\,\mathbf{grad})(\mathbf{v}))$ of the vorticity equation:

$$\frac{\mathrm{D}\boldsymbol{\omega}}{\mathrm{Dt}} = (\boldsymbol{\omega}.\,\mathbf{grad})(\mathbf{v}).$$

As we have shown in the exercises above, this term is not present for 2-dimensional flows, for which vorticity is conserved along flow lines. Therefore, we know that this is the term which brings some complications for 3-dimensional flows. This is interesting to get an intuitive understanding of it. Rephrased in words, $(\boldsymbol{\omega}.\,\mathbf{grad})(\mathbf{v})$ is proportional to the derivative in the direction of $\boldsymbol{\omega}$ along a vortex line: $(\omega.\,\partial/\,\partial s_\omega)(\mathbf{v})$. Where $\delta s_\omega$ is the length of an element of vortex line. We now resolve the vector $\mathbf{v}$ into components $\mathbf{v}_\omega$ parallel to the vortex line (of direction $\boldsymbol{\omega}$) and $\mathbf{v}_\perp$ perpendicular to $\omega$ and hence to $\delta s_\omega$. Projected along the vortex line, we obtain:

$$\frac{1}{\omega}\frac{\mathrm{D}\boldsymbol{\omega}}{\mathrm{Dt}}.\,\delta s_\omega = \frac{\partial(\mathbf{v})}{\partial s_\omega}.\,\delta s_\omega = \frac{\partial(\mathbf{v}_\omega + \mathbf{v}_\perp)}{\partial s_\omega}.\,\delta s_\omega.$$

This gives:

$$\frac{1}{\omega}\frac{D\boldsymbol{\omega}}{Dt}.\delta s_\omega = \frac{\partial(\mathbf{v}_\omega)}{\partial s_\omega}.\delta s_\omega + \frac{\partial(\mathbf{v}_\perp)}{\partial s_\omega}.\delta s_\omega = \{\mathbf{v}_\omega(\mathbf{x}+\delta s_\omega) - \mathbf{v}_\omega(\mathbf{x})\} + \{\mathbf{v}_\perp(\mathbf{x}+\delta s_\omega) - \mathbf{v}_\perp(\mathbf{x})\}.$$

The first term on the right hand side represents the rate of stretching of the element $\delta s_\omega$. The second term represents the rate of turning of the element $\delta s_\omega$. Then, stretching along the length of a vortex line causes relative amplification of the vorticity field, while turning away from the vortex line causes a reduction of the vorticity in that direction, but an increase in the new direction.

<u>Summary of important equations for the vorticity (incompressible ideal fluids):</u>

| | |
|---|---|
| General equivalent equations for the vorticity, also called the 'stretching formulae'. | $\frac{\partial\boldsymbol{\omega}}{\partial t} + (\mathbf{v}.\mathbf{grad})(\boldsymbol{\omega}) = (\boldsymbol{\omega}.\mathbf{grad})(\mathbf{v})$,<br>equivalent to<br>$\boldsymbol{\omega}(\boldsymbol{\Phi}(\mathbf{X},t),t) = \mathbf{grad}_\mathbf{X}(\boldsymbol{\Phi}(\mathbf{X},t)).\boldsymbol{\omega}_\mathbf{0}(\mathbf{X})$<br>with $\boldsymbol{\omega}_\mathbf{0}(\mathbf{X}) = \boldsymbol{\omega}(\boldsymbol{\Phi}(\mathbf{X},0),0) = \boldsymbol{\omega}(\mathbf{X},0)$. |
| The equation using the flow map expresses the fact that vortex lines are carried by the flow. In 2-dimensional spaces, the vorticity is carried along by particle paths, its magnitude unchanged. In 3-dimensional spaces, vorticity is carried as well, but its magnitude is amplified or diminished by the gradient of the flow map. | In 2D, the general vector equation (above) becomes:<br>$\boldsymbol{\omega}(\boldsymbol{\Phi}(\mathbf{X},t),t) = \boldsymbol{\omega}_\mathbf{0}(\mathbf{X})$.<br>Moreover, the vorticity in 2D is a scalar and this last equation can be written as:<br>$\omega(\Phi_x(X_1,X_2,t),\Phi_y(X_1,X_2,t),t)$<br>$= \omega_0(X_1,X_2)$.<br>Or equivalently:<br>$\frac{\partial}{\partial t}\omega + \left(v_x\frac{\partial}{\partial x} + v_y\frac{\partial}{\partial y}\right)(\omega) = 0.$ |

## §9. Potential flows for ideal fluids

We start by restating the property of conservation of circulation (see §8). We have shown that for an ideal incompressible fluid the circulation of the velocity along the closed curve $\mathrm{C_t}$ (moving with the flow) is conserved:

$$\oint_{C_t} \mathbf{v}.\,\mathrm{d}\mathbf{l} = \oint_{C_0} \mathbf{v}.\,\mathrm{d}\mathbf{l}.$$

However, in the course of the proof (in §8), we have used a mathematical trick (with no justification) concerning the inversion of the integral whose boundary depends on time and the material derivative: $\frac{\mathrm{d}}{\mathrm{dt}}\left[\oint_{C_t} \mathbf{v}.\,\mathrm{d}\mathbf{l}\right] = \oint_{C_t} \frac{\mathrm{d}\mathbf{v}}{\mathrm{dt}}.\,\mathrm{d}\mathbf{l}$. In this section we redo this important proof completely. It will allow us to state clearly the minimal assumptions needed to derive this property (theorem).

We are interested by the material derivative of the circulation of velocity on a closed contour: $\oint \mathbf{v}.\,\mathrm{d}\mathbf{l}$. The closed contour is supposed to be drawn in the fluid at some instant of time and we assume that this corresponds to a fluid contour, build up by fluid particles, which lie on this contour. See the figure below for illustration, where only a few fluid particles have been pictured for simplicity.

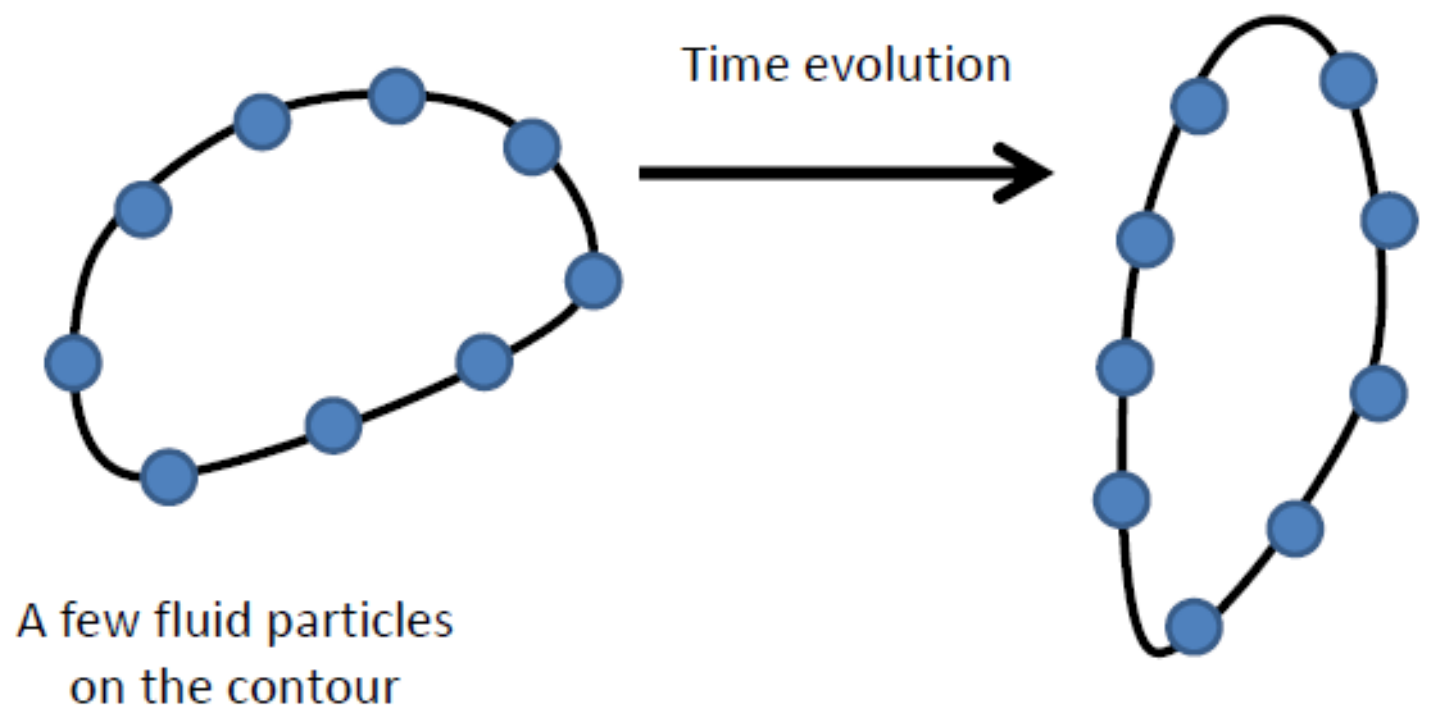

In the course of time, these fluid particles move about with the flow and thus the contour moves with them accordingly. We calculate the material derivative of the velocity circulation bounded by this contour:

$$\frac{\mathrm{d}}{\mathrm{dt}}\left[\oint \mathbf{v}.\,\mathrm{d}\mathbf{l}\right].$$

First, this is really the material derivative that we need to evaluate, as we are interested in the change of the circulation round the fluid contour moving with the flow (displayed in the figure above), and not round a fixed contour in space coordinates.

Then, we write the element of length in the contour as $\mathrm{d}\mathbf{l} = \delta\mathbf{r}$, where $\delta\mathbf{r}$ is the difference between the radius vectors of the points at the ends of the element of the contour $\mathrm{d}\mathbf{l}$. In this proof, we use the symbol $'\delta'$ for the differentiation with respect to space coordinates and $'\mathrm{d}'$ for the differentiation with respect to time.

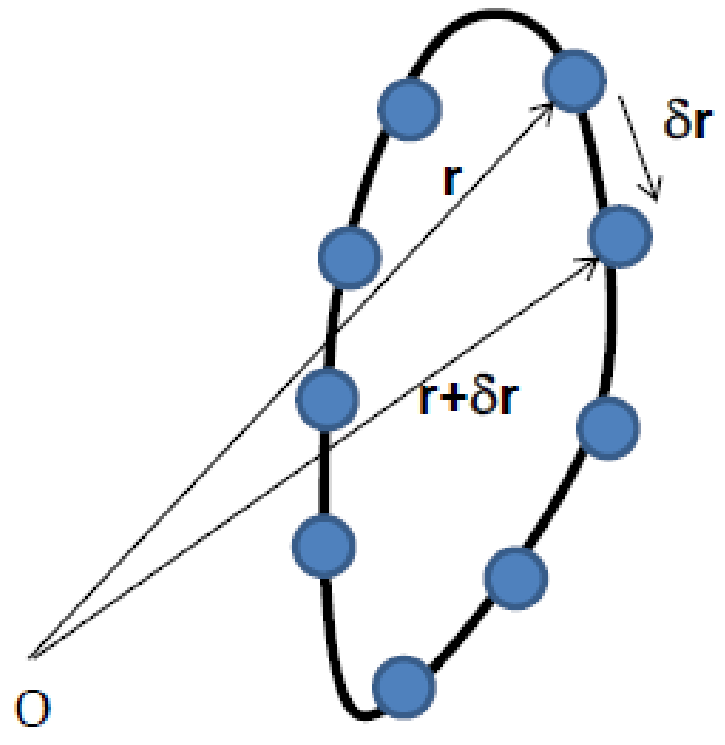

Then, the circulation of velocity can be written as $\oint \mathbf{v}.\mathrm{d}\mathbf{l} = \oint \mathbf{v}.\delta\mathbf{r}$. In differentiating this integral with respect to time, we need to consider that not only the velocity but also the contour itself changes, along the flow. That's why we need to differentiate both **v** and $\delta$**r**. This leads to:

$$\frac{\mathrm{d}}{\mathrm{dt}}\left[\oint \mathbf{v}.\delta\mathbf{r}\right] = \oint \frac{\mathrm{d}\mathbf{v}}{\mathrm{dt}}.\delta\mathbf{r} + \oint \mathbf{v}.\frac{\mathrm{d}\delta\mathbf{r}}{\mathrm{dt}}.$$

The second integral on the right hand side is trivial to compute after writing the integrand in the following form using simple algebra: $\mathbf{v}.\frac{\mathrm{d}\delta\mathbf{r}}{\mathrm{dt}} = \mathbf{v}.\delta\frac{\mathrm{d}\mathbf{r}}{\mathrm{dt}} = \delta\left(\frac{\mathbf{v}^2}{2}\right)$. This gives after integration on a closed contour:

$$\oint \mathbf{v}.\frac{\mathrm{d}\delta\mathbf{r}}{\mathrm{dt}} = 0.$$

This proves the mathematical identity presented previously: $\frac{\mathrm{d}}{\mathrm{dt}}\left[\oint \mathbf{v}.\delta\mathbf{r}\right] = \oint \frac{\mathrm{d}\mathbf{v}}{\mathrm{dt}}.\delta\mathbf{r}$. From Euler's equations for isentropic fluid (not necessarily incompressible), we can write:

$$\frac{\mathrm{d}\mathbf{v}}{\mathrm{dt}} = -\mathbf{grad}(\mathrm{h}).$$

Then, this is easy to conclude that $\oint \frac{d\mathbf{v}}{dt}.\delta\mathbf{r} = 0$, since:

$$\oint \frac{d\mathbf{v}}{dt}.\delta\mathbf{r} = \oint \mathbf{curl}\left(\frac{d\mathbf{v}}{dt}\right).\delta\mathbf{\Sigma} = \oint \mathbf{curl}(-\mathbf{grad}(h)).\delta\mathbf{\Sigma} = 0.$$

We end up with the property of conservation, which completes the proof:

$$\frac{d}{dt}\left[\oint \mathbf{v}.d\mathbf{l}\right] = 0.$$

$$\oint \mathbf{v}.d\mathbf{l} = \text{constant (along the flow)}.$$

Therefore, for an ideal fluid, the velocity circulation round a closed fluid contour is constant in time: this is also called the Kelvin's theorem. We can remark that this property assumes Euler's equations for an isentropic flow. In fact, we need to be able to write $\frac{\mathbf{grad}(p)}{\rho}$ as a gradient of some function. This is the case for an isentropic flow for which the relation s(p,ρ)=constant poses a one to one relation between pressure and mass density.

From the law of conservation of circulation (along the flow), we can derive another essential property concerning vorticity. We consider an infinitely small contour (build up by fluid particles) moving with the flow and we assume that the vorticity is zero at some point along this path. We know (Stokes theorem) that the velocity circulation round this (infinitely small) closed contour is equal to $\mathbf{curl}(\mathbf{v}).d\mathbf{\Sigma} = \boldsymbol{\omega}.d\mathbf{\Sigma}$, where $d\mathbf{\Sigma}$ is the element of area enclosed by this small contour. At the point where $\boldsymbol{\omega} = \mathbf{0}$, the velocity circulation round this small contour is thus also zero. In the course of time, this contour moves with the fluid, always remaining infinitely small. Since the velocity circulation is conserved along the flow, it remains equal to zero for all points of this path, and it follows that the vorticity also must be zero at any point of this path. Therefore, we can state that: if at any point of some trajectory followed by fluid particles the vorticity is zero, the same is true at any point of this trajectory.

Note that if the flow is steady ($\frac{\partial \mathbf{v}}{\partial t} = \mathbf{0}$), streamlines coincide with paths described in the course of time by some fluid particles. In this case, we can consider a small contour that encircles the streamline. In particular, if it encircles the streamline at the point where the vorticity is zero, this property is conserved along the streamline. This means that for steady flow, the previous statement holds for streamlines: if at any point on a streamline, the vorticity is zero, the same is true at all other points on that streamline.

We continue the argument assuming the flow is steady. We consider a steady flow past a material body with the much reasonable hypothesis that the incident flow is uniform at infinity. This means that its velocity is constant at infinity, and thus its vorticity is zero at infinity. Following the previous statement, we conclude that the vorticity is zero along all streamlines and thus in all space. In fact, this is not exactly correct as the proof that vorticity is zero along a streamline is invalid for a line which lies on the surface of the solid body, since the presence of the surface makes it impossible to draw a closed contour encircling such a streamline! Of course, the physical problem of flow past a given body has a well-defined solution. The key point is that ideal fluids do not really exist: any real fluid has a certain viscosity, even small. This viscosity may have in practice no effect on the motion of most of the fluid, but, no matter how small it is, it will become essentially important in a thin layer adjoining the surface of the body. We come back on the mathematical description of how real fluids behave in the following sections. For the moment, we keep the point of view of ideal fluids, knowing that there is a boundary layer around a solid body inside which this (ideal) description does not apply.

A flow for which the vorticity is zero in all space is called a potential flow, or irrotational flow. Rotational flows correspond to flow where the vorticity is not zero everywhere. Following the discussion above, a steady flow past some material body, with a uniform incident flow at infinity must be potential.

Another consequence of the theorem of conservation of circulation is that, if at some instant, the flow is potential throughout the volume of the fluid, we can deduce that this will hold at any future instant. This is also in agreement with the equation for the vorticity, derived from Euler's equation: $\frac{\partial \boldsymbol{\omega}}{\partial \mathrm{t}} = \mathbf{curl}[\mathbf{v} \times \boldsymbol{\omega}]$, which shows that if $\boldsymbol{\omega} = \mathbf{0}$ at time t, it holds at time t+dt.

We now derive some general simple properties of potential flows, which are very useful in practice:

i. First, we have proved the property (theorem) of conservation of circulation, under the assumption that the flow is isentropic. This means that if the flow is not isentropic, this law does not hold. Therefore, if a non-isentropic flow is potential at some instant, the vorticity will in general be non-zero at subsequent instants, and the concept of potential flow is useless. Therefore, all what we discuss next assumes that the flow is isentropic.

ii. For a potential flow: $\oint \ \mathbf{v}.\mathrm{d}\mathbf{l} = 0$. It follows that closed streamlines cannot exist. Only in rotational flows, closed streamlines can be present.

iii. If the vorticity (vector) is zero $\boldsymbol{\omega} = \mathbf{0}$, this implies that there exist a scalar potential such that:

$$\mathbf{v} = \mathbf{grad}(\phi).$$

With the Euler's vector equation, we get:

$$\mathbf{grad}\left(\frac{\partial \phi}{\partial \mathrm{t}} + \frac{\mathrm{v}^2}{2} + \mathrm{h}\right) = \mathbf{0}.$$

Then, the function inside the gradient is not a function of coordinates and only depends on time:

$$\frac{\partial \phi}{\partial \mathrm{t}} + \frac{\mathrm{v}^2}{2} + \mathrm{h} = \mathrm{f}(\mathrm{t}).$$

Here, f(t) is an arbitrary function of time. As the velocity is the space coordinate gradient of the scalar potential: $\phi$, we can add to $\phi$ any function of time without modifying the velocity field. In particular, we can make the substitution:

$$\phi \rightarrow \phi + \int \mathrm{f}\mathrm{d}\mathrm{t}.$$

This means that we can take f(t)=0 without loss of generality in the above equation.

iv. For a steady (and potential) flow, we can simplify the equation given in (iii) with $\frac{\partial \phi}{\partial t} = 0$ and f(t)=constant:

$$\frac{v^2}{2} + h = \text{constant}.$$

This is the Bernoulli's equation. However, there is an important difference with the Bernoulli's equation established in the general case (§5), where the 'constant' on the right hand side is constant along any given streamline, but different for different streamlines. In potential flows, the 'constant' (above) is constant throughout the fluid.

v. An important physics case, where potential flow occurs, concerns small oscillations of an immersed body in a fluid. It can be shown that if the amplitude of oscillations is small compared with the dimension of the body, the flow past the body is approximately potential. The proof is left as an exercise below: the idea is to show that throughout the fluid $\partial \boldsymbol{\omega} / \partial t = \mathbf{0}$ and thus the vorticity in the fluid is constant. In oscillatory motion, the average of the velocity is zero, and then we establish that this constant is zero.

vi. Potential flows for incompressible fluids. We first recall that we define an incompressible fluid by: $\rho = \text{constant}$, throughout the volume of the fluid and its motion. This means that there cannot be noticeable compression or expansion of the fluid. Following the continuity equation, this implies: $\text{div}(\mathbf{v}) = 0$. We finally recall that, for an incompressible fluid, we have: $d\epsilon = 0$ (always with the isentropic hypothesis). This implies that $\epsilon$ is constant, and since constant terms in the energy do not matter, the energy flux density for an incompressible fluid becomes: $\rho \mathbf{v} \left( \frac{1}{2} \mathbf{v}^2 + P/\rho \right)$. Similarly, the enthalpy h can be replaced by $P/\rho$ in the equation established in (iii). This leads to:

$$\frac{\partial \phi}{\partial t} + \frac{v^2}{2} + P/\rho = f(t).$$

Now, we combine the two equations: $\text{div}(\mathbf{v}) = 0$ and $\boldsymbol{\omega} = \mathbf{curl}(\mathbf{v}) = \mathbf{0}$ (or $\mathbf{v} = \mathbf{grad}(\phi)$).

We get:

$$\Delta(\phi) = 0.$$

This is the Laplace's equation for the potential $\phi$. In order to solve this equation, it must be supplemented by boundary conditions (see §6). For example, at fixed solid surfaces, where the fluid meets solid bodies, the fluid velocity component normal to the surface ($v_n$) must be zero. For moving surfaces, it must be equal to the normal component of the velocity of the surface (which can be a function of time). Note that the following relation holds: $v_n = \frac{\partial\phi}{\partial n}$. Therefore, the general expression of boundary conditions is that $\frac{\partial\phi}{\partial n}$ is a given function of coordinates and time at boundaries. We show how to solve the Laplace's equation with specific boundary conditions in some exercises below.

Two other (less simple) consequences of the Kelvin's theorem:

i. Vortex lines move with the fluid: consider a tube of particles, which at some instant forms a vortex tube, which means a tube of particle with a given value of:

$$\oint \mathbf{v}.\,d\mathbf{l} = K.$$

Then, at that time, the circulation of the velocity round any contour C' lying in the tube without embracing the tube is zero, while for any contour embracing the tube (once), the circulation of velocity is equal to K. These values of the velocity circulation do not change moving with the fluid. This means that the vortex tube remains a vortex tube with an invariant: $K = \oint \mathbf{v}.\,d\mathbf{l}$. A vortex line is a limiting case of vortex tube and therefore vortex lines moves with the fluid (under the hypothesis of the Kelvin's theorem).

ii. The direction of vorticity turns as the vortex line turns, and its magnitude increases as the vortex line is stretched: the circulation round a thin vortex tube remains the same. As it stretches the area of the section decreases and thus the vorticity (~circulation/area) increases in proportion to the stretch.

Summary of some important relations for ideal fluids:

| | |
|---|---|
| ρ=constant. | Incompressibility condition. |
| $$\frac{\partial \mathbf{curl}(\mathbf{v})}{\partial t} = \mathbf{curl}[\mathbf{v} \times \mathbf{curl}(\mathbf{v})].$$ | Vorticity equations (general). |
| $$\frac{\partial \boldsymbol{\omega}}{\partial t} + (\mathbf{v}.\mathbf{grad})(\boldsymbol{\omega}) = (\boldsymbol{\omega}.\mathbf{grad})(\mathbf{v}).$$ In 2-dimensional space the term on the right hand side comes to zero and the vector equation is reduced to: $$\frac{d\boldsymbol{\omega}}{dt} = \mathbf{0}.$$ | Vorticity equations (incompressible fluid). |
| The circulation of the velocity along the closed curve $C_t$ (moving with the flow) is conserved (Kelvin's theorem): $$\oint_{C_t} \mathbf{v}.d\mathbf{l} = \oint_{C_0} \mathbf{v}.d\mathbf{l}.$$ | Incompressible (ideal) fluid (isentropic flow). |
| Potential flow: $\mathbf{v} = \mathbf{grad}(\phi)$. $$\frac{\partial \phi}{\partial t} + \frac{v^2}{2} + h = f(t).$$ | Potential flow. |
| Potential and steady flow: $$\frac{v^2}{2} + h = \text{constant}.$$ This 'constant' is constant throughout the fluid. | Potential flow (steady). |
| Potential and incompressible flows: $$\frac{\partial \phi}{\partial t} + \frac{v^2}{2} + P/\rho = f(t).$$ $$\Delta(\phi) = 0.$$ | Potential flow (incompressible). |

## §10. Real fluids and Navier-Stokes equations

We now study the effect of energy dissipation, occurring during the motion of a fluid, on that motion itself. This process is the result of the thermodynamic irreversibility of the motion. This irreversibility is always present to some extent, and is due to internal friction (viscosity) and thermal conduction. In the following, we always assume that the fluid is incompressible ($\rho$=constant).

In order to obtain the equations of motion of a viscous incompressible fluid, we have to include some additional terms in the equations of motion of an ideal fluid.

i. First the equation of continuity, as it is clear from its derivation, is equally valid for any fluid, whether viscous or not.
ii. Then, Euler's equations requires modifications, that are described this section.

We have seen in §5 that Euler's equations can be written in the form:

$$\frac{\partial}{\partial t}\int \rho v_i dV = -\sum_{k=x,y,z} \oint \Pi_{ik} d\Sigma_k.$$

Where $\Pi_{ik}$ is the momentum flux density tensor (of rank 2): it depends on 2 indices i and k, running for x, y, z in 3-dimensional space. For ideal fluids, we have $\Pi_{ik} = p\delta_{ik} + \rho v_i v_k$ (with $\delta_{ik} = 1$ if $i = k$ and 0 otherwise), which represents a completely reversible transfer of momentum, originating from the mechanical transport of the different particles of the fluid from place to place with pressure forces acting in the fluid. As already mentioned, the hypothesis behind ideal fluids is that each particle pushes its neighbors equally in every direction. This is why a single scalar quantity, the pressure (p), is sufficient to describe the force per unit area that a particle exerts on all its neighbors at a given time. Then, the acceleration of the fluid particle results from the pressure differences.

For real fluid, for which energy dissipation in the fluid is not neglected any longer, the viscosity or internal friction is due to another, irreversible, transfer of momentum from points where the velocity is large to those where it is small. This means that if a fluid particle moves faster than the average of its neighbors, then friction slows it down. The equation of motion for real fluids can then be obtained by modifying the momentum flux tensor accordingly $\Pi_{ik} = p\delta_{ik} + \rho v_i v_k + \Pi'_{ik}$, where $\Pi'_{ik}$ designs the part of the flux density due to viscosity. A general form of $\Pi'_{ik}$ can be established using the fact that processes related to internal friction occur in a fluid only when different particles move with different velocities, such that there is a relative motion between various parts of the fluid. Hence, $\Pi'_{ik}$ must depend on the space derivatives of the velocity. As a first approximation, $\Pi'_{ik}$ can thus be written as a linear combination of terms of the form $\frac{\partial v_i}{\partial x_k}$.

This way of thinking can be continued to obtain the general vector equation of motion of incompressible viscous fluid, for which the viscosity is determined by only one coefficient:

$$\frac{d\mathbf{v}}{dt} = \frac{\partial \mathbf{v}}{\partial t} + (\mathbf{v}.\mathbf{grad})(\mathbf{v}) = -\frac{1}{\rho}\mathbf{grad}(p) + \frac{\eta}{\rho}\Delta\mathbf{v}. \tag{6}$$

This is the Navier-Stokes (vector) equation, first established by Navier in 1822 and then by Stokes in 1845. Other derivations have been proposed in between by Cauchy in 1828 and Poisson in 1829, but the history has only kept the names of Navier and Stokes. Here, $\eta > 0$ is called the viscosity coefficient (precisely the dynamic viscosity), while $\nu = \frac{\eta}{\rho}$ is called the kinematic viscosity. In Cartesian coordinates, we can write equations (6) as a set of 3 equations for the 3 components in x, y and z:

$$\left(\frac{\partial}{\partial t} + v_x \frac{\partial}{\partial x} + v_y \frac{\partial}{\partial y} + v_z \frac{\partial}{\partial z}\right)\begin{bmatrix} v_x \\ v_y \\ v_z \end{bmatrix} = -\frac{1}{\rho}\begin{bmatrix} \frac{\partial p}{\partial x} \\ \frac{\partial p}{\partial y} \\ \frac{\partial p}{\partial z} \end{bmatrix} + \frac{\eta}{\rho}\Delta\begin{bmatrix} v_x \\ v_y \\ v_z \end{bmatrix}.$$

Some general comments are in order:

i. In general the viscosity coefficient is a function of pressure and temperature of the fluid. As pressure and temperature may not be constant throughout the fluid, the viscosity coefficient also may not be constant throughout the fluid. Obviously, the viscosity coefficient is much larger for glycerin than for water. Typical values at 20°C are $\frac{\eta}{\rho} = 0.01\ \mathrm{cm^2/s}$ for water, $0.15\ \mathrm{cm^2/s}$ for air and above $6\ \mathrm{cm^2/s}$ for glycerin. Remark that the unit for $\frac{\eta}{\rho}$ is in $\mathrm{cm^2/s}$ as be seen immediately from equations (6). Also, we can mention that the dynamic viscosity ($\eta$) of a gas at a given temperature is independent of the pressure, while the kinematic viscosity ($\frac{\eta}{\rho}$) is inversely proportional to the pressure. Also, it can be shown that the kinematic viscosity of a gas is about: $\Lambda\upsilon$, where $\Lambda$ is the mean free path and $\upsilon$ the thermal speed of molecules (of the gas), of the same order of magnitude as sound speed. It presents the correct unit (in length/time$^2$). With this expression, we understand that viscosity is a vestige of the continuum limit based on the molecular nature of the fluid while Euler's equations, corresponding to zero viscosity, set this molecular length scale to zero (see §2).

ii. The Navier-Stokes equations (6) represent a vector equation (with as many equation as dimensions of space), which generalizes the Euler's vector equation due to the presence of a diffusive term in $\frac{\eta}{\rho}\Delta\mathbf{v}$. The convective term in $(\mathbf{v}.\mathbf{grad})(\mathbf{v})$, which appears in the material derivative, is not affected by the internal friction phenomenon.

iii. The Navier-Stokes equations (6) need to be complemented by the continuity relation for incompressible (real) fluids, div($\mathbf{v}$)=0, as well as initial and boundary values to compute the velocity field for later times $t > 0$ (next sections). Of course, in the presence of external forces, we can add a term in $\frac{1}{\rho}\mathbf{F}_{\mathrm{ext}}$ to equations (6).

iv. A simple derivation of the viscosity term can be proposed, without the use of the tensor formalism. The idea is to make the parallel between two properties: (1) the physics property that frictional forces counteract at each point the deviation of the velocity field from its local average and (2) the mathematical fact that the deviation of a function at a point from its average value on small surrounding spheres is measured by the negative of the Laplace's operator ($\Delta$). This implies trivially that frictional (vector) forces must be proportional to $\Delta\mathbf{v}$. Adding such a term to Euler's vector equation, we obtain the Navier-Stokes vector equation (6) for incompressible fluids. We let the justification of the equivalence between (1) and (2) above as an exercise. The key point here is to see the intricacy between the physical intuition and mathematics: once we have the physics idea (1), then equations (6) is a direct consequence of the mathematical fact (2).

Using equations (6), we can verify that the presence of viscosity results in the dissipation of energy, which is finally transformed into heat: this must be a direct consequence of the Navier-Stokes equations (6). The calculation of energy dissipation is quite simple for an incompressible fluid filling a volume of space Ω. Indeed, the total kinetic energy for the fluid contained in Ω reads:

$$E_K = \rho \int_{\Omega} \frac{1}{2}\mathbf{v}^2 dV \quad (\text{with } \rho = \text{constant})\ .$$

Then, it can be shown easily that $E_K$ is decreasing in time under a flow following the Navier-Stokes equations (6). Precisely, we can compute for a 3-dimensional space:

$$\frac{dE_K}{dt} = -\eta \sum_{i=1}^{3} \int_{\Omega} |\mathbf{grad}(u_i)|^2 dV = -\eta \sum_{i=1}^{3} \int_{\Omega} |\boldsymbol{\nabla} u_i|^2 dV < 0.$$

When the viscosity is zero (Euler's equations), we find that the kinetic energy is conserved for an incompressible fluid (as already shown), while for viscous fluid $\frac{dE_K}{dt} < 0$: friction transforms kinetic energy into heat.

Exercise: For a 3-dimensional space, prove that:

$$\frac{dE_K}{dt} = -\rho\eta \sum_{i=1}^{3} \int_{\Omega} |\mathbf{grad}(u_i)|^2 dV = -\rho\eta \sum_{i=1}^{3} \int_{\Omega} |\boldsymbol{\nabla} u_i|^2 dV < 0.$$

Where $E_K$ is the kinematic energy of the incompressible fluid contained in the volume Ω.
At boundaries of the volume containing the fluid (labeled as S in the following), we assume that the fluid is at rest with $v_n$=0.

Solution: We compute the time derivative of the kinematic energy:

$$\frac{dE_K}{dt} = \rho\frac{d}{dt}[\int_\Omega \frac{1}{2}\mathbf{v}^2dV] = \int_\Omega \mathbf{v}.\rho\left[\frac{\partial\mathbf{v}}{\partial t}\right]dV = \int_\Omega \mathbf{v}.[-\rho(\mathbf{v}.\mathbf{grad})(\mathbf{v}) - \mathbf{grad}(p) + \eta\Delta\mathbf{v}]dV.$$

There are 3 terms to be calculated. The first 2 terms are zero. For the first one, we find using the well-known vector identity for $(\mathbf{v}.\mathbf{grad})(\mathbf{v})$ and integrating by parts:

$$\int_\Omega \mathbf{v}.[(\mathbf{v}.\mathbf{grad})(\mathbf{v})]dV = \int_\Omega \frac{1}{2}\mathbf{v}.\mathbf{grad}(\mathbf{v}^2)dV = -\frac{1}{2}\int_\Omega \operatorname{div}(\mathbf{v})\,\mathbf{v}^2dV + \frac{1}{2}\int_S \mathbf{v}^2\,\mathbf{v}.d\boldsymbol{\Sigma} = 0.$$

Similarly, we have: $\int_\Omega \mathbf{v}.[\mathbf{grad}(p)]dV = -\int_\Omega \operatorname{div}(\mathbf{v})\,pdV + \int_S p\,\mathbf{v}.d\boldsymbol{\Sigma} = 0$. Finally, we are left with the last term proportional to the viscosity:

$$\frac{dE_K}{dt} = \eta\int_\Omega \mathbf{v}.[\Delta\mathbf{v}]dV.$$

The integrand: $\mathbf{v}.[\Delta\mathbf{v}]$ can be expanded as: $v_1.[\Delta v_1] + v_2.[\Delta v_2] + v_3.[\Delta v_3]$. Each of the 3 integrals (like $\int_\Omega v_1.[\Delta v_1]dV$) can be transformed using integration by parts. We obtain:

$$\frac{dE_K}{dt} = -\eta\sum_{i=1}^{3}\int_\Omega |\mathbf{grad}(v_i)|^2dV < 0.$$

Consequently, the kinetic energy is decreasing in time, which reflects the losses due to friction in a viscous flow. It is also possible to write the last formula as:

$$\frac{dE_K}{dt} = -\eta\int_S \mathbf{grad}(v^2).d\boldsymbol{\Sigma}.$$

Finally, we write the vector equation for the vorticity by taking the **curl** of the Navier-Stokes equations. We obtain:

$$\frac{\partial\mathbf{curl}(\mathbf{v})}{\partial t} = \mathbf{curl}[\mathbf{v}\times\mathbf{curl}(\mathbf{v})] + \nu\Delta(\mathbf{curl}(\mathbf{v})).$$

Equivalently, in a form generalizing the stretching formula to (incompressible) viscous fluids:

$$\frac{d\boldsymbol{\omega}}{dt} = \frac{\partial\boldsymbol{\omega}}{\partial t} + (\mathbf{v}.\mathbf{grad})(\boldsymbol{\omega}) = (\boldsymbol{\omega}.\mathbf{grad})(\mathbf{v}) + \nu\Delta(\boldsymbol{\omega}) \text{ with } \boldsymbol{\omega} = \mathbf{curl}(\mathbf{v}).$$

In this expression, we recognize transport, stretch as well as diffusion of vorticity. Note that even if the vorticity is null at initial time, it does not imply that it will last for later time since vorticity can be created by boundary conditions. Also in 2-dimensional space, we obtain:

$$\frac{d\boldsymbol{\omega}}{dt} = \nu\Delta(\boldsymbol{\omega}).$$

Here, only one component of $\boldsymbol{\omega}$ is non zero: the vorticity is thus a scalar quantity. Vorticity is transported through convection and diffuses. Of course, in the absence of viscosity ($\nu = 0$) we find again that vorticity is a conserved quantity in 2-dimensional space.

## §11. Boundary conditions for real fluids

We must also write down the boundary conditions on the equations of motion of a viscous fluid. There are always forces of molecular attraction between viscous fluid and the surface of a solid body, and these forces have the result that the layer of fluid immediately adjacent to the surface is brought completely to rest, and adheres to the surface. Accordingly, the boundary conditions require that the fluid velocity should vanish at fixed solid surfaces: $\mathbf{v} = 0$. It can be emphasized that both the normal and tangential velocity must vanish, whereas for an ideal fluid, the boundary conditions require only that the normal component vanish. Obviously, in the general case of a moving surface, the velocity (vector) $\mathbf{v}$ must be equal to the velocity of the surface.

We consider one example to illustrate how these conditions operate together with the equations of motion for a real fluid. Consider the following viscous incompressible flow between two stationary plates located at y=0 and y=1, with the notations indicated on the figure. This is obviously a 2D configuration, where only x and y component (in Cartesian coordinates) should be considered.

We are looking for a stationary solution of Navier-Stokes equations of the form $\mathbf{v}$=(u(x, y), 0). Indeed, the flow is directed only along the x-axis. With: p=p(x) and p(0)=$p_1$, p(L)=$p_2$. The continuity equation for incompressible fluid gives $\partial_x u = 0$, where we use the standard notation $\partial_x = \partial / \partial x$. Hence u(x,y)=u(y) and **v**=(u(y), 0). The Navier-Stokes equation for the only component to be considered reads:

$$\partial_t u + u\,\partial_x u = -\frac{1}{\rho}\partial_x p + \frac{\eta}{\rho}[\partial_{xx}u + \partial_{yy}u] \text{ where } \partial_{xx} = \left(\frac{\partial}{\partial x}\right)^2 \text{ and } \partial_{yy} = \left(\frac{\partial}{\partial y}\right)^2.$$

Where we have: $\partial_t u = 0$, $\partial_x u = 0$ and $\partial_{xx}u = 0$. We obtain:

$$-\frac{1}{\rho}\partial_x p + \frac{\eta}{\rho}\partial_{yy}u = 0 \text{ or equivalently } \partial_x p = \eta\,\partial_{yy}u.$$

With the boundary conditions: $u(x,0) = 0$ and $u(x,1) = 0$. Since both sides of the equation $\partial_x p = \eta\,\partial_{yy}u$, depends on different variables, we conclude that $\partial_x p = \eta\,\partial_{yy}u = \text{constant}$, which gives trivially the result. The velocity profile is a parabola.

We now add a level of complexity to the previous example. There is one plate at y=0. Assume that the half space $y > 0$ is filled with a fluid, bounded by a plate (xz-plane) at
$y = 0$. The fluid is not moving for $t \leq 0$. The plate at the fluid boundary starts to move at initial time $(t = 0)$ with constant velocity U in the x-direction. We may assume that the fluid starts to flow due to friction. In this case we look for a solution of the Navier Stokes equations of the form $\mathbf{v} = (u(x,y,t), 0, 0)$, where $u(x,y,t) = u(y,t)$ in order to satisfy the continuity relation. The Navier-Stokes equations read:

$$\partial_t u = -\frac{1}{\rho}\partial_x p + \frac{\eta}{\rho}\partial_{yy}u, \text{with } \partial_y p = \partial_z p = 0.$$

This implies that $\partial_x p = \text{constant}$, that we may choose to be zero. Then, we are left a the pure diffusion equation $\partial_t u = \nu\,\partial_{yy}u$ (with $\nu = \frac{\eta}{\rho}$) together with the boundary conditions u(0,t)=U and the initial condition u(y,0)=0. We can pose this problem for the dimensionless variable: $\bar{u} = u/U$. We obtain: $\partial_t \bar{u} = \nu\,\partial_{yy}\bar{u}$, $\bar{u}(y,0) = 0$ and $\bar{u}(0,t) = 1$. Since $\bar{u}$ is dimensionless, it can only be a function of dimensionless variable(s), combining the variables y, ν and t. There is only one such variable that can be built $\frac{y}{\sqrt{\nu t}}$.

Therefore, we can look for u as a function of the variable $\xi = \frac{y}{2\sqrt{\nu t}}$.

Hence, we can write: $\bar{u} = F(\xi)$, with: $\partial_{\xi\xi}F + 2\xi F = 0$, $F(0) = 1$ and $F(\infty) = 0$. The solution is well known: the complementary error function. It reads:

$$F(\xi) = \text{erfc}(\xi) = \frac{2}{\sqrt{\pi}}\int_{\xi}^{\infty} \exp(-s^2)\, ds.$$

Finally, we can transform easily this solution to obtain:

$$u(y,t) = U.\, \text{erfc}\left(\frac{y}{2\sqrt{\nu t}}\right).$$

In particular, we observe that at fixed distance to the plate, the velocity of the fluid will converge to the velocity of the plate at the limit of infinite time.

From these examples, we can remark that the solution of a problem for a viscous fluid for which we take the limit $\nu \to 0$ is not equivalent to the similar problem solved in the ideal case. Indeed, the boundary conditions are different in both cases. Then, even if the equations of motion would be similar in the limit of zero viscosity, the general solution of a problem with given boundaries would not be equivalent.

Finally, we come back to the formulation of the Navier-Stokes equations in the form:

$$\frac{\partial}{\partial t}\int \rho v_i dV = -\sum_{k=x,y,z} \oint \Pi_{ik} d\Sigma_k.$$

Here, the index i represents the Cartesian coordinates x, y and z. The correspondence with equations (6) leads to: $\Pi_{ik} = p\delta_{ik} - \eta\left(\frac{\partial v_i}{\partial x_k} + \frac{\partial v_k}{\partial x_i}\right) + \rho v_i v_k$. Then, the expression above is exactly equivalent to Navier-Stokes equations (6). The interest of the last formula is that it is easy to write down from it an expression of the force acting on a solid surface bounding the fluid. The force acting on an element of the surface is just the momentum flux through this element. The momentum flux through surface $d\mathbf{\Sigma}$ is $\sum_{k=x,y,z} \Pi_{ik} d\Sigma_k$. We write the element of surface along the k-axis as $d\Sigma_k = d\Sigma.\, n_k$, where $\mathbf{n}$ is the unit vector along the normal (along the outward normal to the fluid).

Then, as **v**=**0** at the solid surface, we conclude that the force (along the i-axis) acting on a unit surface of the body is equal to:

$$\sum_{k=x,y,z}\left[p\delta_{ik}-\eta\left(\frac{\partial v_i}{\partial x_k}+\frac{\partial v_k}{\partial x_i}\right)\right]n_k = pn_i-\eta\sum_{k=x,y,z}\left(\frac{\partial v_i}{\partial x_k}+\frac{\partial v_k}{\partial x_i}\right)n_k.$$

The first term is the ordinary pressure of the fluid acting on the surface of the body, while the second is the force of friction, due to viscosity.

Note also that if we have a surface of separation between 2 immiscible fluids, the conditions at the surface are that the velocities of the fluids must be equal and the forces which they exert on each other must be equal and opposite. This is a generalization to the case of real fluids of the condition of the continuity of pressure for the ideal case.

### §12. Reynolds number and related properties

We define a macroscopic (dimensionless) number corresponding to the ratio of the strength of the non-linear effects to the strength of the linear viscous effects. In order to define this quantity, we need to introduce a characteristic scale U for the velocity and a characteristic length scale L of the flow. In addition, we write the characteristic time scale as T=L/U. Then, we can pose the dimensionless parameter we are looking for, the Reynolds number, Re=UL/ν. It is obvious to check that it corresponds to the ratio mentioned above. We remark that the (kinematic) viscosity ν has dimension $L^2/T$, using these characteristic scales. This corresponds to the diffusivity of the velocity: namely, in time T, velocity diffuses over a distance of order $\sqrt{\nu T}$. For example, for water at ambient temperature and pressure, $\nu\sim 1mm/s^2$, so direct viscous effects diffuse velocity at a distance of order 1mm in one second. Also, if we consider the flow of water with a speed of 1 m/s in a tank of 1 m, we obtain: $Re=10^6$.

We now introduce the dimensionless variables: (x', y', z')=(x/L, y/L, z/L), t'=t/T, **v'**=**v**/U. The Navier-Stokes equations can then be re-written using the prime (dimensionless) variables in the form:

$$\frac{\partial \mathbf{v}'}{\partial t'} + (\mathbf{v}'.\mathbf{grad}')(\mathbf{v}') = -\mathbf{grad}'(p') + \frac{1}{Re}\Delta'\mathbf{v}'.$$

Here, we can check that $p' = p/(\rho U^2)$. With this expression, we can see that solutions present a scaling invariance $\mathbf{v} \to \lambda \mathbf{v}(\lambda \mathbf{x}, \lambda^2 t)$, up to the limit of the continuum hypothesis. This leads to the law of similarity: flows which can be obtained from one another by simply changing the unit of measurement of coordinates and velocities are said similar. Thus flows of the same type and same Reynolds numbers are similar.

As the Reynolds number is a standard referenced quantity for various flows, this last expression is useful in order to make some comparisons between the relevant terms of the equation:

i. For example, we may think that we can neglect the viscous term the Navier-Stokes equations in comparison with the convective term when the Reynolds number is sufficiently large (at small viscosity). However, the Navier-Stokes equations correspond to a singular perturbation of the Euler equations, since the viscosity multiplies the term that contains the highest-order spatial derivatives. As a result, this is not always possible to operate this simplification. The high Reynolds number limit of the Navier-Stokes equations is a very difficult problem, where turbulent effects may dominate the dynamic of the flow.
ii. On the other hand, the limit of small Reynolds numbers is particularly simple. For steady flow of incompressible viscous fluid, the Navier-Stokes equations read:

$$(\mathbf{v}.\mathbf{grad})(\mathbf{v}) = -\frac{1}{\rho}\mathbf{grad}(p) + \frac{\eta}{\rho}\Delta\mathbf{v}.$$

Using the same notations for characteristic scales as defined above, the convective term is of the order $U^2/L$,. The diffusive term is of the order of magnitude $\frac{\eta}{\rho}U/L^2$. The ratio of the two is the Reynolds number (by definition). Hence, the convective term may be neglected if the Reynolds number is small, and the equations become linear: $-\mathbf{grad}(\mathrm{p}) + \eta\Delta\mathbf{v} = 0$. Together with the continuity equation $\mathrm{div}(\mathbf{v}) = 0$, it completely determines the motion of the fluid. It is useful to note that:

$$-\mathbf{grad}(\mathrm{p}) + \eta\Delta\mathbf{v} = 0 \text{ implies } \Delta\mathbf{curl}(\mathbf{v}) = 0.$$

Using these formulae, it is possible to determine the force exerted on a 'fixed' sphere (of radius R) by a moving fluid, having a given velocity **U** constant at infinity, far away from the sphere body. This force is also called the drag force (F). As already shown in previous sections, the solution of this problem with the Euler's formalism gives a resulting null force. Using the Navier-Stokes formalism, in the small Reynolds number approximation, we expect that the answer is non-zero due to the presence of the viscosity coefficient. Indeed, it can be shown that $\mathrm{F} = 6\pi\eta\mathrm{R}||\mathbf{U}||$, with a force parallel to the velocity direction **U**. This formula is called the Stokes formula. After a trivial change of reference frame, it gives the drag force on a sphere moving slowly in a fluid at rest at infinity. We notice that the drag is proportional to the first power of the velocity and viscosity coefficient, as well as linear in the dimension of the body. These are general properties, whatever the exact shape of the body considered.

Finally, we discuss an interesting physics effect for real fluids: boundary layers that we have briefly mentioned in §8. We consider a simple example first. Assume that the half space $\mathrm{y} > 0$ is filled with a fluid, bounded by a plate (xz-plane) at $\mathrm{y} = 0$. We consider that the plate is fixed and the fluid moves at constant velocity U along the x-axis at $\mathrm{y} = \infty$. We have solved a similar problem in the previous section. Here, only boundary conditions are changed compared to our previous discussion. They read: $\mathrm{u}(\mathrm{y} = \infty, \mathrm{t}) = \mathrm{U}$ and $\mathrm{u}(0, \mathrm{t}) = 0$.
The solution follows:

$$\mathrm{u}(\mathrm{y}, \mathrm{t}) = \mathrm{U}.\,\mathrm{erf}\left(\frac{\mathrm{y}}{2\sqrt{\nu \mathrm{t}}}\right) \text{ where } \mathrm{erf}(\xi) = \frac{2}{\sqrt{\pi}}\int_0^{\xi} \exp(-\mathrm{s}^2)\,\mathrm{ds}$$

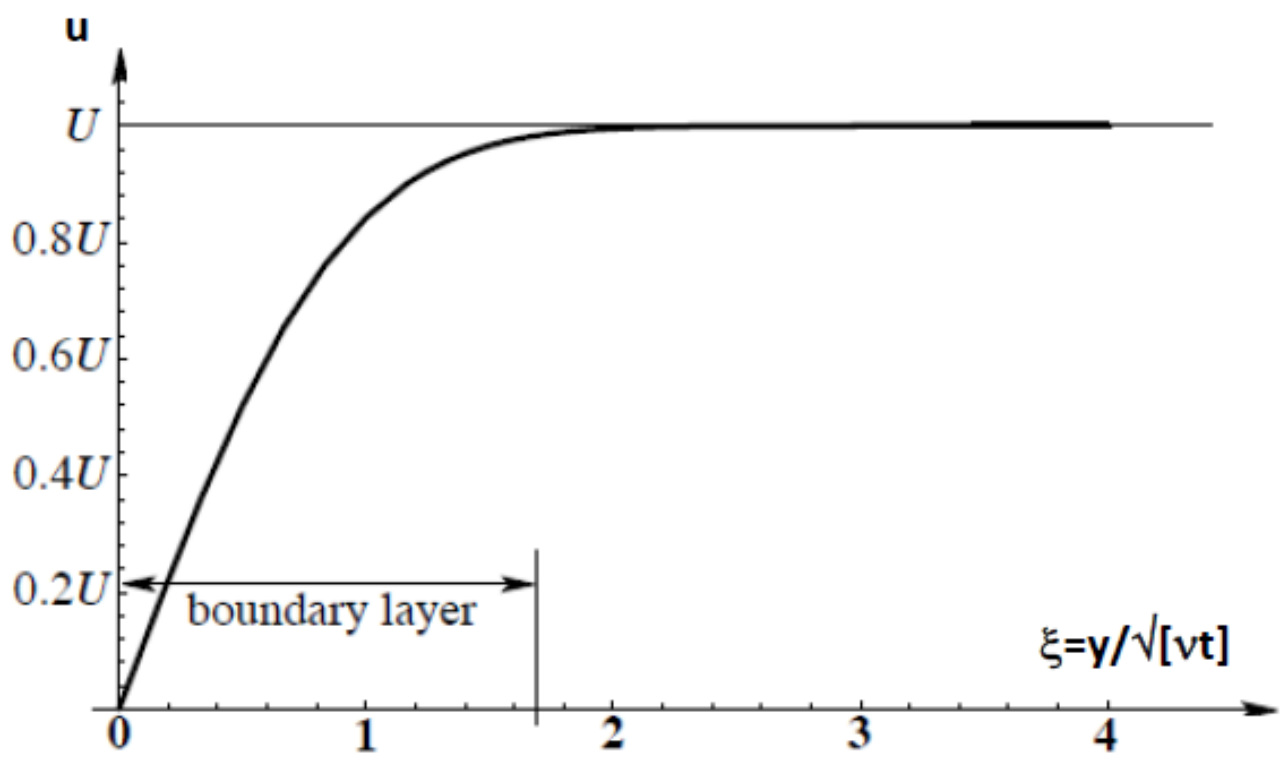

The region in which the velocity field departs significantly from the constant flow U is called boundary layer: according to the expression of $u(y,t)$, this layer is proportional to $\sqrt{\nu t}$. Hence, for fixed time, the boundary layer decreases as $1/\sqrt{\mathrm{Re}}$.

We summarize the differences between incompressible Euler's and Navier-Stokes equations:

| Navier-Stokes: | Euler: |
|---|---|
| ρ=constant, taken to be 1 (in proper units) | ρ=constant, taken to be 1 (in proper units) |
| $\frac{\partial \mathbf{v}}{\partial t} + (\mathbf{v}.\mathbf{grad})(\mathbf{v}) = -\mathbf{grad}(p) + \frac{1}{\mathrm{Re}}\Delta\mathbf{v}$ | $\frac{\partial \mathbf{v}}{\partial t} + (\mathbf{v}.\mathbf{grad})(\mathbf{v}) = -\mathbf{grad}(p)$ |
| $\mathrm{div}(\mathbf{v})=0$ | $\mathrm{div}(\mathbf{v})=0$ |
| **v**=**0** at fixed boundary | **v.n**=0 at fixed boundary |

We assume that both types of flows coincide at initial time and that both flows are irrotational at initial time. Under Euler's formalism, the flow stays irrotational at all times. Under Navier-Stokes equations, the effect observed in the example above can be generalized: the flow is drastically modified (compared to the ideal case) near the boundary in a region with thickness proportional to $1/\sqrt{\mathrm{Re}}$. In addition, this (local) effect may be a source of vorticity.

## §13. The millennium problem of the Clay Institute

With all the knowledge of the previous sections, we can pose the millennium problem related to Navier-Stokes equations.

Given the Navier-Stokes equations for incompressible fluids in 3-dimensional space:

$$\frac{\partial \mathbf{v}}{\partial \mathrm{t}} + (\mathbf{v}.\mathbf{grad})(\mathbf{v}) = -\frac{1}{\rho}\mathbf{grad}(\mathrm{p}) + \frac{\eta}{\rho}\Delta\mathbf{v} \text{ and } \mathrm{div}(\mathbf{v}) = 0$$

(NSI)

*for Navier − Stokes − Incompressible*

with $\mathbf{v} = \mathbf{v}(\mathbf{x}, \mathrm{t} \geq 0)$ and $\mathrm{p} = \mathrm{p}(\mathbf{x}, \mathrm{t} \geq 0)$.

Given smooth (divergence free and infinitely differentiable) initial conditions:

$$\mathbf{v_0}(\mathbf{x}) = \mathbf{v}(\mathbf{x}, \mathrm{t} = 0).$$

Given boundary conditions at infinity: the statement of the problem indicates that for physically reasonable solutions, such that the velocity field does not grow large as $\|\mathbf{x}\| \to \infty$, the space of initial conditions is restricted to functions for which the norm of space derivatives of any order (written schematically as: $\|\partial_{\mathbf{x}}^{\alpha}\mathbf{v_0}\|$) is bounded. Precisely, for any order $\alpha$ and for any constant K, there exists a constant $\mathrm{C}(\alpha, \mathrm{K})$ such that:

$$\|\partial_{\mathbf{x}}^{\alpha}\mathbf{v_0}(\mathbf{x})\| \leq \frac{\mathrm{C}(\alpha, \mathrm{K})}{(1 + \|\mathbf{x}\|)^{\mathrm{K}}}.$$

Here the norm of a function $\mathrm{f}: \Omega \to \mathrm{IR}$ has to be understood as $\|\mathrm{f}\| = \left(\int_{\Omega} |\mathrm{f}(\mathrm{x})|^2 \mathrm{dx}\right)^{\frac{1}{2}}$.

Then, prove that:

> Option 1: there exists $\mathbf{v}(\mathbf{x}, \mathrm{t} \geq 0)$ and $\mathrm{p}(\mathbf{x}, \mathrm{t} \geq 0)$ solutions of (NSI) which are infinitely differentiable in space and time coordinates (for $\mathrm{t} \geq 0$), with bounded kinetic energy over the all space. Precisely, there exists a constant such that: $\int \|\mathbf{v}(\mathbf{x}, \mathrm{t})\|^2 \mathrm{d}^3\mathbf{x} <$ constant for $\mathrm{t} \geq 0$.

Option 2: there exists a divergence free and infinitely differentiable vector field $\mathbf{v_0}(\mathbf{x}, \mathrm{t} = 0)$ satisfying the boundary conditions and a smooth external force vector field, such that there exists no solution of (NSI) infinitely differentiable in space and time coordinates (for $\mathrm{t} \geq 0$) and with bounded energy over the all space. In this option, the force vector field $\mathbf{f}(\mathbf{x},\mathrm{t})$ is simply added to the vector equation (NSI). It is assumed to be bounded at infinity. Precisely, for any orders α and β and for any constant K, there exists a constant C(α, β, K) such that:

$$\left\|\partial_{\mathbf{x}}^{\alpha}\,\partial_{\mathrm{t}}^{\beta}\mathbf{f}(\mathbf{x},\mathrm{t})\right\| \leq \frac{\mathrm{C}(\alpha,\beta,\mathrm{K})}{(1+\|\mathbf{x}\|+\mathrm{t})^{\mathrm{K}}}.$$

Options 1 or 2 are equally good to solve the problem stated by the Clay mathematical institute (CMI) and get the 1 million dollars prize.

Alternatively, instead of taking boundary conditions with strong decay at infinity, we can consider periodic boundary conditions in space coordinates for $\mathbf{v_0}$, $\mathbf{v}$ and $\mathbf{f}$, and one additional condition for the external force: $\left\|\partial_{\mathbf{x}}^{\alpha}\,\partial_{\mathrm{t}}^{\beta}\mathbf{f}(\mathbf{x},\mathrm{t})\right\| \leq \frac{\mathrm{C}(\alpha,\beta,\mathrm{K})}{(1+\mathrm{t})^{\mathrm{K}}}$. Then, options 1 and 2 can be posed in a similar way as done above.

At the end, with the 2 types of boundary conditions, we have 4 possible statements of the problem. The CMI asks for the proof of one of them to get the prize.

In short, this problem raises the question of whether arbitrary smooth (also called regular) solutions of the incompressible Navier-Stokes equations in 3-dimensional space can be continued globally from smooth initial data or not. Either, one needs to prove that initially smooth solutions with strong decay conditions at infinity (or with periodic boundary conditions) remain smooth for all times, or one needs to find at least one solution which blows up in finite time.

If this problem is solved positively, it would imply:

i. The existence of solutions: the system described by the equations must have a way to evolve in the future.
ii. The uniqueness: there must not be arbitrary choices for this evolution.
iii. And the continuous dependence on the initial state: any future state of the system is determined, to arbitrary finite precision, by the initial conditions to a sufficient finite precision.

Obviously, a complete answer to this problem is still open. However, what is known is that this problem can be solved locally! Starting out from divergence free and infinitely differentiable (smooth) initial conditions $\mathbf{v_0}(\mathbf{x})$, solutions for $\mathbf{v}(\mathbf{x}, \mathrm{t} \geq 0)$ and $\mathrm{p}(\mathbf{x}, \mathrm{t} \geq 0)$ are unique, depend continuously on the initial conditions, and remain smooth for at least, possibly short, interval of time: $[0, \mathrm{T}^*]$, where T* depends on $\mathbf{v_0}(\mathbf{x})$. It is not known if the solutions exist for $\mathrm{t} \geq \mathrm{T}^*$. This last statement holds for Navier-Stokes equations as well as Euler's equations. We prove rigorously this result in the next section.
Then, either a given solution (from a given initial condition) can be continued for all times or it exists only up to T* and the norm of the solution diverges when $\mathrm{t} \rightarrow \mathrm{T}^*$: the solution is blowing up at T*. Then, schematically, proving the problem globally reduces to finding a bound on the norm of the solution for all times.

Some interesting partial results are known:

1. In 2-dimensional space, the problem has been solved and there exists smooth and globally defined solutions. Very roughly, in 2-dimensional space, the dissipation of energy due to friction is sufficiently strong to prevent blow-up of finite energy solutions. Note that for the Euler's formalism, the same conclusion holds, for a simple reason that we discuss in detail later: in 2-dimensional space, vorticity is conserved as a scalar quantity along flow lines. The key point is that for Euler's formalism, it can be shown that any singularity of the velocity field or its derivatives of any order is necessarily a singularity in the vorticity. Then, the conservation of vorticity in 2-dimensional space is sufficient to prevent blow-up of finite energy solutions.

2. In 3-dimensional space, if $\mathbf{v}(\mathbf{x}, \mathrm{t} \geq 0)$ is sufficiently small, the problem is also solved positively for Navier-Stokes equations and there exists smooth and globally defined solutions. Much effort has been spent on characterizing smallness, in terms of smallness of initial conditions, of the viscosity being large, or of the solution being in some sense close to some known special solution. Physically, the effect of the diffusion term $\frac{\eta}{\rho}\Delta\mathbf{v}$ is so strong that any perturbation coming from the convective term $(\mathbf{v}.\mathbf{grad})(\mathbf{v})$ is damped away before it could lead to singularities. Intuitively, if fluid equations for water are in danger of developing singularities, we replace water by honey, sufficiently viscous, and no singularity can develop. Obviously, these kinds of results are not available for Euler's equation, where the viscosity is absent.
3. Finally, another class of partial results known to hold for Navier-Stokes equations, but not for Euler's formalism, concerns the existence of the so-called weak solutions (J. Leray 1933). In fact, solutions of Navier-Stokes equations can be continued past the time of their first possible singularity as weak or generalized solutions. We do not want to describe this mathematical issue here. This means that equations are satisfied by weak solutions in an average sense but not point by point. Weak solutions exist globally in time. However, they are not known to be unique!

From these first comments, we understand that the difficulty of the problem comes from the relative balance between the quadratic term and the diffusive term in equations while the evolution in time is realized. Similar observations have been done in §7 for much simpler dynamical systems. Indeed, for the one dimensional equation: $\frac{du}{dt} = -u + u^2$, we have shown rigorously that there is a global existence of solutions with small initial data and local existence (in time) of solutions with large initial data. Also, for the one dimensional equation: $\frac{\partial u}{\partial t} + \frac{\partial(\frac{u^2}{2})}{\partial x} = \frac{\partial u}{\partial t} + u\frac{\partial(u)}{\partial x} = 0$, we have noticed that a (unique) global smooth solution may not exist in general.

Clearly also, when the quadratic term can be neglected in Navier-Stokes equations (NSI), we obtain a much simpler system following:

$$\frac{\partial \mathbf{v}}{\partial \mathrm{t}} = -\mathbf{grad}(\mathrm{p}) + \nu\Delta\mathbf{v} \quad [\text{with } \mathrm{div}(\mathbf{v}) = 0].$$

Where we have posed ρ=1. Taking the divergence of the last expression, we obtain a single equation for the pressure field $\Delta \mathrm{p} = 0$, which decouples the pressure field dependence in space from the velocity field.

This system of equations (for $\mathbf{v} = \mathbf{v}(\mathbf{x}, \mathrm{t} \geq 0)$ and $\mathrm{p} = \mathrm{p}(\mathbf{x}, \mathrm{t} \geq 0)$) can be solved according to initial and boundary conditions as given, which justifies the conclusion of point (2) mentioned above.

For the general form of the Navier-Stokes equations, it is also possible to take the divergence of the equations. Then, the Laplace's operator of the pressure is more complex. Indeed we need to consider the divergence of the quadratic (convective) term. With simple algebra, we obtain:

$$-\Delta \mathrm{p} = \sum_{\mathrm{i,j}=1}^{3} \frac{\partial \mathrm{v_i}}{\partial \mathrm{x_j}} \frac{\partial \mathrm{v_j}}{\partial \mathrm{x_i}}.$$

We have used the notation: $(\mathrm{x_i})_{\mathrm{i}=1,2,3} = (\mathrm{x}, \mathrm{y}, \mathrm{z})$. This means that the pressure field is a given function of the velocities at the same instant time. Any change in the velocity field at a position $\mathbf{x}$ affects the pressure field immediately everywhere else. This is the subtle reason why no initial and boundary conditions have been mentioned for the pressure field in the statement of the problem at the beginning of this section. Otherwise the problem would be ill-posed. This effect is a direct consequence of the incompressible condition, which assumes that the sound speed is infinite compared to the magnitude of the flow speed. This implies that sound waves can carry any perturbations of the pressure field (δp) instantaneously across the entire volume of the fluid.

Finally, we conclude this presentation of the millennium problem by a remark concerning the interplay between Euler's and Navier-Stokes equations. The statement of the millennium problem is related only to Navier-Stokes equations in 3-dimensional space. However, as we have seen, the case of small Reynolds numbers is not difficult, when the quadratic term can be neglected, or at least do not dominate the evolution in time of the velocity field, while the case of high Reynolds numbers is really hard. Indeed, in this case, turbulent effects can appear that could lead to divergent solutions: that's what we need to understand. That's the reason why the Euler's formalism can be thought of as a good 'laboratory' to approach the Navier-Stokes one in the latter case. Indeed, we can expect that, under some conditions, a result which is valid for any finite, but very large, Reynolds number is also compatible with results concerning infinite Reynolds number (Euler's formalism). This is not a general statement but it justifies that we can gain ideas from the infinite Reynolds number scenario to enrich the discussion of the millennium problem.

### §14. Bounds and partial proofs

As it is clear from the statement of the millennium problem (§14), we can gain a deeper view of how to approach it by proper definitions of bounds on the norms of functions or their derivatives. Assuming that we have such definitions, it may be possible to apply this formalism to (NSI) equations in order to find upper limit to the magnitude of velocity field and its derivatives, and to relate such results to energy dissipation. This is the purpose of this section to give a precise content to the formalism and answer as much as possible the last open points.

We recall some basic definitions and notations for continuous real valued functions defined on a set $\Omega$ (that can be the set of real numbers IR or an interval [a, b] of IR), $f: \Omega \rightarrow$ IR.
All the mathematics we introduce in this section is useful for the millennium problem. We present them in the simplest way, avoiding the technical jargon as much as possible. In order to simplify the presentation, we assume first that f depends on only one variable x (a real number contained in Ω). When f is measurable on Ω, we can define a norm of the function $f\colon \Omega \rightarrow IR$ as:

$$\|f\|_{L^p(\Omega)} = \left(\int_\Omega |f(x)|^p dx\right)^{\frac{1}{p}} \text{ with } p \geq 1.$$

This defines in parallel the set of functions for which $\|f\|_{L^p(\Omega)}$ is finite:

$$L^p(\Omega) = \{f\colon \Omega \rightarrow IR, \|f\|_{L^p(\Omega)} < \infty\}.$$

Note that $\|f\|_{L^p(\Omega)}$ is really a norm in the mathematical sense, with $L^p(\Omega)$ being a Hilbert set. It is called a Lebesgue's set. We do not need to enter into the details of these notions here. In words, it means that $L^p(\Omega)$ is an abstract vector set (of functions) that allows length and angle to be measured. In this sense, the norm defined above represents a length in $L^p(\Omega)$. In what follows, what we need is to consider mainly the case: p=2, functions which are square-integrable. Then, we get for the norm of $f\colon \Omega \rightarrow IR$ (squared):

$$\|f\|^2_{L^2(\Omega)} = \int_\Omega |f(x)|^2 dx.$$

The related set of square-integrable functions is:

$$L^2(\Omega) = \{f\colon \Omega \rightarrow IR, \|f\|_{L^2(\Omega)} < \infty\}.$$

We see that these definitions of $L^2(\Omega)$ and $\|f\|^2_{L^2(\Omega)}$ represent a kind of generalization of what is well known for the algebra of vectors. These definitions can be extended trivially to functions that depend on 2 or 3 or even more real variables, defined on spaces Ω of 2, 3 or more dimensions.

In the context of the millennium problem, we are searching for velocity fields (solutions of partial differential equations if they exist) that must belong to the set of functions $L^2(\Omega)$, where Ω represents the 3-dimensional space of coordinates $IR^3$. Indeed, the kinetic energy is required to be bounded over the all space, which means that $\int \|\mathbf{v}(\mathbf{x},t)\|^2 d^3\mathbf{x}$ is finite for $t \geq 0$, where

$$\int \|\mathbf{v}(\mathbf{x},t)\|^2 d^3\mathbf{x} = \int \left[v_x^2(\mathbf{x},t) + v_y^2(\mathbf{x},t) + v_z^2(\mathbf{x},t)\right] d^3\mathbf{x}.$$

We write the integral over space of the velocity vector squared ($\int \|\mathbf{v}(\mathbf{x},t)\|^{2d^3}\mathbf{x}$) as $\|\mathbf{v}\|^2_{L^2(\Omega)}$ according to the definition above, extended to the 3-dimensional case. In addition to the fact that $\|\mathbf{v}\|^2_{L^2(\Omega)}$ must remain finite, as the equations (NSI) contain also derivatives of the velocity field over space coordinates, these derivatives for regular solutions, if they exist, must also remain of finite magnitude ( length or norm in $L^2(\Omega)$) up to a certain order that we discuss later. We write generically $\partial^m v_I$ the derivative of order m for the component i of the velocity field. Precisely, let $(m_1, m, m_3)$ be a set of 3 integers, such that
$m = m_1 + m_2 + m_3$.
Then:

$$\partial^m v_i = \partial^m_{\mathbf{x}} v_i = \partial_x^{m_1}\,\partial_y^{m_2}\,\partial_z^{m_3} v_i.$$

Where we use the standard notation: $\partial_x^{m_1} = \left(\frac{\partial}{\partial x}\right)^{m_1}$. In the following, we use also: $\partial_i = \frac{\partial}{\partial x_i}$.

- For m=0, this is simply the component i of the velocity field $\partial^0 v_i = v_i$ and:

$$\|\partial^0\mathbf{v}\|^2_{L^2(\Omega)} = \int \left[v_x^2 + v_y^2 + v_z^2\right] d^3\mathbf{x}.$$

- For m=1:

$$\|\partial^1\mathbf{v}\|^2_{L^2(\Omega)} = \int \sum_{i,j=1}^{3} \left|\partial_i v_j\right|^2 d^3\mathbf{x}.$$

  This represents the length (norm) squared of the first derivative (1-gradient) of the velocity field in $L^2(\Omega)$. Clearly, this length (or magnitude) must remain finite for all times for a regular solution of (NSI) following the millennium problem.

- In general the norm in $L^2(\Omega)$ of the derivative of order m (m-gradient) of the velocity field reads:

$$\|\partial^m\mathbf{v}\|^2_{L^2(\Omega)} = \int \sum_{j=1}^{3} \sum_{m_1,m_2,m_3} \left|\partial_x^{m_1}\,\partial_y^{m_2}\,\partial_z^{m_3} v_j\right|^2 d^3\mathbf{x} \;\; (\text{with } m_1 + m_2 + m_3 = m)$$

Therefore, following these definitions, it is useful to define special sets of functions (here the velocity fields $\mathbf{v}(\mathbf{x},\mathrm{t})\colon \Omega \times [0,\infty[\to \mathrm{IR}^3)$ in terms of the norms in $\mathrm{L}^2(\Omega)$ of their derivatives, namely a category of what Sobolev's sets:

$$\mathrm{H}^{\mathrm{s}}(\Omega) = \{\mathbf{v}\colon \Omega \times [0,\infty[\to \mathrm{IR}^3 \text{ such that } \mathbf{v} \in \mathrm{L}^2(\Omega) \text{ and } \forall \mathrm{m} \leq \mathrm{s}, \partial^{\mathrm{m}}\mathbf{v} \in \mathrm{L}^2(\Omega)\}.$$

Here, $\Omega$ is the 3-dimensional space $\mathrm{IR}^3$ or a subset of it. This definition corresponds to Sobolev's sets of rank s (integer), defined on $\mathrm{L}^2(\Omega)$, with the definition of the derivatives as written above. A norm (squared) in $\mathrm{H}^{\mathrm{s}}(\Omega)$ can then be expressed as:

$$\|\mathbf{v}\|^2_{\mathrm{H}^{\mathrm{s}}(\Omega)} = \sum_{\mathrm{k}\leq \mathrm{s}} \left\|\partial^{\mathrm{k}}\mathbf{v}\right\|^2_{\mathrm{L}^2(\Omega)}$$

We write explicitly some simple examples for single real valued functions (with $\Omega$=IR):

- $\mathrm{H}^0(\Omega) = \mathrm{L}^2(\Omega)$
- $\mathrm{H}^1(\Omega) = \{\mathrm{f}\colon \Omega \to \mathrm{IR} \text{ such that } \mathrm{f} \in \mathrm{L}^2(\Omega) \text{ and } \partial\mathrm{f} \in \mathrm{L}^2(\Omega)\}$
- $\mathrm{H}^2(\Omega) = \{\mathrm{f}\colon \Omega \to \mathrm{IR} \text{ such that } \mathrm{f} \in \mathrm{L}^2(\Omega), \partial\mathrm{f} \in \mathrm{L}^2(\Omega) \text{ and } \partial^2\mathrm{f} \in \mathrm{L}^2(\Omega)\}$
- $\mathrm{H}^{\mathrm{m}}(\Omega) = \{\mathrm{f}\colon \Omega \to \mathrm{IR} \text{ such that } \mathrm{f} \in \mathrm{L}^2(\Omega), \partial\mathrm{f} \in \mathrm{L}^2(\Omega), \dots, \partial^{\mathrm{m}}\mathrm{f} \in \mathrm{L}^2(\Omega)\}$ (m integer)
- $\|\mathrm{f}\|^2_{\mathrm{H}^{\mathrm{m}}(\Omega)} = \sum_{\mathrm{k}=0}^{\mathrm{m}} \left\|\partial^{\mathrm{k}}\mathrm{f}\right\|^2_{\mathrm{L}^2(\Omega)} = \|\partial^0\mathrm{f}\|^2_{\mathrm{L}^2(\Omega)} + \|\partial^1\mathrm{f}\|^2_{\mathrm{L}^2(\Omega)} + \cdots + \|\partial^{\mathrm{m}}\mathrm{f}\|^2_{\mathrm{L}^2(\Omega)}$

Note that $\mathrm{H}^{\mathrm{s}}(\Omega)$ can also be defined for real values of s. In this case, we need to generalize the formula above. For simplicity, we provide the new definition for single real valued function:

$$\|\mathrm{f}\|^2_{\mathrm{H}^{\mathrm{s}}(\Omega)} = \int (1 + |\xi|^2)^{\mathrm{s}} \left|\hat{\mathrm{f}}(\xi)\right|^2 \mathrm{d}\xi.$$

Here, $\hat{\mathrm{f}}$ is the Fourier transform of $\mathrm{f}\colon \Omega \to \mathrm{IR}$. Extension of the definition to dimension 3 is immediate.

Exercise: We consider a single real valued function, $\mathrm{f}\colon \Omega = \mathrm{IR} \to \mathrm{IR}$. We note $\hat{\mathrm{f}}$ its Fourier transform. Prove that there exist 2 real numbers $c_1$ and $c_2$ such that:

$$c_1 \int (1 + |\xi|^2)^{\mathrm{s}} \left|\hat{\mathrm{f}}(\xi)\right|^2 \mathrm{d}\xi \leq \|\mathrm{f}\|^2_{\mathrm{H}^{\mathrm{s}}(\Omega)} \leq c_2 \int (1 + |\xi|^2)^{\mathrm{s}} \left|\hat{\mathrm{f}}(\xi)\right|^2 \mathrm{d}\xi \quad \text{with s integer.}$$

Remark: based this property, it is then possible to prove that $\int(1+|\xi|^2)^s\left|\hat{f}(\xi)\right|^2 d\xi$ is equivalent to the standard norm (squared) in $H^s(\Omega)$ even when s is a real number.

Solution: a few hints. With $\widehat{\partial^\alpha}f(\xi) = i^\alpha\xi^\alpha\hat{f}(\xi)$, we can show that:

$$\|f\|_{H^s(\Omega)}^2 = \int\{\sum_{\alpha\leq s}|\xi^\alpha|^2\}\left|\hat{f}(\xi)\right|^2 d\xi.$$

Then, we need to find upper and lower bounds of the term $\sum_{\alpha\leq s}|\xi^\alpha|^2$. They are given by terms proportional to $(1+|\xi|^2)^s$.

A final important definition, clearly needed to approach the millennium problem, is the set of bounded and measurable velocity fields on a set $\Omega = \mathrm{IR}^3$ at a given time t. We define:

$$\|\mathbf{v}\|_{L^\infty(\Omega)} = \{\text{smallest } C \geq 0, \|\mathbf{v}(\mathbf{x},t)\| \leq C \text{ for almost every } \mathbf{x} \text{ in } \Omega\} = \|\mathbf{v}\|_{L^\infty(\Omega)}(t).$$

Then, the set of velocity fields bounded in $\Omega$ (for a given instant t), $L^\infty(\Omega)$, is defined as the set of velocity fields such that: $\|\mathbf{v}\|_{L^\infty(\Omega)}(t) < \infty$. Obviously, the same definition holds also for any derivative of the velocity field. For the gradient of the velocity fields, it reads:

$$\|\partial\mathbf{v}\|_{L^\infty(\Omega)} = \{\text{smallest } C \geq 0, \|\partial\mathbf{v}(\mathbf{x},t)\| \leq C \text{ for almost every } \mathbf{x} \text{ in } \Omega\} = \max_{\mathbf{x}\in\mathrm{IR}^3}\|\partial\mathbf{v}(\mathbf{x},t)\|.$$

In practice, Sobolev's sets of functions are useful because they *inform us* that a function and its derivatives up to a certain order belongs to $L^2(\Omega)$. Then, it would be interesting to conclude from this property that the functions belonging to a particular Sobolev's set are (continuously) differentiable up to a certain order, to be related to the rank of the Sobolev's set. This kind of statement (if possible) corresponds to what is called embedding. In general, a set of functions X is said to be embedded in the set of functions Y (written as

$X \subset Y$) if all elements (functions) in X are also in Y. We start with a trivial example for single real valued (measurable) functions: $L^\infty(\Omega) \subset L^1(\Omega)$.

Then, $L^\infty(\Omega)$ is embedded in $L^1(\Omega)$, where $L^1(\Omega)$ is the Lebesgue's set of functions such that $\left(\int_\Omega |f(x)|dx\right)$ is finite. This means also that $L^1(\Omega)$ contains $L^\infty(\Omega)$. An important consequence that derives from this definition is that if $X \subset Y$, then there exists a constant $C > 0$ such that:

$\|f\|_Y \leq C\ \|f\|_X$. The last inequality can also be taken as a definition of $X \subset Y$. In the following, it will be clear how to use this notion in very efficient ways.

Some important embedding properties are stated below:

(1) $s < t$ implies that: $H^t(\Omega) \subset H^s(\Omega)$

(2) $H^{s+k}(\Omega) \subset C^k(\Omega)$ for $s > \frac{D}{2}$ (where $k \geq 1$ and D is the dimension of space $\Omega$),

$C^k(\Omega)$ = set of functions $k-$times differentiable in $\Omega$, with

$$\|\mathbf{v}\|_{C^k(\Omega)} = \sum_{p=1}^{k} \max_{\mathbf{x}\in\Omega} \|\partial^p \mathbf{v}(\mathbf{x})\|.$$

<u>Summary of the mathematical definitions and properties for 3-dimensional set $\Omega$:</u>

$$L^2(\Omega) = \{\mathbf{v}: \Omega \times [0,\infty[\to IR^3, \|\mathbf{v}\|_{L^2(\Omega)} < \infty\} \text{ with } \|\mathbf{v}\|^2_{L^2(\Omega)} = \int_\Omega \|\mathbf{v}(\mathbf{x},t)\|^2 d^3\mathbf{x}.$$

$$H^s(\Omega) = \begin{Bmatrix} \mathbf{v}: \Omega \times [0,\infty[\to IR^3 \text{ such that } \mathbf{v} \in L^2(\Omega) \text{ and } \forall\ m = m_1 + m_2 + m_3\ \leq s, \\ \partial^m \mathbf{v} = \partial_x^{m_1} \partial_y^{m_2} \partial_z^{m_3} \mathbf{v} \in L^2(\Omega) \end{Bmatrix}$$

with $\|\mathbf{v}\|^2_{H^s(\Omega)} = \sum_{k\leq s} \left\|\partial^k \mathbf{v}\right\|^2_{L^2(\Omega)}$.

$\|\mathbf{v}\|_{L^\infty(\Omega)}(t)$ = the upper value of $\|\mathbf{v}(\mathbf{x},t)\|$ in $\Omega$.

Embedding:

$X \subset Y$ if all elements (functions) in X are also in Y. If $X \subset Y$,

then there exists a constant $C > 0$ such that: $\|f\|_Y \leq C\ \|f\|_X$.

$H^{s+k}(\Omega) \subset C^k(\Omega)$ for $s > \frac{3}{2} = \frac{D}{2}$ (where $k \geq 1$ and D is the dimension of space $\Omega$).

It is time to use the mathematical notions exposed above in order to progress in the understanding of the millennium problem. However, this would be too hard to start directly with Navier-Stokes equations. That's the reason why we apply first the above definitions and properties to the case of (incompressible) Euler's equations. We discuss (mathematically) 2 issues related to regularity of solutions of the Euler's equations: (i) the local existence and uniqueness of such solutions and (ii) the role of vorticity. And then, we discuss some mathematical issues for the case of the Navier-Stokes equations (in the presence of viscosity). We pose: $\Omega=IR^3$ (unless explicitly stated otherwise) and $\rho$=constant=1.

Also, we write $L^2$ for $L^2(\Omega)$ and similarly for other sets of functions used hereafter.

A last comment is needed before starting the discussion. We often need to estimate differential inequalities like $\frac{dX}{dt} \leq b(t)X(t)$, where X and b are positive functions. The last differential inequality can be solved, it gives: $X(t \geq 0) \leq X(0)\exp[\int_0^t b(s)ds]$. In particular if b(t) is a constant (b) strictly positive: $X(t \geq 0) \leq X(0)\exp[bt]$. In addition, if we know that X(t) is bounded for all times, the only possible solution is then $X(t \geq 0) = 0$.

(i) We discuss the proof of the local existence and uniqueness of solutions of the (incompressible) Euler equations in 3-dimensional space, under the hypothesis that the initial condition is smooth and regular with strong decay at infinity (in space). This is not a simple proof. We intend to show the important steps. The interest is that it clearly illustrates how to make useful the above definitions and some properties that come with them.

1. First, from Euler's equations, it is possible to show that there exists a constant $C > 0$ such that:

$$\frac{d}{dt}\left\|\partial^k \mathbf{v}\right\|_{L^2}^2 \leq C\, \|\partial \mathbf{v}\|_{L^\infty} . \left\|\partial^k \mathbf{v}\right\|_{L^2}^2 \text{ (for k} \geq 1).$$

   We let the proof of this relation as this exercise (below). For k=0, $\frac{d}{dt}\|\partial^0 \mathbf{v}\|_{L^2}^2$ is the time derivative of the kinetic energy which is constant in time (§9 and 10) and thus:

$$\frac{d}{dt}\|\partial^0 \mathbf{v}\|_{L^2}^2 = 0.$$

2. For m>1+3/2=1+D/2, where D is the dimension of space, we have already mentioned that $H^m(\Omega)$ is embedded in $C^1(\Omega)$, the set of continuous and differentiable functions on Ω. This implies that there exists a constant $A > 0$ such that: $\|\mathbf{v}\|_{C^1} \leq A\ \|\mathbf{v}\|_{H^m}$. Knowing that we also have in general (from a trivial embedding) $\|\partial \mathbf{v}\|_{L^\infty} \leq B\ \|\mathbf{v}\|_{C^1}$ (with the constant $B > 0$), we can write:
$$\|\partial \mathbf{v}\|_{L^\infty} \leq B\|\mathbf{v}\|_{C^1} \leq A.B\ \|\mathbf{v}\|_{H^m}.$$
3. Then, using the definition of the norm in $H^m$, $\|\mathbf{v}\|_{H^m}^2 = \sum_{k\leq m}\left\|\partial^k \mathbf{v}\right\|_{L^2}^2$, as well as items (1) and (2), it can easily be seen that there exists a constant $C' > 0$ such that:
$$\frac{1}{2}\frac{d}{dt}\|\mathbf{v}\|_{H^m}^2 \leq C'\ \|\mathbf{v}\|_{H^m}^3 \text{ with } m > 1 + 3/2 = 1 + D/2.$$
Equivalently, we can write:
$$\frac{d}{dt}\|\mathbf{v}\|_{H^m} \leq C'\ \|\mathbf{v}\|_{H^m}^2.$$
4. This implies that $\mathbf{v}(\mathbf{x}, t)$ (solution of the Euler's equations) is bounded in $H^m \times [0, T]$. Its upper bound is given by (see §7 applied to differential inequalities):
$$\frac{\|\mathbf{v}_0\|_{H^m}}{1 - C'T\|\mathbf{v}_0\|_{H^m}} \text{ for } T \leq T^* = \frac{1}{C'\|\mathbf{v}_0\|_{H^m}}, m > 1 + 3/2 = 1 + D/2.$$
This proves the local existence.
5. The uniqueness can be shown by a simple reasoning. Assuming that there exist 2 solutions of the Euler's equations $\mathbf{u}_1$ and $\mathbf{u}_2$, it is then possible to show that there exists a constant $K > 0$ such that:
$$\frac{d}{dt}\|\mathbf{u}_1 - \mathbf{u}_2\|_{L^2}^2 \leq K\ [\|\mathbf{u}_1\|_{H^m} + \|\mathbf{u}_2\|_{H^m}]\|\mathbf{u}_1 - \mathbf{u}_2\|_{L^2}^2.$$
As we know that $K\ [\|\mathbf{u}_1\|_{H^m} + \|\mathbf{u}_2\|_{H^m}] > 0$, this implies that $\mathbf{u}_1 = \mathbf{u}_2$.

(ii) Next, we discuss the role of vorticity on the regularity of solutions of the Euler's equations. We start from the inequality (i.1):

$$\frac{d}{dt}\left\|\partial^k \mathbf{v}\right\|_{L^2}^2 \leq C\,\|\partial \mathbf{v}\|_{L^\infty}\,\left\|\partial^k \mathbf{v}\right\|_{L^2}^2 \text{ (for k} \geq 1).$$

Or equivalently:

$$\frac{d}{dt}\|\mathbf{v}\|_{H^m} \leq K\,\|\partial \mathbf{v}\|_{L^\infty}\,\|\mathbf{v}\|_{H^m}\ \ (K > 0).$$

Here $m \geq 1$, as we do not need the embedding relation between $\|\partial \mathbf{v}\|_{L^\infty}$ and $\|\mathbf{v}\|_{H^m}$ which requires $m > 1 + 3/2$ [see (i.2)].

Note that $\|\partial \mathbf{v}\|_{L^\infty} = \|\partial \mathbf{v}\|_{L^\infty}(t)$ and we have no way to guarantee that $\|\partial \mathbf{v}\|_{L^\infty}(t) < \infty$ at all times. The only statement that we have made (i.4) is that there is a finite time up to which the velocity field is regular (and not blowing up). Indeed, we obtain from the above the inequality:

$$\|\mathbf{v}\|_{H^m} \leq \|\mathbf{v}_0\|_{H^m} \exp\left[K\int_0^t \|\partial \mathbf{v}\|_{L^\infty}\,dt\right].$$

Thus, as long as $\|\partial \mathbf{v}\|_{L^\infty} < \infty$ in [0, T*], the velocity field $\mathbf{v}(\mathbf{x}, t)$, solution of Euler's equations, exists up to T*. Starting from these relations, it is possible to prove another important theorem concerning solutions of Euler's equations (from Beale, Kato and Majda in 1984):

> We consider the velocity field $\mathbf{v}(\mathbf{x}, t)$, solution of Euler's equations in 3-dimensional space, such that $\mathbf{v}(\mathbf{x}, t)$ belongs to $H^m$ (m>1+3/2) for all $t \in [0, T^*]$. We write $\boldsymbol{\omega} = \boldsymbol{\nabla} \times \mathbf{v}$, the vorticity of this velocity field. Then, $T^*$ is the maximum time for the velocity field to be in the above function class if and only if $\int_0^{T^*} \|\boldsymbol{\omega}\|_{L^\infty}\,dt = \infty$. In other words, $\int_0^{T^*} \|\boldsymbol{\omega}\|_{L^\infty}\,dt = \infty$ is a necessary and sufficient condition for blow-up at time T* for 3-dimensional Euler's equations. We let the proof of this theorem as an exercise (below).
>
> An immediate consequence of this theorem is that there is no finite time blow-up in 2-dimensional Euler's equations assuming that the initial vorticity field is bounded. This can be shown in various ways: we let this as an exercise (below). This implies the existence of global regular solutions in this context. According to §8 or §9, in 2D, the vorticity is conserved as a scalar.

<u>Exercise</u>: Consider the velocity field $\mathbf{v}(\mathbf{x},t)$ solution of the (incompressible) Euler equations in 3-dimensional space, under the hypothesis that the initial condition is smooth and regular with strong decay at infinity (in space).

Prove the inequality (i.1): $\frac{d}{dt}\left\|\partial^k \mathbf{v}\right\|_{L^2}^2 \leq C\, \|\partial \mathbf{v}\|_{L^\infty}.\left\|\partial^k \mathbf{v}\right\|_{L^2}^2$ (for $k \geq 1$).

A similar proof can easily be done in the case of periodic boundary conditions.

<u>Solution</u>: We start with k=1. We re-write $\|\partial^1 \mathbf{v}\|_{L^2}^2$ as:

$$\|\partial^1 \mathbf{v}\|_{L^2}^2 = \int \sum_{i,j=1}^{3} \left|\partial_i v_j\right|^2 dV = \sum_{i,j=1}^{3} \int \left|\partial_i v_j\right|^2 dV.$$

The integral is taken over the volume of the 3-dimensional space ($\Omega$), knowing that the velocity field is strongly decaying at infinity (in space). Then, we take the time derivatives of this quantity:

$$\frac{1}{2}\frac{d}{dt}\|\partial^1 \mathbf{v}\|_{L^2}^2 = \sum_{i,j=1}^{3} \int \partial_i v_j\, \partial_i \frac{\partial}{\partial t} v_j\, dV.$$

The term in $\frac{\partial}{\partial t} v_j$ can be obtained from Euler's equations in the absence of external field (in which we take $\rho$=constant=1):

$$\frac{\partial}{\partial t} v_j + \sum_{\alpha=1}^{3} v_\alpha\, \partial_\alpha\, v_j + \partial_j p = 0.$$

In the following, we do not write the sum symbol for repeated indices, this is implicit. This means that $\sum_{\alpha=1}^{3} v_\alpha\, \partial_\alpha$ is written (simply) as $v_\alpha\, \partial_\alpha$.

Euler's equations become: $\frac{\partial}{\partial t} v_j + v_\alpha\, \partial_\alpha v_j + \partial_j p = 0$. Next, we introduce $\frac{\partial}{\partial t} v_j$ in the expression of the quantity $\frac{1}{2}\frac{d}{dt}\|\partial^1 \mathbf{v}\|_{L^2}^2$.

We obtain:

$$\frac{1}{2}\frac{d}{dt}\|\partial^1\mathbf{v}\|_{L^2}^2 = \sum_{i,j=1}^{3} -\int \partial_i v_j\, \partial_i\left(v_\alpha\, \partial_\alpha v_j + \partial_j p\right) dV$$

$$= -\int \partial_i v_j\, \partial_i\left(v_\alpha\, \partial_\alpha v_j + \partial_j p\right) dV. \quad \text{(i and j are also repeated indices)}$$

With:

$$\int \partial_i v_j\, \partial_i\left(v_\alpha\, \partial_\alpha v_j + \partial_j p\right) dV = \int \left[\partial_i v_j\, \partial_i v_\alpha\, \partial_\alpha v_j + \partial_i v_j v_\alpha\, \partial_i\, \partial_\alpha v_j + \partial_i v_j\, \partial_i\, \partial_j p\right] dV.$$

By integration by parts, with the condition of incompressibility $\partial_\alpha v_\alpha = 0$ and the fact that the velocity field (at thus its derivatives) is zero at infinity (boundary surface), we can show very easily that:

$$\int \left[\partial_i v_j v_\alpha\, \partial_i\, \partial_\alpha v_j + \partial_i v_j\, \partial_i\, \partial_j p\right] dV = 0.$$

Explicitly for one term:

$$\int \left[\partial_i v_j v_\alpha\, \partial_i\, \partial_\alpha v_j\right] dV = -\int \left[\partial_\alpha [\partial_i v_j v_\alpha]\, \partial_i v_j\right] dV = -\int \left[\partial_\alpha [\partial_i v_j]\; v_\alpha\, \partial_i v_j\right] dV$$

$$= -\frac{1}{2}\int \left[\partial_\alpha [(\partial_i v_j)^2]\; v_\alpha\right] = \frac{1}{2}\int \left[[(\partial_i v_j)^2]\;\; \partial_\alpha v_\alpha\right] = 0.$$

Then, we are left with the relation following Euler's equations:

$$\frac{1}{2}\frac{d}{dt}\|\partial^1\mathbf{v}\|_{L^2}^2 = -\int \left[\partial_i v_j\, \partial_i v_\alpha\, \partial_\alpha v_j\right] dV.$$

The term $\int \left[\partial_i v_j\, \partial_i v_\alpha\, \partial_\alpha v_j\right] dV$ looks complicated. Also, it does not have a definite sign. However, we can always write:

$$-\int \left[\partial_i v_j\, \partial_i v_\alpha\, \partial_\alpha v_j\right] dV \leq \left|\int \left[\partial_i v_j\, \partial_i v_\alpha\, \partial_\alpha v_j\right] dV\right|.$$

It follows that:

$$\frac{1}{2}\frac{d}{dt}\|\partial^1\mathbf{v}\|_{L^2}^2 \leq \left|\int \left[\partial_i v_j\, \partial_i v_\alpha\, \partial_\alpha v_j\right] dV\right|.$$

In order to conclude the proof for k=1, it remains to show that $\|\partial\mathbf{v}\|_{L^\infty}.\, \|\partial^1\mathbf{v}\|_{L^2}^2$ is a proper upper bound for the term on the right hand side. This term is built up by 3 gradients of the

velocity: we can bound 2 of them by their Euclidian lengths and one by $\|\partial \mathbf{v}\|_{L^\infty}$ (by definition of $\|.\|_{L^\infty}$). Then, the term $\|\partial \mathbf{v}\|_{L^\infty}$ comes out of the integral. Altogether, we obtain:

$$\left|\iint [\partial_i v_j \, \partial_i v_\alpha \, \partial_\alpha v_j] \, dV\right| \leq A. \|\partial \mathbf{v}\|_{L^\infty}. \sum_{i,j=1}^{3} \int |\partial_i v_j|^2 \, dV = A. \|\partial \mathbf{v}\|_{L^\infty}. \|\partial^1 \mathbf{v}\|_{L^2}^2.$$

Here, A is a strictly positive constant. This completes the proof for k=1.

We now consider the general case: $k \geq 1$. The proof follows exactly the same development as the case k=1. Indeed, we have:

$\|\partial^k \mathbf{v}\|_{L^2}^2 = \int \sum_{j=1}^{3} \sum_{k_1,k_2,k_3} |\partial^k v_j|^2 \, dV$ with $\partial^k = \partial_x^{k_1} \, \partial_y^{k_2} \, \partial_z^{k_3}$ (and $k = k_1 + k_2 + k_3$).

It is easier to keep the generic notation $\partial^k$ in the formula. Then, we can easily show:

$$\frac{1}{2}\frac{d}{dt}\|\partial^k \mathbf{v}\|_{L^2}^2 \leq \left|\iint [\partial^k v_j \, \partial^k v_\alpha \, \partial_\alpha v_j] \, dV\right|.$$

Again in a similar way as before, we can bound the 2 k-gradients by their Euclidian lengths and the 1-gradient by $\|\partial \mathbf{v}\|_{L^\infty}$. We obtain finally the inequality:

$\frac{d}{dt}\|\partial^k \mathbf{v}\|_{L^2}^2 \leq C \, \|\partial \mathbf{v}\|_{L^\infty}. \|\partial^k \mathbf{v}\|_{L^2}^2$ (for $k \geq 1$), which completes the proof ($k \geq 1$).

<u>Exercise</u>: Justify the result from Beale, Kato, and Majda in 1984: we consider the velocity field $\mathbf{v}(\mathbf{x}, t)$, solution of Euler's equations in 3-dimensional space, such that $\mathbf{v}(\mathbf{x}, t)$ belongs to $H^m$ ($m>1+3/2$) for all $t \in [0, T^*]$.
We write the vorticity $\boldsymbol{\omega} = \boldsymbol{\nabla} \times \mathbf{v}(., t)$. Then, $T^*$ is the maximum time for the velocity field to be in the above function class if and only if $\int_0^{T^*} \|\boldsymbol{\omega}\|_{L^\infty} \, dt = \infty$.

Hint: we assume that the following relation holds:

$$\|\partial \mathbf{v}\|_{L^\infty} \leq \|\partial \mathbf{v}_0\|_{L^\infty} \exp\left[C \int_0^t \|\partial \boldsymbol{\omega}\|_{L^\infty} \, dt\right] \quad \text{(C a positive constant).}$$

Solution: The first part of the proof is trivial: if $\int_0^{T^*} \|\boldsymbol{\omega}\|_{L^\infty}\, dt = \infty$,

this implies that $\|\boldsymbol{\omega}\|_{L^\infty}(t) \xrightarrow[t\to T^*]{} \infty$, and obviously the velocity function cannot belong to $H^m$ (m>1+3/2) for all $t \in [0, T^*]$. For the second part of the proof, we need to justify that as long as $\|\boldsymbol{\omega}\|_{L^\infty} < \infty$ in $[0, T^*]$ (or $\int_0^{T^*} \|\boldsymbol{\omega}\|_{L^\infty}\, dt < \infty$), then $\mathbf{v}(\mathbf{x}, t)$ solution of Euler's equations exists up to T*.

From the inequality (i.1): $\|\mathbf{v}\|_{H^m} \leq \|\mathbf{v}_0\|_{H^m} \exp\left[C \int_0^t \|\partial \mathbf{v}\|_{L^\infty}\, dt\right]$. We conclude immediately that as long as $\|\partial \mathbf{v}\|_{L^\infty} < \infty$ in $[0, T^*]$, the velocity field $\mathbf{v}(\mathbf{x}, t)$, solution of Euler's equations, exists up to T*.

To complete the second part of the proof, we need to show the following relation: $\int_0^{T^*} \|\boldsymbol{\omega}\|_{L^\infty}\, dt < \infty \Rightarrow \int_0^{T^*} \|\partial \mathbf{v}\|_{L^\infty}\, dt < \infty$. This is a direct consequence of the relation given in the statement of the exercise. This completes the proof.

Exercise: Prove that in 2-dimensional space Euler's equations admit global regular solutions $\mathbf{v}(\mathbf{x}, t)$ under initial and boundary conditions (with strong decay at infinity).

Solution: This exercise is a direct consequence from the 2 previous ones. There are 3 simple methods to achieve the proof. (1) In 2D, we have the great simplification: $\int \left[\partial_i v_j\, \partial_i v_\alpha\, \partial_\alpha v_j\right] dV = 0$. It is equivalent to the absence of the vortex stretching term in the vorticity equation presented in previous sections. This implies that $\|\partial^1 \mathbf{v}\|^2_{L^2(IR^2)}$ is conserved (and thus bounded once the initial condition is), as well as $\|\mathbf{v}\|^2_{L^2(IR^2)}$. This makes the proof. (2) Equivalently, we can use the previous result on the behavior of $\|\boldsymbol{\omega}\|_{L^\infty}(t)$.

Once the vorticity is finite at initial time, there is no possibility for the existence of a blow-up time: $\|\boldsymbol{\omega}\|_{L^\infty}(t) \xrightarrow[t\to T^*]{} \infty$. This completes the proof. (3) Finally, it is immediate to see that the $L^p(IR^2)$ norm of the vorticity $\|\boldsymbol{\omega}\|^2_{L^p(IR^2)}$ is preserved for $p \geq 2$. This is also a guarantee that

$||\boldsymbol{\omega}||_{L^\infty}(t)$ remains finite for all times. This ensures the existence of global regular solution of Euler's equations.

In summary, we have seen how to apply the mathematical notions introduced at the beginning of this section in order to search for bounds on the velocity field or its derivatives in the case of (incompressible) Euler's equations. This has allowed us to prove some fundamental theorems related to Euler's equations. We come back to the case of the Navier-Stokes equations (i.e. in the presence of viscosity), in the context of the millennium problem. The elements of reasoning follow what we have developed concerning the Euler's equations. Namely, we need to check whether $||\mathbf{v}(t)||_{L^2}^2$ and $||\partial^1\mathbf{v}(t)||_{L^2}^2$ are bounded for all times or not. In the case of Euler's equations in 3-dimensional space, we have seen that $||\partial^1\mathbf{v}(t)||_{L^2}^2$ is only locally bounded up to a critical time T*. However, we could expect that, even if solutions of Euler's equations could blow up, solutions of Navier-Stokes would not due to the viscosity term. The idea is that this term could be sufficient to control the non-linearity of the equations. That's the reason why the proofs done for the Euler's case do not directly imply similar results for the Navier-Stokes equations. Therefore we need dedicated proofs for the Navier-Stokes case and thus the millennium problem.

We start with the norm of the velocity field (squared): $||\mathbf{v}(t)||_{L^2}^2$. We have already discussed the time evolution of $||\mathbf{v}(t)||_{L^2}^2$ in §10, where we have shown that:

$$\frac{1}{2}\frac{d}{dt}||\mathbf{v}||_{L^2}^2 = -\eta\int\sum_{i,j=1}^{3}\left|\partial_i v_j\right|^2 dV.$$

We recall briefly the key points of the proof. First, we have the trivial identity:

$\frac{1}{2}\frac{d}{dt}||\mathbf{v}||_{L^2}^2 = \int v_j \frac{\partial}{\partial t} v_j \, dV$ (summation implicit on repeated indices).

The Navier-Stokes equation for the component j of the velocity reads in the absence of external force:

$$\frac{\partial}{\partial t}v_j + v_\alpha\,\partial_\alpha v_j + \partial_j p = \eta\,\partial_\alpha\,\partial_\alpha v_j.$$

We could take into account a regular external force. This would not modify the line of the proof and the conclusions would be similar. Including $\frac{\partial}{\partial t}v_j$ into $\frac{1}{2}\frac{d}{dt}\|\mathbf{v}\|_{L^2}^2$ leads to:

$$\frac{1}{2}\frac{d}{dt}\|\mathbf{v}\|_{L^2}^2 = \int v_j\left[-v_\alpha\,\partial_\alpha v_j - \partial_j p + \eta\,\partial_\alpha\,\partial_\alpha v_j\right] dV.$$

After integrations by parts with vanishing velocity at boundaries and the incompressibility condition, it is trivial to show that:

$$\int \left[-v_j v_\alpha\,\partial_\alpha v_j - v_j\,\partial_j p\right] dV = 0.$$

Again, after integration by parts, the last term becomes:

$$\int \left[\eta v_j\,\partial_\alpha\,\partial_\alpha v_j\right] dV = -\eta \int \sum_{\alpha,j=1}^{3} \left[\partial_\alpha v_j\right]^2 dV.$$

This gives the result. This term represents the rate of dissipation of the energy. Therefore the cumulative energy dissipation up to time T is equal to: $\eta \int_{t=0}^{T} \int \sum_{i,j=1}^{3} \left|\partial_i v_j\right|^2 dV\,dt$.

We obtain:

$$\frac{1}{2}\|\mathbf{v}\|_{L^2}^2(T) = E(0) - \eta \int_{t=0}^{T} \int \sum_{i,j=1}^{3} \left|\partial_i v_j\right|^2 dV\,dt = E(0) - \eta \int_{t=0}^{T} \|\partial^1 \mathbf{v}\|_{L^2}^2\,dt.$$

In conclusion, $\frac{1}{2}\|\mathbf{v}\|_{L^2}^2(t)$ is bounded by the initial energy E(0) for all times as well as the cumulative energy dissipation (as $E(0) - \eta \int_{t=0}^{T} \int \sum_{i,j=1}^{3} \left|\partial_i v_j\right|^2 dV\,dt \geq 0$). Note that for Euler's equations, the energy is a constant of motion, thus there is no implied bound on the space time integral of $\sum_{i,j=1}^{3}\left|\partial_i v_j\right|^2$.

The situation looks more favorable for Navier-Stokes equations at this step.

We now discuss the other point: the time evolution of the quantity $\frac{d}{dt}\|\partial^1\mathbf{v}\|_{L^2}^2$. The beginning of the computation is easy:

$$\frac{1}{2}\frac{d}{dt}\|\partial^1\mathbf{v}\|_{L^2}^2 = \int \partial_i v_j\, \partial_i \frac{\partial}{\partial t} v_j \, dV = \int \partial_i v_j\, \partial_i\left[-v_\alpha\, \partial_\alpha v_j - \partial_j p + \eta\, \partial_\alpha\, \partial_\alpha v_j\right] dV.$$

Following similar arguments as above, we can simplify this expression into:

$$\begin{aligned}\frac{1}{2}\frac{d}{dt}\|\partial^1\mathbf{v}\|_{L^2}^2 &= -\int \partial_i v_j\, \partial_i v_\alpha\, \partial_\alpha v_j \, dV - \eta \int \partial_i\, \partial_i v_j\, \partial_\alpha\, \partial_\alpha v_j \, dV \\ &= -\int \partial_i v_j\, \partial_i v_\alpha\, \partial_\alpha v_j \, dV - \eta\|\nabla^2\mathbf{v}\|_{L^2}^2.\end{aligned}$$

With the notation:

$$\|\nabla^2\mathbf{v}\|_{L^2}^2 = \int \partial_i\, \partial_i v_j\, \partial_\alpha\, \partial_\alpha v_j \, dV = \int \|\nabla^2\mathbf{v}\|^2 \; dV \text{ (can also be written as } \int \|\Delta\mathbf{v}\|^2 \; dV\text{)}.$$

Here, the norm under the integral is the standard vector norm. We obtain at this step of the calculations:

$$\frac{1}{2}\frac{d}{dt}\|\partial^1\mathbf{v}\|_{L^2}^2 + \eta\|\nabla^2\mathbf{v}\|_{L^2}^2 = -\int \partial_i v_j\, \partial_i v_\alpha\, \partial_\alpha v_j \, dV.$$

The integral on the right hand side is similar to the one obtained in the developments done for Euler's equations. There is a new term (on the left hand side) related to the viscosity.

We first make a remark concerning the 2-dimensional case: in 2D, $\int \partial_i v_j\, \partial_i v_\alpha\, \partial_\alpha v_j \, dV = 0$.

Then we obtain (in 2D): $\frac{1}{2}\frac{d}{dt}\|\partial^1\mathbf{v}\|^2_{L^2(IR^2)} + \eta\|\nabla^2\mathbf{v}\|^2_{L^2(IR^2)} = 0$. We conclude that $\|\partial^1\mathbf{v}\|^2_{L^2(IR^2)}(t)$ and also $\|\nabla\times\mathbf{v}\|^2_{L^2(IR^2)}(t)$ are uniformly bounded at all times. This is the proof of the existence of global regular solutions in 2D. This is a consequence from the fact that the term $\int \partial_i v_j\, \partial_i v_\alpha\, \partial_\alpha v_j\, dV$ is directly related to the vortex stretching of the vorticity, which disappears in 2D (see previous sections).

In dimension 3 (3D), this is more complicated. After some algebra that we admit here, we can write:

$$-\int \partial_i v_j\, \partial_i v_\alpha\, \partial_\alpha v_j\, dV \le C\, \|\partial^1\mathbf{v}\|^{\frac{3}{2}}_{L^2}\left\|\nabla^2\mathbf{v}\right\|^{\frac{3}{2}}_{L^2} \le \frac{\eta}{2}\left\|\nabla^2\mathbf{v}\right\|^2_{L^2} + C_\eta\|\partial^1\mathbf{v}\|^6_{L^2} \quad \left(C \text{ and } C_\eta > 0\right).$$

Note that the algebra needed in order to derive the last formula is presented in the appendix to this document. It requires the successive use of Hölder's, Sobolev's and scaled Young's inequalities.

Then:

$$\frac{1}{2}\frac{d}{dt}\|\partial^1\mathbf{v}\|^2_{L^2} \le -\frac{\eta}{2}\left\|\nabla^2\mathbf{v}\right\|^2_{L^2} + C_\eta\|\partial^1\mathbf{v}\|^6_{L^2}.$$

This inequality can be transformed using the relation:

$$\|\partial^1\mathbf{v}\|^2_{L^2} = \int \partial_i v_j\, \partial_i v_j\, dV = -\int v_j\, \partial_i\, \partial_i v_j\, dV \le \|\mathbf{v}\|_{L^2}\left\|\nabla^2\mathbf{v}\right\|_{L^2}.$$

Finally:

$$\frac{1}{2}\frac{d}{dt}\|\partial^1\mathbf{v}\|^2_{L^2} \le -\frac{\eta}{2}\frac{\|\partial^1\mathbf{v}\|^4_{L^2}}{\|\mathbf{v}\|^2_{L^2}} + C_\eta\|\partial^1\mathbf{v}\|^6_{L^2}.$$

In particular, the last relation leads to the inequality:

$$\frac{1}{2}\frac{d}{dt}\|\partial^1\mathbf{v}\|^2_{L^2} \le C_\eta\|\partial^1\mathbf{v}\|^6_{L^2}.$$

This implies:

$$\|\partial^1 \mathbf{v}\|_{L^2}^2(t) \leq \frac{\|\partial^1 \mathbf{v}\|_{L^2}^2(0)}{\left[1 - 2C_\eta \|\partial^1 \mathbf{v}\|_{L^2}^4(0)t\right]^{\frac{1}{2}}}.$$

This proves the local existence of solutions of the Navier-Stokes equations (similarly to the previous result obtained for Euler's equations) up to the critical time T* such that:

$$T^* = \frac{1}{\left[2C_\eta \|\partial^1 \mathbf{v}\|_{L^2}^4(0)\right]}.$$

Note that this is because the positive term $C_\eta \|\partial^1 \mathbf{v}\|_{L^2}^6$ may be too large compared to $-\frac{\eta}{2}\frac{\|\partial^1 \mathbf{v}\|_{L^2}^4}{\|\mathbf{v}\|_{L^2}^2}$ that we may observe a finite-time blow-up of $\|\partial^1 \mathbf{v}\|_{L^2}^2(t)$.

In summary, we have a set of coupled inequalities:

| | |
|---|---|
| $\frac{1}{2}\|\mathbf{v}\|_{L^2}^2(T) = E(0) - \eta \int_{t=0}^{T} \|\partial^1 \mathbf{v}\|_{L^2}^2 \, dt$ <br> $\leq E(0)$ | $\frac{1}{2}\frac{d}{dt}\|\partial^1 \mathbf{v}\|_{L^2}^2 \leq -\frac{\eta}{2}\frac{\|\partial^1 \mathbf{v}\|_{L^2}^4}{\|\mathbf{v}\|_{L^2}^2} + C_\eta \|\partial^1 \mathbf{v}\|_{L^2}^6.$ |

It is important to notice that if there exists a time T for which:

$-\frac{\eta}{2}\frac{\|\partial^1 \mathbf{v}\|_{L^2}^4}{\|\mathbf{v}\|_{L^2}^2}(T) + C_\eta \|\partial^1 \mathbf{v}\|_{L^2}^6(T) > 0$. Then $\|\partial^1 \mathbf{v}\|_{L^2}^2(t)$ can only decrease for $t \geq T$, and the following inequality holds:

$$-\frac{\eta}{2}\frac{\|\partial^1 \mathbf{v}\|_{L^2}^4}{\|\mathbf{v}\|_{L^2}^2}(t \geq T) + C_\eta \|\partial^1 \mathbf{v}\|_{L^2}^6(t \geq T) > 0.$$

This means that if the initial conditions are such that:

$$-\frac{\eta}{2}\frac{\|\partial^1 \mathbf{v}\|_{L^2}^4}{\|\mathbf{v}\|_{L^2}^2}(t = 0) + C_\eta \|\partial^1 \mathbf{v}\|_{L^2}^6(t = 0) > 0.$$

Then the solution stays regular for all times. Stated otherwise, for small enough initial conditions $\|\mathbf{v}_0\|_{L^2}^2 \|\partial^1 \mathbf{v}_0\|_{L^2}^2$ (given by the above inequality), $\|\partial^1 \mathbf{v}\|_{L^2}^2(t)$ is finite for all times, this ensures the existence (and uniqueness) of global regular solutions.

Note that instead of searching for inequalities for $\|\partial^1 \mathbf{v}\|_{L^2}^2$, we can work with the norm of the vorticity $\|\boldsymbol{\omega} = \boldsymbol{\nabla} \times \mathbf{v}\|_{L^2}^2$. The results are absolutely equivalent, but it is interesting to prove this rigorously. Following a similar developments as previously, we can show:

$$\frac{1}{2}\frac{d}{dt}\|\boldsymbol{\omega}\|_{L^2}^2 \leq -\frac{\eta}{2}\|\partial^1 \boldsymbol{\omega}\|_{L^2}^2 + C_\eta \|\boldsymbol{\omega}\|_{L^2}^6.$$

Using the inequality $\|\boldsymbol{\omega}\|_{L^2}^4 \leq B\,\|\mathbf{v}\|_{L^2}^2 \|\partial^1 \boldsymbol{\omega}\|_{L^2}^2$ $(B > 0)$ (see appendix to this document), we get:

$$\frac{1}{2}\frac{d}{dt}\|\boldsymbol{\omega}\|_{L^2}^2 \leq -\frac{\eta}{2}\|\partial^1 \boldsymbol{\omega}\|_{L^2}^2 + B.C_\eta \|\boldsymbol{\omega}\|_{L^2}^2 \|\mathbf{v}\|_{L^2}^2 \|\partial^1 \boldsymbol{\omega}\|_{L^2}^2.$$

Altogether:

$$\frac{1}{2}\frac{d}{dt}\|\boldsymbol{\omega}\|_{L^2}^2 \leq \left[B.C_\eta \|\boldsymbol{\omega}\|_{L^2}^2 \|\mathbf{v}\|_{L^2}^2 - \frac{\eta}{2}\right]\|\partial^1 \boldsymbol{\omega}\|_{L^2}^2.$$

Then, the following inequality holds $\frac{1}{2}\frac{d}{dt}\|\boldsymbol{\omega}\|_{L^2}^2 \leq 0$ provided that:

$$B.C_\eta \|\boldsymbol{\omega}_0\|_{L^2}^2 \|\mathbf{v}_0\|_{L^2}^2 \leq \frac{\eta}{2}.$$

Since $\|\mathbf{v}\|_{L^2}^2(t) \leq \|\mathbf{v}_0\|_{L^2}^2$, the inequality $\frac{1}{2}\frac{d}{dt}\|\boldsymbol{\omega}\|_{L^2}^2 \leq 0$ implies that $\|\boldsymbol{\omega}\|_{L^2}^2(t) \leq \|\boldsymbol{\omega}_0\|_{L^2}^2$ for all times. Obviously, as $\|\boldsymbol{\omega}_0\|_{L^2}^2 = \|\partial^1 \mathbf{v}_0\|_{L^2}^2$, the above inequality is exactly equivalent to the conclusion on the smallness of $\|\mathbf{v}_0\|_{L^2}^2 \|\partial^1 \mathbf{v}_0\|_{L^2}^2$.

Finally, we discuss what would be needed to obtain regular solutions for any arbitrary large initial conditions. We already know that $\|\mathbf{v}\|_{L^2}^2(t)$ is bounded for all times. We check at which conditions, we could find that $\|\partial^1\mathbf{v}\|_{L^2}^2$ is also bounded at all times. For this, we can express the upper bound of the 3 gradients of the velocity in a specific way. As previously, we start with:

$$\frac{1}{2}\frac{d}{dt}\|\partial^1\mathbf{v}\|_{L^2}^2 + \eta\|\boldsymbol{\nabla}^2\mathbf{v}\|_{L^2}^2 = -\int \partial_i v_j\, \partial_i v_\alpha\, \partial_\alpha v_j\, dV.$$

After some algebra (admitted here), we can obtain the following inequality:

$$-\int \partial_i v_j\, \partial_i v_\alpha\, \partial_\alpha v_j\, dV \leq \frac{\eta}{2}\left\|\boldsymbol{\nabla}^2\mathbf{v}\right\|_{L^2}^2 + C_\eta\|\partial^1\mathbf{v}\|_{L^2}^2\|\mathbf{v}\|_{L^4}^8 \quad (C_\eta > 0).$$

Again, we recall that the algebra needed in order to derive the last formula is presented in the appendix to this document. It requires the successive use of Hölder's, Sobolev's and scaled Young's inequalities.

Explicitly, we can write:

$$\|\mathbf{v}\|_{L^4}^8 = \|\mathbf{v}\|_{L^4}^8(t) = \left[\int \|\mathbf{v}(\mathbf{x},t)\|^4 d^3\mathbf{x}\right]^2.$$

If we could prove that this term is bounded for all times, $\|\mathbf{v}\|_{L^4}^8(t) \leq K$, then we would get:

$$\frac{1}{2}\frac{d}{dt}\|\partial^1\mathbf{v}\|_{L^2}^2 + \eta\|\boldsymbol{\nabla}^2\mathbf{v}\|_{L^2}^2 \leq \frac{\eta}{2}\left\|\boldsymbol{\nabla}^2\mathbf{v}\right\|_{L^2}^2 + C_\eta.K\,\|\partial^1\mathbf{v}\|_{L^2}^2.$$

Or:

$$\frac{1}{2}\frac{d}{dt}\|\partial^1\mathbf{v}\|_{L^2}^2 \leq -\frac{\eta}{2}\left\|\boldsymbol{\nabla}^2\mathbf{v}\right\|_{L^2}^2 + C_\eta.K\,\|\partial^1\mathbf{v}\|_{L^2}^2.$$

This would imply immediately that $\|\partial^1\mathbf{v}\|_{L^2}^2$ also is bounded for all times! And consequently we would conclude on the regularity of solutions of the Navier-Stokes equations (under the

hypothesis of the millennium problem). In fact, this is what we miss in the partial proofs. We would need to find a bound on the norm of the velocity field, not only in $L^2(IR^3)$, but also in $L^4(IR^3)$. An intuitive explanation is given below.

Beyond all these mathematical inequalities, we try to find an intuitive reason why this is so hard to prove the global regularity of solutions of the Navier-Stokes equations in the context of the millennium problem.

A key point is to recall that the Navier-Stokes equations are invariant under the transformation:

$$\mathbf{v}_\lambda(\mathbf{x}, t) = \lambda \mathbf{v}(\lambda \mathbf{x}, \lambda^2 t) \quad \text{and} \quad p_\lambda(\mathbf{x}, t) = \lambda^2 p_\lambda(\lambda \mathbf{x}, \lambda^2 t).$$

This means that if the velocity field $\mathbf{v}$ is a solution of the equations, then the velocity field $\mathbf{v}_\lambda$ is another acceptable solution (by construction). Also, this implies $\lambda \|\mathbf{v}_\lambda\|_{L^2}^2 = \|\mathbf{v}\|_{L^2}^2$ and therefore $\|\mathbf{v}_{1/\lambda}\|_{L^2}^2 = \lambda \|\mathbf{v}\|_{L^2}^2$. We can think of this transformation, with $\lambda \gg 1$, as taking the fine-scale behavior of the velocity field $\mathbf{v}$ and matching it with an identical (but rescaled and slowed down) coarse scale behavior of $\mathbf{v}_{1/\lambda}$.

Along the mathematical arguments given previously, our goal was to find upper bounds on the maximum energy $\max_{t \leq T} \|\mathbf{v}\|_{L^2}^2(t)$ and the cumulative energy dissipation $\int_{t=0}^{T} \|\partial^1 \mathbf{v}\|_{L^2}^2(t)\, dt$. We assume that such bounds exist: we label them respectively as M and C for the field velocity $\mathbf{v}$. Obviously, for $\mathbf{v}_{1/\lambda}$, these bounds become: $\lambda M$ and $\lambda C$. The last statement means that each time we have a solution of Navier-Stokes equations with bounds M and C, then the solution $\mathbf{v}_{1/\lambda}$ is possible, with worsened bounds $\lambda M$ and $\lambda C$. Blow-up can occur when the solution of Navier-Stokes equations shifts its energy into increasingly finer and finer scales, thus evolving more and more rapidly and eventually reaching a singularity in which the scales (in both space and time) tend towards zero. In such a scenario, we lose obviously the effectiveness of the bounds (and consequently the control) on the maximum energy and cumulative energy dissipation. For example, this is possible that at some time, a solution of the equations shifts its energy from a spatial scale $1/\lambda$ to $1/2\lambda$ in a time of order $1/\lambda^2$. And if this behavior repeats over and over again, this is clear that the solution is divergent. This simple argument gives an intuitive view of the problem.

At this point, we can come back to what would happen if we could get a bound on $\|\mathbf{v}\|_{L^4}^4(t)$ for all times. With the relation $\|\mathbf{v}_{1/\lambda}\|_{L^4}^4 = \frac{1}{\lambda}\|\mathbf{v}\|_{L^4}^4$, we understand that the bound in $L^4(IR^3)$ would become increasingly better for the velocity field $\mathbf{v}_{1/\lambda}$ when the parameter $\lambda$ is increased. Under such conditions, it would be possible to understand intuitively that blow-up could not occur.

### §15. Fluid mechanics in relativistic Heavy-Ions collisions

We recall that modern aspects of fluid mechanics cover more materials than liquid and gases in the non-relativistic approximation. As already stated, the idea of exploiting the laws of ideal fluid mechanics to describe the expansion of the strongly interacting nuclear matter that is formed in high energetic hadronic collisions was proposed in 1953 by Landau. Of course, in this context of particles moving at extremely high velocities, fluid mechanics has to be understood in its relativistic limit, which brings some subtleties to the calculations.

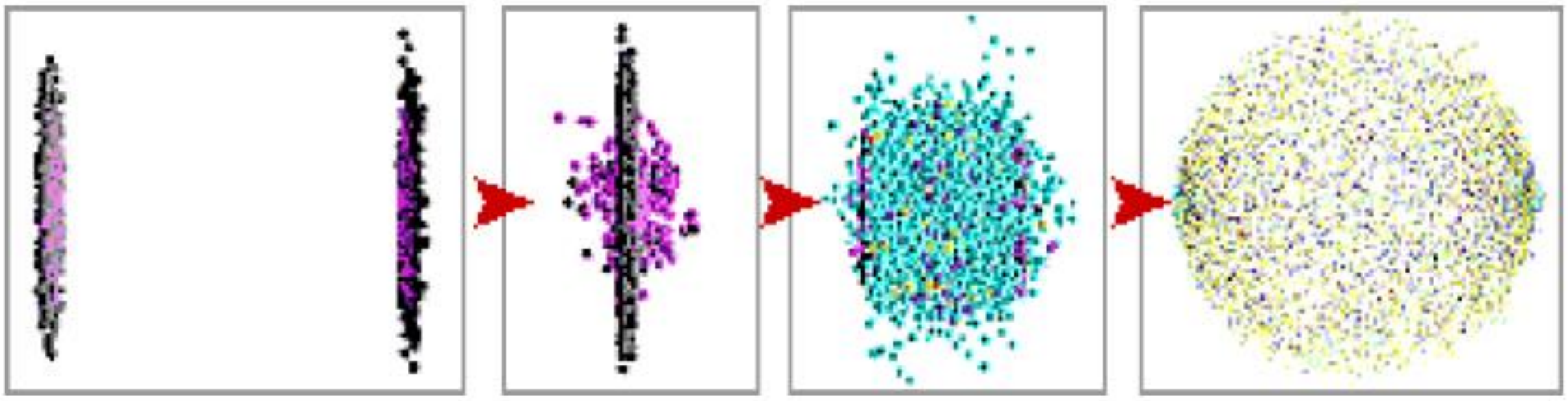

When colliding 2 fast moving heavy ions, built up with many charged particles (picture 1), the idea is that a zone of high density of charges will be formed (picture 2). Assuming local equilibrium and depending on the equation of state of this nuclear medium, this overlap region may reach the conditions that are described by an average (high) density and (high) temperature. The local thermal equilibrium means that the microscopic collision time scale is much shorter than any macroscopic evolution time scales. The interest of such collisions is that, having stored part of the available initial energy in compression and thermal excitation, the collisions produce unique conditions of the nuclear matter not accessible otherwise. In particular, this hot and dense matter is thought to be composed of strongly interacting quarks and gluons. The next stage of the reaction is the relaxation of the energy density. The central system is undergoing expansion, into the direction of largest gradients in density and temperature (pictures 3 and 4), thus reducing its temperature and density.

Recent experimental results (from 2005 to 2014) provide conclusive evidences that the created hot and dense matter produced in these collisions behaves collectively and has properties resembling that of a nearly ideal relativistic fluid, following the theoretical ideas formulated by Landau in 1953. We can write equations that describe the early stages of the expansion (pictures 2 and 3). The most important assumption under these equations is that the system can reach local thermodynamic equilibrium in a very short time. All thermodynamic quantities (see §3) of a fluid element can then be defined under this hypothesis, in the rest frame of the fluid element. The frame of reference is thus the local rest frame. Next, we assume that the energy density (E/V) and pressure admit a first order development around their initial values:

$$\epsilon(\mathrm{t},\mathbf{x}) = \epsilon_0(\mathrm{t},\mathbf{x}) + \delta\epsilon(\mathrm{t},\mathbf{x}) \text{ and } \mathrm{p}(\mathrm{t},\mathbf{x}) = \mathrm{p}_0(\mathrm{t},\mathbf{x}) + \delta\mathrm{p}(\mathrm{t},\mathbf{x}).$$

Neglecting second order terms, it can be shown that the relativistic equations of motion read (admitted here):

$$\frac{\partial \mathbf{v}}{\partial \mathrm{t}} = -\frac{1}{\epsilon + \mathrm{p}}\boldsymbol{\nabla}(\mathrm{p}).$$

The conservation of energy reads:

$$\frac{\partial \epsilon}{\partial \mathrm{t}} = -(\epsilon + \mathrm{p})\ \boldsymbol{\nabla}.\mathbf{v}.$$

The energy density is decreasing as the velocity field is diverging ($\boldsymbol{\nabla}.\mathbf{v} > \mathbf{0}$). Similarly, the charge density (n) conservation reads:

$$\frac{\partial \mathrm{n}}{\partial \mathrm{t}} = -(\mathrm{n})\, \boldsymbol{\nabla}.\mathbf{v}.$$

We have 6 unknowns (p, $\epsilon$, n, **v**) for 5 equations (stated above). We are missing the 6th equation, the equation of state:

$$\mathrm{p} = \mathrm{p}(\mathrm{n}, \epsilon).$$

This closes the system. In practice, this last equation can be expressed using the speed of sound:

$$\mathrm{c_S} = \frac{\partial \mathrm{p}(\mathrm{n}, \epsilon)}{\partial \epsilon}.$$

The last equation is not complete as we need to precise under which line the partial derivative is taken. This is a subtle issue. However, following ideas developed in the non-relativistic case, we can show that these lines correspond to: s/n=constant, where s is the entropy density (S/V). We can re-write the equations of motion as:

$$\frac{\partial \mathbf{v}}{\partial \mathrm{t}} = -\frac{\mathrm{c_s^2}}{1 + \mathrm{c_s^2}} \frac{\boldsymbol{\nabla}(\mathrm{p})}{\mathrm{p}}.$$

In fact, the ideal fluid behavior is only approximate: a (small) degree of viscosity is required for a proper description of the data (pictures 2 and 3) and the late stage of the expansion needs a dedicated treatment as it appears to be too dissipative for a macroscopic fluid dynamical approach. It must be described microscopically. The influence of the dissipative stage on measured observables is large as the system in expansion spends a large fraction of its history in this dissipative stage. Obviously, this makes the measurements quite difficult to interpret but this is another story.

## References (books)

Obviously, our presentation of fluid mechanics cannot be exhaustive, in particular concerning real fluids.

To our view, the best reference ever written on fluid mechanics is: L. D. Landau and E. M. Lifshitz [1968] Fluid Mechanics, Pergamon.

Another useful presentation of general fluid mechanics can be found in: G. K. Bachelor [1974] An Introduction to Fluid Dynamics, Cambridge University Press, London. For what concerns the regularity of solutions of Euler's or Navier Stokes equations, many developments can be found in: P. Constantin and C. Foias [1988] Navier-Stokes Equations, University of Chicago Press, Chicago and London; R. Temam [1984] Navier-Stokes Equations, volume 2, Elsevier Science Publishing Company, Amsterdam; or A. Bertozzi and A.J. Majda [2002] Vorticity and incompressible flow, Cambridge texts in applied mathematics.

## Appendix: useful mathematical inequalities

### Basic inequalities

For real numbers: $a \geq 0, b \geq 0$, integers p and $q > 0$:

$$\frac{1}{p} + \frac{1}{q} = 1 \rightarrow ab \leq \frac{a^p}{p} + \frac{b^q}{q} \text{ (Young's inequality).}$$

<u>Proofs</u>: We write: $ab = \exp[\log(a) + \log(b)]$, $a \geq 0, b \geq 0$. Then, by convexity of the exponential function $[e^{[\alpha x + (1-\alpha)y]} \leq \alpha e^x + (1-\alpha)e^y]$, we get:

$$ab = \exp\left[\frac{1}{p}.\log(a^p) + \frac{1}{q}.\log(b^q)\right] \leq \frac{1}{p}.\exp[\log(a^p)] + \frac{1}{q}.\exp[\log(b^q)], \text{ with } \frac{1}{q} = 1 - \frac{1}{p} \text{ and } a \geq 0, b \geq 0.$$

This completes the proof.

Another simple way to make the proof is to consider the real function for positive real numbers: $f(x) = \frac{x^p}{p} + \frac{1}{q} - x$. We can easily show that $f(x) \geq f(1) = 0$. Then, using the relation $x = a.b^{1/(p-1)}$, this completes the proof, under the condition $\frac{1}{p} + \frac{1}{q} = 1$.

From the Young's inequality, we can deduce immediately the following relation ($a, b \geq 0$): For any real number $\delta > 0$, there exists $C_\delta$ such that $ab \leq \delta a^p + C_\delta . b^q$, $C_\delta = \frac{1}{q}[\delta p]^{-\frac{q}{p}}$. We can write equivalently: for any $\varepsilon > 0$, $ab \leq \varepsilon \frac{a^p}{p} + [\varepsilon]^{-\frac{q}{p}} \frac{b^q}{q}$ $(a, b \geq 0)$.

From the Young's inequality, we can derive the important <u>Hölder's inequality</u>. For sets of real numbers $(a_i)$ and $(b_i)$, and positive integers p and q, we have:

$$\frac{1}{p} + \frac{1}{q} = 1 \rightarrow \sum_{i=1}^{n} |a_i b_i| \leq \left(\sum_{i=1}^{n} |a_i|^p\right)^{\frac{1}{p}} \left(\sum_{i=1}^{n} |b_i|^q\right)^{\frac{1}{q}} \text{ (Hölder's inequality).}$$

<u>Proof</u>: We define: $A = |a_i| / (\sum_{i=1}^{n} |a_i|^p)^{\frac{1}{p}}$ and $B = |b_i| / (\sum_{i=1}^{n} |b_i|^q)^{\frac{1}{q}}$. Then, we apply the Young's inequality to the product AB and sum over the index i. This gives the result, under the condition: $\frac{1}{p} + \frac{1}{q} = 1$.

Obviously, this holds also for measurable real valued functions f and g ($\Omega \to IR$):

$$\|f.g\|_{L^1(\Omega)} = \int |f.g|\, dx \le \left(\int |f|^p\, dx\right)^{\frac{1}{p}} \left(\int |g|^q\, dx\right)^{\frac{1}{q}} = \|f\|_{L^p(\Omega)} \|g\|_{L^q(\Omega)} \quad \left(\text{with } \frac{1}{p} + \frac{1}{q} = 1\right).$$

Here the proof relies on the definitions of $A = |f| / \|f\|_{L^p(\Omega)}$ and $B = |g| / \|g\|_{L^p(\Omega)}$.

The Hölder's inequality, for p=q=2, corresponds to the well-known Cauchy-Schwarz inequality. It can be formulated explicitly:

$$(a_1^2 + a_2^2 + \cdots + a_n^2)(b_1^2 + b_2^2 + \cdots + b_n^2) \ge (a_1 b_1 + \cdots a_n b_n)^2.$$

For this particular case, a simple geometric proof can be done. We note **u** the vector $(a_1, a_2, \dots, a_n)$ and $\mathbf{v}(b_1, b_2, \dots, b_n)$, then:

$$\|\mathbf{u}.\mathbf{v}\|^2 = \|\mathbf{u}\|^2 \|\mathbf{v}\|^2 \cos(\mathbf{u}, \mathbf{v}) \le \|\mathbf{u}\|^2 \|\mathbf{v}\|^2.$$

This gives the result. Similarly, for real valued functions, we obtain (p=q=2):

$$\left(\int f.g\, dx\right)^2 \le \left(\int f^2 dx\right) \left(\int g^2 dx\right) .$$

### Applications of the Hölder's inequality

For any measurable real valued functions f and g ($\Omega \to IR$), the Hölder's inequality reads:

$$\|f.g\|_{L^1(\Omega)} = \int |f.g|\, dx \le \left(\int |f|^p\, dx\right)^{\frac{1}{p}} \left(\int |g|^q\, dx\right)^{\frac{1}{q}}$$
$$= \|f\|_{L^p(\Omega)} \|g\|_{L^q(\Omega)} \quad \left(\text{with } p, q \ge 1: \frac{1}{p} + \frac{1}{q} = 1\right).$$

The integrals are taken over the set $\Omega$. In the most general case, $\Omega$ can be equal to the set $IR^n$, where n is the dimension of space, provided that the integrals are well-defined. The set $\Omega$ can also be a bounded subset of $IR^n$. For n dimensions of space, the integral $\int |f.g|\, dx$ becomes $\int |f.g|\, dV$, where $dV = dx_1 \dots dx_n$ is the infinitesimal volume element $\Omega$. Also, we have: $\|f\|_{L^p(\Omega)} = (\int |f|^p\, dV)^{\frac{1}{p}}$. Note that the inequality holds also for the particular case $p = \infty$, then $q = 1$. We obtain:

$$\|f.g\|_{L^1(\Omega)} \leq \|f\|_{L^\infty(\Omega)} \|g\|_{L^1(\Omega)}.$$

Here, $\|f\|_{L^\infty(\Omega)} = \{\text{smallest } C \geq 0, f \leq C \text{ for (almost) every point in } \Omega\}$. This means that $\|f\|_{L^\infty(\Omega)}$ is the upper bound of the function f in $\Omega$. The above inequality is thus obvious.

The Hölder's inequality can be immediately generalized to different cases, assuming that $\Omega = IR^n$ or a bounded subset of it.

(i) $\frac{1}{p} + \frac{1}{q} + \frac{1}{r} = 1$ (with $p, q, r \geq 1$), $f \in L^p(\Omega)$, $g \in L^q(\Omega)$ and $h \in L^r(\Omega)$, then

$$\|f.g.h\|_{L^1(\Omega)} \leq \|f\|_{L^p(\Omega)} \|g\|_{L^q(\Omega)} \|h\|_{L^r(\Omega)}.$$

(ii) $\frac{1}{p} + \frac{1}{q} = \frac{1}{r}$ (with $p, q, r \geq 1$), $f \in L^p(\Omega)$ and $g \in L^q(\Omega)$, then $f.g \in L^r(\Omega)$ and

$$\|f.g\|_{L^r(\Omega)} \leq \|f\|_{L^p(\Omega)} \|g\|_{L^q(\Omega)}.$$

(iii) $\frac{\theta}{p} + \frac{1-\theta}{q} = \frac{1}{r}$ (with $p \leq r \leq q$ and $0 \leq \theta \leq 1$), $f \in L^p(\Omega) \cap L^q(\Omega)$, then $f \in L^r(\Omega)$, and

$$\|f\|_{L^r(\Omega)} \leq \|f\|_{L^p(\Omega)}^{\theta} \|f\|_{L^q(\Omega)}^{1-\theta}.$$

This means that we can interpolate between higher and lower $L^p(\Omega)$ sets to get something in between. That's the reason why we can call this relation an interpolation inequality for $L^p(\Omega)$ sets.

The proof of the last relation follows from the standard Hölder's inequality after writing:

$$\|f\|^r_{L^r(\Omega)} = \int_\Omega |f|^r\, dV = \int_\Omega |f|^{\theta r}\, |f|^{(1-\theta)r} dV \leq \left\||f|^{\theta r}\right\|_{L^{\frac{p}{\theta r}}(\Omega)} \left\||f|^{(1-\theta)r}\right\|_{L^{\frac{q}{(1-\theta)r}}(\Omega)}.$$

This gives the result.

When the set $\Omega$ is bounded (in $IR^n$), these inequalities become very useful to derive important links between $L^p(\Omega)$ norms (integrability) and regularity of functions. For example, if $\Omega$ is an interval [a, b] in IR (n=1), then we get the following properties.

(i) If $q \geq p$, then there exists a constant K which depends on $\Omega, p, q$ such that:

$$\|f\|_{L^p(\Omega)} \leq K\, \|f\|_{L^q(\Omega)} \quad (\, q \geq p \geq 1).$$

This can also be written as:

$$L^q(\Omega) \subset L^p(\Omega) \quad (\, q \geq p \geq 1).$$

Explicitly, $L^q(\Omega)$ is embedded in $L^p(\Omega)$. A set of functions X is said to be embedded in the set of functions Y (written as $X \subset Y$) if all elements (functions) in X are also in Y. In addition, the relation between the norms in both sets X and Y is as written above.

Namely, $L^q(\Omega) \subset L^p(\Omega)$ $(\, q \geq p \geq 1)$ is equivalent to the inequality

$$\|f\|_{L^p(\Omega)} \leq K\, \|f\|_{L^q(\Omega)},$$

for some constant $K(\Omega, p, q)$. $L^q(\Omega)$ is embedded in $L^p(\Omega)$ can also be stated as: $L^p(\Omega)$ contains $L^q(\Omega)$, or $L^p(\Omega) \supset L^q(\Omega)$ $(\, q \geq p \geq 1)$. We let this proof as an exercise (below).

This relation means that higher $L^p(\Omega)$ norms can control lower ones ($\|f\|_{L^p(\Omega)} \leq K\, \|f\|_{L^{q\geq p}(\Omega)}$). In particular, this implies that functions that belong to higher $L^p(\Omega)$ spaces are more regular. This is only correct for a bounded set $\Omega$.

(ii) This also implies that, for any bounded set $\Omega$:

$$L^1(\Omega) \supset L^{p\geq 1}(\Omega) \supset L^{q\geq p}(\Omega) \supset L^\infty(\Omega).$$

$L^1(\Omega)$ is the larger set. In particular for a bounded set $L^1(\Omega) \supset L^\infty(\Omega)$, which means that once $\int_\Omega |f|\, dV$ is finite, then the function $f: \Omega \to IR^n$ is bounded up.

Exercise: Prove that, for a bounded set $\Omega$ and $q \geq p \geq 1$, then there exists a constant K which depends on $\Omega$, p and q such that: $\|f\|_{L^p(\Omega)} \leq K \|f\|_{L^q(\Omega)}$ (for $q \geq p$).

Solution: We have:

$$\|f\|_{L^p(\Omega)} = \left(\int_\Omega 1.\, |f|^p\, dV\right)^{\frac{1}{p}} \leq \left(\int_\Omega 1^r\, dV\right)^{\frac{1}{r}} \left(\int_\Omega |f|^q\, dV\right)^{\frac{1}{q}}$$

with $\frac{1}{p} - \frac{1}{q} = \frac{1}{r}$ (Hölder's inequality).

This gives the result, with $K = \left(\int_\Omega 1\, dV\right)^{\frac{1}{p}-\frac{1}{q}}$ (since $\Omega$ is bounded).

**First Sobolev's inequalities**

We now intend to present some links between the integrability and the differentiability of these functions. This is an essential task as, at the end, we aim to study solutions of partial differential equations (for example in the context of fluid mechanics). The general idea is that it seems to be more difficult to obtain the differentiability than the integrability. We need to develop this idea quantitatively with precise definitions. That's the reason why we will define some sets of functions which involve and index for differentiability and another index for integrability.

We recall some basic definitions. In order to present first the formulae using simple notations, we start with single real valued functions $h: IR \to IR$. We assume that the derivatives of h up to order k are well-defined. We note $\partial^m h$ $(m \leq k)$, these derivatives.

Then, we define what is called a Sobolev's set $W^{k,p}(IR)$ as the set of functions $h: IR \rightarrow IR$ such that $h \in L^p(IR)$ and $\partial^{m\leq k} h \in L^p(IR)$. The second condition means that the derivatives of h, up to order k, belong to $L^p(IR)$. As mentioned above, one index (k) stands for the differentiability and one index (p) for the integrability.

We note:

$$W^{k,p}(IR) = \{h: IR \rightarrow IR \text{ such that } \partial^m h \in L^p(IR) \text{ for all } 0 \leq m \leq k\}. \quad (\text{with } \partial^0 h = h)$$

In parallel, we can define a norm on this set as:

$$\|h\|^p_{W^{k,p}(IR)} = \sum_{m\leq k} \|\partial^m h\|^p_{L^p(IR)} \,.$$

Next, this definition can be extended immediately to functions $f: IR^n \rightarrow IR$ (where n –the dimensions of space- can be larger than 1). We write the formulae for the particular case n=3, corresponding to 3-dimensional spaces (3D).

Let $(m_1, m, m_3)$ be a set of 3 integers, such that $m = m_1 + m_2 + m_3$. We assume that the derivatives of the function f are well-defined up to rank k, using the notation:

$$\partial^m f = \partial^m_{\mathbf{x}} f = \partial^{m_1}_x \partial^{m_2}_y \partial^{m_3}_z f \qquad \left\{ \text{with } \partial^{m_1}_x = \left(\frac{\partial}{\partial x}\right)^{m_1} \right\}.$$

The Sobolev's set $W^{k,p}(IR^3)$ of such functions is defined as:

$$W^{k,p}(IR^3) = \{f: IR^3 \rightarrow IR \text{ such that } \partial^m f \in L^p(IR^3) \text{ for all } 0 \leq m = m_1 + m_2 + m_3 \leq k\}.$$

In parallel, we can define a norm on this set as:

$$\|f\|^p_{W^{k,p}(IR^3)} = \sum_{m_1+m_2+m_3\leq k} \|\partial^m f\|^p_{L^p(IR^3)} \,.$$

With:

$$\|\partial^M f\|^p_{L^p(IR^3)} = \int_{IR^3} \sum_{m=m_1+m_2+m_3=M} |\partial^m f|^p \; dV$$

Obviously, we observe that the definitions for dimension 3 (or more generally n) follow exactly those for dimension 1. Note that the case $p = \infty$, is also well-defined with the definition:

$$\|f\|_{W^{k,\infty}(IR^n)} = \max_{m_1+\cdots+m_n \le k} \|\partial^m f\|_{L^\infty(IR^n)} \,.$$

The particular case p=2 is essential for the following discussion. We write:

$$H^k(IR^n) = W^{k,2}(IR^n)$$

Explicitly:

$$H^k(IR^n) = \left\{ \begin{matrix} f\colon IR^n \to IR \text{ such that } f \in L^2(IR^n) \text{ and } \partial^m f \in L^2(IR^n) \\ \text{for all } 1 \le m = m_1 + m_2 + \cdots + m_n \le k \end{matrix} \right\}.$$

We also use the name Sobolev's set for $H^k(IR^n)$. Moreover, $H^k(IR^n)$ is a Hilbert's space with the inner product:

$$\langle u, v\rangle_{H^k(IR^n)} = \int_{IR^n} dV \left[ \sum_{m_1+\cdots+m_n \le k} \partial^m u \ \partial^m v \right].$$

Where u and v are functions that belong to $H^k(IR^n)$.

Note that $H^s(\Omega)$ can also be defined for real values of s. In this case, we need to generalize the formula above. For simplicity, we provide the new definition for single real valued function:

$$\|f\|^2_{H^s(IR)} = \int (1 + |\xi|^2)^s \left|\hat{f}(\xi)\right|^2 d\xi.$$

Here, $\hat{f}$ is the Fourier transform of the function f. Then the definition of $H^s(IR)$ can (also) be written as:

$$H^s(IR) = \left\{f\colon IR \to IR \text{ such that } f \in L^2(IR) \text{ and } (1 + |\xi|^2)^{s/2}\hat{f} \in L^2(IR)\right\}.$$

Extension to dimension 3 (or n) is immediate.

We give below 4 important Sobolev's inequalities that follow the previous definitions and properties. In general, any inequality which trades differentiability and integrability of functions is called a Sobolev's inequality. We understand below that this trade is one way: we cannot generally sacrifice integrability to gain differentiability (as already mentioned). In the following, we always assume that the functions are vanishing at the boundary of $\Omega$. If $\Omega = \mathrm{IR}^n$, this means that functions are assumed to vanish at infinity. The theorems are not correct otherwise. Also, we systematically mention whether the inequalities are given for $\Omega = \mathrm{IR}^n$ or for $\Omega = \{\text{bounded subset of } \mathrm{IR}^n\}$.

i. From the relation $(1+|\xi|^2)^r\left|\hat{f}(\xi)\right|^2 \leq (1+|\xi|^2)^s\left|\hat{f}(\xi)\right|^2 \quad (s \geq r)$, it follows the first embedding theorem for Sobolev's sets: $H^s(\Omega) \subset H^r(\Omega) \quad (s \geq r)$.

ii. For any functions, $u \in W^{1,p}(\mathrm{IR}^n) \ (1 \leq p < n)$, then $u \in L^{np/(n-p)}(\mathrm{IR}^n)$ and there exists a constant $C(n,p)$ such that:

$$\|u\|_{L^{np/(n-p)}(\mathrm{IR}^n)} \leq C(n,p)\,.\,\|\partial u\|_{L^p(\mathrm{IR}^n)}.$$

This implies also the embedding relation:

$$\|u\|_{L^{np/(n-p)}(\mathrm{IR}^n)} \leq C(n,p)\,.\,\|u\|_{W^{1,p}(\mathrm{IR}^n)}.$$

Or equivalently:

$$W^{1,p}(\mathrm{IR}^n) \ \subset L^{\frac{np}{n-p}}(\mathrm{IR}^n).$$

iii. More generally, if $u \in W^{1,p}(\mathrm{IR}^n) \ (p < n)$ and $p \leq q \leq np/(n-p)$, then $u \in L^q(\mathrm{IR}^n)$ and there exists a constant $C(n,q,p)$ such that:

$$\|u\|_{L^q(\mathrm{IR}^n)} \leq C(n,q,p)\,.\,\|\partial u\|_{L^p(\mathrm{IR}^n)}.$$

For $q < \frac{np}{n-p}$, the embedding relation holds only for a bounded subset of $\mathrm{IR}^n$ $(\Omega)$:

$$\|u\|_{L^q(\Omega)} \leq C(n,q,p)\,.\,\|u\|_{W^{1,p}(\Omega)}.$$

Or equivalently:

$$W^{1,p}(\Omega) \ \subset L^q(\Omega).$$

iv. The last relation can be generalized to $W^{k,p}(IR^n)$ with the updated result: for any functions, $u \in W^{k,p}(IR^n)$ $(kp < n)$, then $u \in L^{np/(n-kp)}(IR^n)$ and

$$\|u\|_{L^{np/(n-kp)}(IR^n)} \leq C(n,k,p) \,.\, \|\partial^k u\|_{L^p(IR^n)}.$$

This implies also the embedding relation, for any function $u \in W^{k,p}(IR^n)$:

$$\|u\|_{L^{np/(n-kp)}(IR^n)} \leq C(n,p) \,.\, \|u\|_{W^{k,p}(IR^n)}.$$

Or equivalently:

$$W^{k,p}(IR^n) \subset L^{\frac{np}{n-kp}}(IR^n).$$

A simple view of the above Sobolev's inequalities can be given assuming a certain behavior for the function u. Let us assume that u is a bump of size H and length L and this for all dimensions of the space $IR^n$. This means that:

$$\|u\|_{L^q(IR^n)} \sim [H^q L^n]^{1/q} \quad \text{and} \quad \|\partial u\|_{L^p(IR^n)} \sim [(H/L)^p L^n]^{1/p}.$$

Then, we can verify at which conditions we can get a Sobolev's inequality (involving $\|u\|_{L^q(IR^n)}$ and $\|\partial u\|_{L^p(IR^n)}$) for all L and H:

$$\|u\|_{L^q(IR^n)} \leq C \,.\, \|\partial u\|^a_{L^p(IR^n)}.$$

Equivalently, we can write:

$$HL^{n/q} \leq C.\ H^a L^{(n-p)a/p}.$$

This implies that: a=1 and $\frac{1}{q} = \frac{1}{p} - \frac{1}{n}$ or $q = pn/(n-p)$.

This method provides a general intuitive trick to re-derive quickly Sobolev's inequalities.

Exercise: Prove item (ii). For any functions, $u \in W^{1,p}(IR^n)$ $(1 \leq p < n)$, then $u \in L^{np/(n-p)}(IR^n)$ and there exists a constant $C(n, p)$ such that:

$$\|u\|_{L^{np/(n-p)}(IR^n)} \leq C(n, p) \, . \, \|\partial u\|_{L^p(IR^n)}.$$

Solution: Essentially, the proof uses the Hölder's inequality several times. This kind of proof is not difficult *per se* but requires a good organization. First, we need to show the result in the simpler case: p=1. Then, it will be possible to generalize (see later) to the general case for all p values. For the case p=1, we need to show that, for any function $u \in W^{1,1}(IR^n)$ $(1 < n)$, there exists a constant $C(n)$ such that:

$$\|u\|_{L^{n/(n-1)}(IR^n)} \leq C(n) \, . \, \|\partial u\|_{L^1(IR^n)}.$$

Explicitly, we need to derive the formula:

$$\int |u|^{n/(n-1)} \, dV \leq C(n) \left( \int |\partial u| \, dV \right)^{\frac{n}{n-1}}.$$

Where the integrals are taken over the all space $IR^n$ and thus, dV is a volume element of $IR^n$. The first step is to make a derivative appear using a simple identity:

$$u(x_1, \dots, x_n) = \int_{-\infty}^{x_j} \partial_j u(\dots, a_j, \dots) da_j.$$

The derivative inside the integral is taken on the $j^{th}$ variable of the function, and this is this variable which is integrated over. Then, taking the absolute values on both sides, the inequality follows:

$$|u(x_1, \dots, x_n)| \leq \int_{-\infty}^{x_j} \left|\partial_j u(\dots, a_j, \dots)\right| da_j.$$

Repeating this process n-times for all variables of the function $u(x_1, \dots, x_n)$, we obtain a product from j=1 to n. At this stage, we get:

$$|u(x_1, \dots, x_n)|^{n/(n-1)} \leq \left[ \prod_{j=1}^{n} \int_{-\infty}^{x_j} \left|\partial_j u(\dots, a_j, \dots)\right| da_j \right]^{1/(n-1)}.$$

As the integrand on the right hand side is positive (absolute value), all integrals can be extended up to $+\infty$. Hence, we can write:

$$|u|^{n/(n-1)} \leq \left[\prod_{j=1}^{n} \int_{-\infty}^{\infty} |\partial_j u| da_j\right]^{\frac{1}{n-1}} = \prod_{j=1}^{n} \left[\int_{-\infty}^{\infty} |\partial_j u| da_j\right]^{1/(n-1)}.$$

The next step of the proof consists in integrating this inequality, over $x_1, x_{2,\dots,} x_n$, in order to derive a bound for $\int |u|^{n/(n-1)}\, dV$. We start by integrating over $x_1$:

$$\int_{-\infty}^{\infty} |u|^{n/(n-1)}.dx_1 \leq \int_{-\infty}^{\infty} \prod_{j=1}^{n} \left[\int_{-\infty}^{\infty} |\partial_j u| da_j\right]^{\frac{1}{n-1}}.dx_1.$$

But, for what concerns the integral over $x_1$of the right hand side part of the inequality, all terms that do not depend on $x_1$are thus constant of $x_1$, and can then be taken out of the integral over $x_1$. This is obviously the case for $\left[\int_{-\infty}^{\infty} |\partial_1 u| da_1\right]^{\frac{1}{n-1}}$ in which the variable $x_1$is integrated over. We obtain:

$$\int_{-\infty}^{\infty} \prod_{j=1}^{n} \left[\int_{-\infty}^{\infty} |\partial_j u| da_j\right]^{\frac{1}{n-1}}.dx_1 = \left[\int_{-\infty}^{\infty} |\partial_1 u| da_1\right]^{\frac{1}{n-1}} \int_{-\infty}^{\infty} \prod_{j=2}^{n} \left[\int_{-\infty}^{\infty} |\partial_j u| da_j\right]^{\frac{1}{n-1}}.dx_1.$$

This is the stage where the Hölder's inequality can be applied to $\int_{-\infty}^{\infty} \prod_{j=2}^{n} \left[\int_{-\infty}^{\infty} |\partial_j u| da_j\right]^{\frac{1}{n-1}}.dx_1$. We recall that:

$$\left|\int f_1.f_2 \dots f_n\, dx\right| \leq \left(\int |f_1|^p\, dx\right)^{\frac{1}{p_1}} \dots \left(\int |f_n|^q\, dx\right)^{\frac{1}{p_n}} \text{ with } \frac{1}{p_1} + \dots + \frac{1}{p_n} = 1.$$

This leads to:

$$\int_{-\infty}^{\infty} \prod_{j=2}^{n} \left[\int_{-\infty}^{\infty} |\partial_j u| da_j\right]^{\frac{1}{n-1}}.dx_1 \leq \left[\prod_{j=2}^{n} \iint_{-\infty}^{\infty} |\partial_j u| da_j\, dx_1\right]^{\frac{1}{n-1}}.$$

All together at this stage, we have:

$$\int_{-\infty}^{\infty} |u|^{n/(n-1)}.dx_1 \leq \left[\int_{-\infty}^{\infty} |\partial_1 u| da_1\right]^{\frac{1}{n-1}}.\left[\prod_{j=2}^{n} \iint_{-\infty}^{\infty} |\partial_j u|\, dx_1 da_j\right]^{\frac{1}{n-1}}.$$

Following the same strategy, as mentioned above, we now integrate the last inequality over $x_2$:

$$\int_{-\infty}^{\infty} |u|^{n/(n-1)} . dx_1 dx_2$$

$$\leq \left[\iint_{-\infty}^{\infty} |\partial_2 u| dx_1 da_2\right]^{\frac{1}{n-1}} . \int_{-\infty}^{\infty} \left[\int_{-\infty}^{\infty} |\partial_1 u| da_1\right]^{\frac{1}{n-1}} \prod_{j=3}^{n} \left[\iint_{-\infty}^{\infty} |\partial_j u| dx_1 da_j\right]^{\frac{1}{n-1}} dx_2.$$

Again, for what concerns the integral over $x_2$of the right hand side part of the inequality, all terms that do not depend on $x_2$are thus constant of $x_2$, and can then be taken out of the integral over $x_2$. This is the case for $[\iint_{-\infty}^{\infty} |\partial_2 u| dx_1 da_2]^{\frac{1}{n-1}}$ in which the variable $x_2$ is integrated over.

Be careful: this is not $[\iint_{-\infty}^{\infty} |\partial_1 u| da_1 dx_2]^{\frac{1}{n-1}}$. The term $[\int_{-\infty}^{\infty} |\partial_1 u| da_1]^{\frac{1}{n-1}}$ is still in the integral over $x_2$ as this is a function which depends on $x_2$. However, following the same line of process as before, we can apply the Hölder's inequality to:

$\int_{-\infty}^{\infty} [\int_{-\infty}^{\infty} |\partial_1 u| da_1]^{\frac{1}{n-1}} \prod_{j=3}^{n} [\iint_{-\infty}^{\infty} |\partial_j u| dx_1 da_j]^{\frac{1}{n-1}} dx_2$.

Then obtain:

$$\int_{-\infty}^{\infty} |u|^{n/(n-1)} . dx_1 dx_2$$

$$\leq \left[\iint_{-\infty}^{\infty} |\partial_2 u| dx_1 da_2\right]^{\frac{1}{n-1}} . \left[\iint_{-\infty}^{\infty} |\partial_1 u| da_1 dx_2\right]^{\frac{1}{n-1}} . \left[\prod_{j=3}^{n} \iint_{-\infty}^{\infty} |\partial_j u| \, dx_1 dx_2 da_j\right]^{\frac{1}{n-1}} .$$

We can follow exactly the same process by integrating over all the other variables. We end up with:

$$\int |u|^{n/(n-1)} \, dV \leq \left(\int |\partial u| \, dV\right)^{\frac{n}{n-1}}.$$

This completes the proof for p=1. For any p values, we need to prove that:

$$\left[\int |u|^{pn/(n-p)}\, dV\right]^{(n-p)/(np)} \leq C(n,p)\left(\int |\partial u|^p\, dV\right)^{1/p}.$$

The idea is to simply apply the relation derived for p=1, not to the function u, but to $|u|^{\gamma}$.

The following inequality holds:

$$\int |u|^{\gamma n/(n-1)}\, dV \leq \left(\int |\partial |u|^{\gamma}|\, dV\right)^{\frac{n}{n-1}} = \left(\gamma \int |u|^{\gamma-1}|\partial u|\, dV\right)^{\frac{n}{n-1}}.$$

Using the Hölder's inequality, we can express:

$$\gamma \int |u|^{\gamma-1}|\partial u|\, dV \leq \gamma \left[\int |u|^{(\gamma-1)q}\, dV\right]^{\frac{1}{q}} \left[\int |\partial u|^p\, dV\right]^{\frac{1}{p}} \text{ with } \frac{1}{p}+\frac{1}{q}=1.$$

By choosing $\frac{\gamma n}{n-1} = (\gamma-1)q$, we have:

$$\left[\int |u|^{\frac{\gamma n}{n-1}}\, dV\right]^{1-n/q(n-1)} \leq \gamma^{\frac{n}{n-1}} \left[\int |\partial u|^p\, dV\right]^{\frac{n}{p(n-1)}}.$$

Using the relations $\frac{1}{p}+\frac{1}{q}=1$ and $\frac{\gamma n}{n-1} = (\gamma-1)q$, wecan see immediately that the last inequality is equivalent to:

$$\left[\int |u|^{pn/(n-p)}\, dV\right]^{(n-p)/(np)} \leq C(n,p)\left(\int |\partial u|^p\, dV\right)^{1/p}.$$

Note that it is standard to write:

$$p^* = \frac{pn}{n-p}.$$

Then, we can write:

$$\left[\int |u|^{p^*}\, dV\right]^{1/p^*} \leq C(n,p)\left(\int |\partial u|^p\, dV\right)^{1/p}.$$

**Poincaré's inequality**

For any functions, $u \in W^{1,p}(\Omega)$ ($\Omega$ bounded subset of $IR^n$), there exists a constant $C(n,p)$ such that:

$$\|u\|_{L^p(\Omega)} \leq C(n,p) \,.\, \|\partial u\|_{L^p(\Omega)}.$$

This is called the Poincaré's inequality. The general idea is that this inequality gives a way to estimate the $L^p(\Omega)$ norm of a function in terms of the $L^p(\Omega)$ norm of its derivative. Explicitly, this relation reads:

$$\int_\Omega |u(x_1, \dots, x_n)|^p \, dV \leq C(n,p) \,.\, \int_\Omega \sum_{i=1}^{n} \left|\frac{\partial u}{\partial x_i}\right|^p \, dV \,.$$

**Further Sobolev's theorems**

We have already presented an interpolation inequality for $L^p(\Omega)$ sets:
for any function $f \in L^p(\Omega) \cap L^q(\Omega)$, then $f \in L^r(\Omega)$ and $\|f\|_{L^r(\Omega)}$ verifies:

$$\|f\|_{L^r(\Omega)} \leq \|f\|_{L^p(\Omega)}^{\theta} \|f\|_{L^q(\Omega)}^{1-\theta} \text{ provided that } \frac{\theta}{p} + \frac{1-\theta}{q} = \frac{1}{r}$$

with $1 \leq p \leq r \leq q < \infty$ and $0 \leq \theta \leq 1$.

Similarly, there exists an interpolation inequality for Sobolev's sets $H^p(IR^n)$. For any real values $0 < s_1 \leq s_2 < \infty$, and for any $s \in [s_1, s_2]$

such that $s = \theta s_1 + (1-\theta)s_2$ with $0 \leq \theta \leq 1$, we have:

If $u \in H^{s_1}(IR^n) \cap H^{s_2}(IR^n)$, then $u \in H^s(IR^n)$, and

$$\|u\|_{H^s(IR^n)} \leq \|u\|_{H^{s_1}(IR^n)}^{\theta} \|u\|_{H^{s_2}(IR^n)}^{1-\theta} \qquad (s = \theta s_1 + (1-\theta)s_2)$$

This means that we can interpolate between higher and lower $H^s(\Omega)$ sets to get something in between.

We now discuss a final useful inequality, usually called the Gagliardo-Nirenberg-Sobolev's inequality. This is also an interpolation inequality, complementary of the previous relation between sets $H^s(IR^n)$, $H^{s_1}(IR^n)$ and $H^{s_2}(IR^n)$. It follows closely the interpolation inequality for $L^p(\Omega)$ sets, recalled above.

For any function $u \in W^{l,p}(IR^n) \cap W^{j,q}(IR^n)$, then $u \in W^{k,r}(IR^n)$ and there exists a universal constant C>0 (independent of u) such that:

$$\|\partial^k u\|_{L^r(IR^n)} \leq C.\, \|\partial^l u\|_{L^p(IR^n)}^{\theta} \|\partial^j u\|_{L^q(IR^n)}^{1-\theta} \text{ provided that } \frac{1}{r} - \frac{k}{n} = \theta\left(\frac{1}{p} - \frac{l}{n}\right) + (1-\theta)\left(\frac{1}{q} - \frac{j}{n}\right).$$

This is called the Gagliardo-Nirenberg-Sobolev's inequality or simply the interpolation inequality. Remark that it holds for $0 \leq \theta \leq 1, j \leq k < l$ and $1 \leq p \leq r \leq q < \infty$.
For a bounded subset of $IR^n$ $(\Omega)$, we have in addition:

$$\|u\|_{W^{k,r}(\Omega)} \leq C\|u\|_{W^{l,p}(\Omega)}^{\theta} \|u\|_{W^{j,q}(\Omega)}^{1-\theta}.$$

We can apply this interpolation inequality to a particular (interesting) case:

$$\|\partial^1 u\|_{L^p(IR^n)} \leq C\|\partial^2 u\|_{L^p(IR^n)}^{\frac{1}{2}} \|u\|_{L^p(IR^n)}^{\frac{1}{2}}.$$

Here, we can use the scaled Young's inequality to $C\|\partial^2 u\|_{L^p(IR^n)}^{\frac{1}{2}} \|u\|_{L^p(IR^n)}^{\frac{1}{2}}$.

$$\text{For any } \varepsilon > 0, \quad ab \leq \varepsilon \frac{a^p}{p} + [\varepsilon]^{-\frac{q}{p}} \frac{b^q}{q} \qquad (a, b \geq 0).$$

Hence, there exists a constant δ>0 such that:

$$\|\partial u\|_{L^p(IR^n)} \leq \delta\|\partial^2 u\|_{L^p(IR^n)} + C_\delta \|u\|_{L^p(IR^n)}.$$

This means that the $L^p$ norm of $\partial u$ is controlled by the $L^p$ norms of $\partial^2 u$ and u.